\documentclass[%
 reprint,
superscriptaddress,
 amsmath,amssymb,
 aps,
pra,
prb,
rmp,
prstab,
prstper,
]{revtex4-1}

\usepackage{graphicx}
\usepackage{dcolumn}
\usepackage{bm}
\usepackage{placeins}
\usepackage{xcolor}
\usepackage{ulem}

\newcommand{\beginsupplement}{%
        \setcounter{equation}{0}
        \renewcommand{\theequation}{S\arabic{equation}}%
        \setcounter{table}{0}
        \renewcommand{\thetable}{S\arabic{table}}%
        \setcounter{figure}{0}
        \renewcommand{\thefigure}{S\arabic{figure}}%
     }

\begin{document}

\preprint{APS/123-QED}

\title{Topological Crystalline Phases in a Disordered Inversion-Symmetric Chain}

\author{Saavanth Velury}
\author{Barry Bradlyn}
\author{Taylor L. Hughes}
\affiliation{Department of Physics and Institute of Condensed Matter Theory, University of Illinois at Urbana-Champaign, IL 61801, USA}

\date{\today}

\begin{abstract}
When translational symmetry is broken by bulk disorder, the topological nature of states in topological crystalline systems may change depending on the type of disorder that is applied. In this work, we characterize the phases of a one-dimensional (1D) chain with inversion and chiral symmetries, where every disorder configuration is inversion-symmetric. By using a basis-independent formulation for the inversion topological invariant, chiral winding number, and bulk polarization, we are able to construct phase diagrams for these quantities when disorder is present. We show that unlike the chiral winding number and bulk polarization, the inversion topological invariant can fluctuate when the bulk spectral gap closes at strong disorder. Using the position-space renormalization group, we are able to compare how the inversion topological invariant, chiral winding number and bulk polarization behave at low energies in the strong disorder limit. We show that with inversion symmetry-preserving disorder, the value of the inversion topological invariant is determined by the inversion eigenvalues of the states at the inversion centers, while quantities such as the chiral winding number and the bulk polarization still have contributions from every state throughout the chain. We also show that it is possible to alter the value of the inversion topological invariant in a clean system by occupying additional states at the inversion centers while keeping the bulk polarization fixed. We discuss the implications of our results for topological crystalline phases in higher-dimensional electronic systems, and discuss potential experimental realizations in ultra-cold atomic systems.
\end{abstract}

\pacs{Valid PACS appear here}
\maketitle

\section{Introduction}
\vspace{-0.35cm}
\indent The relationship between symmetry and topology has been of fundamental importance in establishing the classification of symmetry-protected topological phases (SPTs) \cite{hasan2010colloquium,qi2011topological}. In non-interacting systems, the ten-fold periodic table provides a classification of strong topological phases based on the presence or absence of charge-conjugation, time-reversal, and/or chiral symmetries \cite{schnyder2008classification,kitaev2009periodic,qi2008topological,ryu2010topological}. This was later extended to systems with crystalline symmetries, leading to a classification of topological crystalline phases (TCPs) \cite{fu2011topological,fang2012bulk,hsieh2012topological,chiu2013classification,fang2013entanglement,benalcazar2014classification,fang2014new,shiozaki2014topology,fang2015new,fang2017rotation,huang2017building,kruthoff2017topological,shiozaki2017many,song2017topological,po2017symmetry,bradlyn2017topological,cano2018building,khalaf_symmetry_2018,po2018fragile,song2018real,song2018quantitative,liu2019shift,ono2019refined,po2020symmetry,song2020twisted}. Recent methods \cite{po2017symmetry,bradlyn2017topological,cano2018building,ono2019refined,po2020symmetry} in classifying these phases in clean systems have involved examining connectivity of electronic bands between high symmetry momenta within the Brillouin zone (BZ). 
These approaches identify the classes of (obstructed) atomic limit states compatible with the crystal symmetries, and expressible as a linear combination of symmetric, localized orbitals (Wannier functions) on the position-space lattice. Topological crystalline bands can then be identified as bands that cannot be characterized by a configuration of localized, symmetric Wannier functions. 

 An underlying aspect of these methods is the use of symmetry properties of localized atomic orbitals on the position-space lattice in addition to the symmetry properties of Bloch states across in momentum space. While most TCPs have been studied in the presence of translation and point group symmetry, it is interesting to consider what features survive when translation symmetry is broken, but point group symmetry remains. To address this, one may ask if it is possible to obtain an understanding of TCPs without having to rely on any momentum-space description (i.e., by only using the information provided by the symmetries of the position-space lattice). Such a question arises in disordered systems for example, where translation symmetry is no longer present, though generic disorder will break the point-group symmetry as well. However, it is possible to apply correlated disorder in such a manner that preserves crystalline symmetries which may stabilize some topological features. For generic disorder, previous works have indicated that the surface states of TCPs are robust as long as the disorder protects the spatial symmetry on average \cite{fu2012topology,fang2012bulk,fang2013entanglement,fulga2014statistical,song2015quantization,mondragon2019robust,song2020twisted,diez2015extended}. This suggests that crystalline symmetries in TCPs play a fundamental role in preserving the topological properties of these systems, even if translation symmetry is absent. In this work, we seek to obtain an understanding of 1D TCPs (or more precisely, obstructed atomic limits) in the presence of disorder that preserves the point-group symmetry, but not translation symmetry. Specifically, we will study the bulk topology of a 1D chain with inversion (and chiral) symmetry in the presence of correlated disorder that is symmetric around a fixed inversion center. This type of correlated disorder may not naturally occur in electronic solid state systems, but it can be straightforwardly engineered in cold-atomic gasses in optical lattices\cite{meier2018observation}. Additionally, our results apply to cases when translation symmetry is broken by any kind of spatially dependent potential, not just a disordered one, as long as inversion symmetry is preserved (e.g., a harmonic trap). 

 In order to analyze the topological properties of this system, we compare the behavior of three bulk topological invariants (more details below): a $\mathbb{Z}_2$-valued inversion symmetry indicator topological invariant $\Delta_{\mathcal{X}}$, the $\mathbb{Z}$-valued chiral winding number $\nu$, and the quantized, $\mathbb{Z}_2$-valued bulk electric polarization $P_0$.  The inversion symmetry indicator in 1D is determined by the parity of the inversion eigenvalues at certain inversion-symmetric momenta in the BZ, and the chiral winding number characterizes 1D strong topological phases in the chiral-symmetric BDI and AIII classes. In the clean limit there is a remarkable relationship between the three quantities. Both the inversion topological invariant and the chiral winding number can be directly related to the bulk polarization via\begin{equation}
     \Delta_{\mathcal{X}}= 2\nu=4P_0\,\,\, ({\rm mod}\, 4).\label{eq:topologicalinvariantrelationship}
 \end{equation}
 
 In clean systems these three quantities are usually determined by evaluating integrals over the momentum-space BZ. However, when disorder is present, we utilize an alternative formulation that does not rely on a specific choice of basis. Ultimately we compute these three bulk topological invariants in position-space and study their respective phase diagrams as a function of model parameters and disorder strength. When comparing the phase diagrams of the different invariants we find that at a special value of the disorder strength (which we denote as the fluctuation onset (FO) value),   a boundary emerges in the phase diagram of the inversion topological invariant past which it begins to fluctuate between a set of integer values, while the chiral winding number and the bulk polarization remain constant. We characterize the nature of these fluctuations by examining the localized states at the inversion centers. We find that, unlike the other two invariants which vary only in the presence of delocalized states, the inversion symmetry indicator is sensitive to the closure of the spectral gap, regardless of whether those states are localized or delocalized. These findings are further substantiated by an asymptotically exact analytic calculation using a position-space renormalization group (RG) technique to determine  the inversion topological invariant, bulk polarization, and chiral winding number when disorder is present. Using these results and additional analyses we explore the fate of the relations Eq. (\ref{eq:topologicalinvariantrelationship}) in the presence of strong disorder.

 This paper is organized as follows: In Sec. \ref{sec:cleanlimit}, we introduce the model Hamiltonian when no disorder is present, and review the properties of this Hamiltonian as well as the inversion topological invariant, bulk polarization, and chiral winding number in both momentum-space and position-space. In Sec. \ref{sec:disorderedregime} we discuss how disorder is applied to this model so that the inversion symmetry is preserved, and present phase diagrams for the inversion topological invariant, chiral winding number, and bulk polarization. In Sec. \ref{sec:fluctuations} we provide a quantitative understanding of the behavior of the inversion topological invariant at strong disorder. In Sec. \ref{sec:simplifiedpositionspaceinvariants}, utilizing the results of the position-space RG calculation, we compare the behavior of the inversion topological invariant, the chiral winding number, and bulk polarization in strong disorder. Finally, in Sec. \ref{sec:fillingadjustment}, we discuss how filling additional localized states at the inversion centers affects the inversion topological invariant and the bulk polarization differently, and thus allows for violations of Eq. (\ref{eq:topologicalinvariantrelationship}) in the absence of translation symmetry.

\section{Clean Limit}\label{sec:cleanlimit}

\subsection{Momentum-Space Representation}\label{subsec:momentumspacerepresentation}
\begin{figure}[t!]
\includegraphics[scale=0.35]{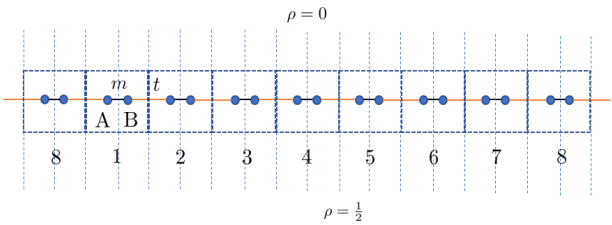}
\caption{Illustration of the 1D inversion-symmetric chain with periodic boundary conditions for $N=8$ sites in the clean limit. The dashed boxes indicate the unit cells labeled by $x=1,\ldots,8$ with two orbitals denoted as $A$ and $B$ respectively. $m$ is the intra-cell hopping and $t$ is the inter-cell hopping. The dashed blue lines indicate the possible inversion centers in the clean limit, distinguished by $\rho=0$ and $\rho=\frac{1}{2}$.}
\label{fig:cleanmodel}
\end{figure}

We consider a tight binding model with $N$ lattice sites, and two degenerate orbitals per lattice site denoted by A and B. We use nearest-neighbor hopping, with an inter-cell hopping amplitude $t$, and an intra-cell hopping amplitude $m,$ as illustrated in Fig. \ref{fig:cleanmodel}. In the clean limit the model has the  Bloch Hamiltonian:
\begin{equation}\begin{split}\label{eq:blochhamiltonian}
h(k_{x})=(m+t\cos(k_{x}))\sigma_{1}+t\sin(k_{x})\sigma_{2},  
\end{split}\end{equation}
\noindent where $\sigma_{\alpha}$, for $\alpha=1,2,3$ denotes the Pauli matrices acting on the two orbitals $\{A,B\}$ in each unit cell. The full Hamiltonian in momentum space is expressed in terms of the Bloch Hamiltonian as $H=\sum\limits_{k_{x}}c_{k_{x}}^{\dagger}h(k_{x})c_{k_{x}}$ where $c_{k_{x}}^{\dagger}=\begin{pmatrix} c_{k_{x},A}^{\dagger} & c_{k_{x},B}^{\dagger} \end{pmatrix}$. The energy eigenvalues of (\ref{eq:blochhamiltonian}) are $\epsilon_{\pm}=\pm\sqrt{(m+t\cos(k_{x}))^{2}+t^{2}\sin^{2}(k_{x})}$. Thus, the spectrum of the Hamiltonian is gapped for all $|m|\neq |t|$.

This model is in the BDI class\cite{altland1997nonstandard} since it has a chiral symmetry that acts on the Bloch Hamiltonian as
\begin{equation}\begin{split}\label{eq:chiraloperator1}
\sigma_{3}h(k_{x})\sigma_{3}^{-1}=-h(k_{x})    
\end{split}\end{equation} in addition to time-reversal symmetry ($h^*(k_{x})=h(-k_{x})$)  and a particle-hole symmetry ($\sigma_{3}h^*(k_{x})\sigma_{3}^{-1}=-h(-k_{x})$). Thus this system has a strong topological invariant: the winding number $\nu$. In the basis where the chiral operator is diagonal, $\nu$ is given by the winding of the off-diagonal block of the matrix $h(k_{x})$\cite{schnyder2009lattice}:
\begin{equation}\begin{split}\label{eq:windingnumber1}
&\nu=\int\limits_{0}^{2\pi}\partial_{k_{x}}\log(te^{ik_{x}}+m)\hspace{0.05cm}\frac{dk_{x}}{2\pi i}= \begin{cases} 
      1 & |m|<t \\
      0 & |m|>t 
   \end{cases}.
\end{split}\end{equation}
\noindent $\nu$ is always integer-valued (i.e., $\nu\in\mathbb{Z}$), and is gauge-invariant under a change of phase of the Bloch states. It also specifies a bulk-edge correspondence in the system,  where $\nu$ determines the difference between the number of zero-energy end states having positive and negative chirality\cite{teo2010topological,mondragon2014topological}.

This model is also inversion-symmetric. When the system is translation invariant, any position $\mathcal{X}=x+\rho$ can serve as the inversion center, where $x$ denotes a fixed lattice site (i.e., $x=1,\ldots,N$) and $\rho\in\{0,\frac{1}{2}\}$ (e.g., $x+\frac{1}{2}$ is the midpoint between lattice sites $x$ and $x+1$), as shown in Fig. \ref{fig:cleanmodel}. When $N$ is even, there are $2N$ total possible inversion centers corresponding to the number of distinct values of $\mathcal{X}$. The inversion symmetry acts on the Bloch Hamiltonian as
\begin{equation}\begin{split}\label{eq:inversionoperator1}
\hat{I}_{\mathcal{X}}(k_{x})h(k_{x})(\hat{I}_{\mathcal{X}}(k_{x}))^{-1}=h(I_{\mathcal{X}}k_{x})=h(-k_{x}),
\end{split}\end{equation} where the momentum-space representation of the inversion operator is given as $\hat{I}_{\mathcal{X}}=e^{-2ik_{x}\rho}\sigma_{1}$. At the inversion-invariant momenta $k_{x}=0,\pi$, the constraint Eq.~(\ref{eq:inversionoperator1}) can be recast as $[\hat{I}_{\mathcal{X}}(k_{x}=0,\pi),h(k_{x}=0,\pi)]=0$. Since the inversion operator and the Hamiltonian commute at the inversion-invariant momenta, the occupied Bloch states at these special points can be labeled by the inversion eigenvalues $\xi_{\mathcal{X}}(0)=\pm 1$ and $\xi_{\mathcal{X}}(\pi)=e^{2i\pi\rho}=\pm 1$. 

Using these results one can define an inversion topological invariant as \cite{hughes2011inversion,turner2012quantized,fang2013entanglement}:
\begin{equation}\begin{split}\label{eq:inversioninvariant1}
\Delta_{\mathcal{X}}=\sum\limits_{k_{x}=0,\pi}[n_{\mathcal{X}}^{(+)}(k_{x})-n_{\mathcal{X}}^{(-)}(k_{x})],
\end{split}\end{equation}
\noindent where $n_{\mathcal{X}}^{(\alpha)}(k_{x})$ denotes the number of occupied Bloch states at the inversion invariant momenta $k_{x}=0,\pi$ with inversion eigenvalue $\alpha=\pm 1$ ($\pm$ as shorthand) for the fixed inversion center $\mathcal{X}=x+\rho$. For $|m|<t$, one has $\{\Delta_{\mathcal{X}=x}=0,\Delta_{\mathcal{X}=x+\frac{1}{2}}=-2\}$, which denotes the topological phase, and for $|m|>t$, one has $\{\Delta_{\mathcal{X}=x}=-2,\Delta_{\mathcal{X}=x+\frac{1}{2}}=0\}$ which denotes the trivial phase.\\

\indent Finally, the third quantity we consider is the bulk polarization $P_{0}$, which is determined by the Berry phase of the occupied energy bands  \cite{qi2008topological,king1993theory,ortiz1994macroscopic,hughes2011inversion,turner2012quantized,benalcazar2017quantized}:
\begin{equation}\begin{split}\label{eq:polarization1}
P_{0}=\frac{1}{2\pi}\int\limits_{0}^{2\pi}dk_{x}\hspace{0.05cm}A(k_{x}),
\end{split}\end{equation}
where $A(k_{x})$ is the Berry connection
\begin{equation}\begin{split}\label{eq:berryconnection}
A(k_{x})=i\sum\limits_{n\in\text{occ.}}\langle u_{n}(k_{x})|\partial_{k_{x}}|u_{n}(k_{x})\rangle,
\end{split}\end{equation}
and $|u_{n}(k_{x})\rangle$ denotes the Bloch state with band index $n$. $A(k_{x})$ is explicitly not gauge-invariant, and the polarization shifts by an integer under large gauge transformations of the occupied Bloch states. The chiral winding number and the polarization are related to each other by $\nu=2P_{0}\hspace{0.05cm}(\text{mod}\hspace{0.1cm}2)$ \cite{mondragon2014topological}. Furthermore, the polarization is related to the inversion topological invariant as $\Delta_{\mathcal{X}=x+\frac{1}{2}}=4P_{0}\hspace{0.05cm}(\text{mod}\hspace{0.1cm}4)$ (this is proven in Supplemental Material (SM) A). Therefore, the relationship between the inversion topological invariant, chiral winding number, and bulk polarization in the clean limit is given by Eq. (\ref{eq:topologicalinvariantrelationship}).

\subsection{Position-Space Representation}\label{subsec:positionspacerepresentation}
We now express the Hamiltonian, inversion topological invariant, chiral winding number, and bulk polarization in the position-space basis, which will be used when translation-symmetry breaking disorder is present. To determine the clean limit of the Hamiltonian in position-space, we perform a Fourier transform on the Bloch Hamiltonian via $c_{j}=\frac{1}{\sqrt{N}}\sum\limits_{k_{x}}e^{ik_{x}j}c_{k_{x}}$ (setting the lattice constant $a=1$) to yield
\begin{equation}\begin{split}\label{eq:realspacehamiltonian}
H=\sum\limits_{j=1}^{N}\left[mc_{j}^{\dagger}\sigma_{1}c_{j}+\left(\frac{1}{2}tc_{j}^{\dagger}(\sigma_{1}-i\sigma_{2})c_{j+1}+\text{h.c.}\right)\right],
\end{split}\end{equation}
\noindent where $c_{j}^{\dagger}=\begin{pmatrix} c_{j,A}^{\dagger} & c_{j,B}^{\dagger} \end{pmatrix}$ creates orbitals of type $A$ and $B$ at the same lattice site $j$. The position-space basis states are given by $c_{j,\sigma}^{\dagger}|0\rangle=|j,\sigma\rangle=|j\rangle\otimes|\sigma\rangle$ where $\sigma\in\{A,B\}$. We use periodic boundary conditions such that for $j\in\{1,\ldots,N\}$, $c_{N+j}^{\dagger}\equiv c_{j}^{\dagger}$. In the clean limit, any position $\mathcal{X}=x+\rho$ for $x\in\{1,\ldots,N\}$ and $\rho\in\left\{0,\frac{1}{2}\right\}$ can be an inversion center. When $N$ is even the index $\rho$ distinguishes two classes of inversion centers: the class of inversion centers with $\rho=0$ corresponds to reflections leaving a pair of lattice sites fixed, whereas the class of inversion centers with $\rho=\frac{1}{2}$ corresponds to  reflections leaving a pair midpoints between lattice sites fixed. These two distinct classes of inversion centers result in $\frac{N}{2}$ unique operators $I_{\mathcal{X}}$ for each $\rho\in\left\{0,\frac{1}{2}\right\}$ for a total of $N$ unique inversion centers. This is because each inversion center $\mathcal{X}=x+\rho$ can also be identified as $\mathcal{X}=\frac{N}{2}+x+\rho$, resulting in only $\frac{N}{2}$ unique inversion centers for each value of $\rho$. The index $\rho$ is not needed
when $N$ is odd since the number of inversion centers is simply $N$, and labeled by $\mathcal{X}=x$ for $x\in\{1,\ldots,N\}$. Each inversion operator $I_\chi$ fixes one lattice site and one midpoint in this case. Throughout this work, we will consider $N$ to be even for simplicity. The inversion operator for any position $\mathcal{X}$ can be expressed in position-space as follows:
\begin{equation}\begin{split}\label{eq:inversionoperator2}
I_{\mathcal{X}=x+\rho}=\sum\limits_{j=1}^{N}c_{N+2(x+\rho)-j}^{\dagger}\sigma_{1}c_{j}.
\end{split}\end{equation}
 Given the position-space representation of the inversion operator in (\ref{eq:inversionoperator2}), it can be shown that $[I_{\mathcal{X}},H]=0$. The chiral symmetry that acts on the Hamiltonian in (\ref{eq:realspacehamiltonian}) can be expressed in position-space as
\begin{equation}\begin{split}\label{eq:chiraloperator2}
S=\sum\limits_{j=1}^{N}c_{j}^{\dagger}\sigma_{3}c_{j},
\end{split}\end{equation}
and it follows that $\{S,H\}=0$.\\
 
The topological invariants discussed above can also be recast in a formulation that is independent of the choice of basis. This will prove very useful when considering translation-symmetry breaking disorder. Here, the inversion topological invariant given by (\ref{eq:inversioninvariant1}) can be expressed as \cite{mondragon2019robust}:
\begin{equation}\begin{split}\label{eq:inversioninvariant2}
\Delta_{\mathcal{X}}=\text{Tr}[\bar{I}_{\mathcal{X}}],  
\end{split}\end{equation}
\noindent where $\bar{I}_{\mathcal{X}}=P_{\text{occ}}\hat{I}_{\mathcal{X}}P_{\text{occ}}$ is the inversion operator projected onto the subspace of occupied states, with $P_{\text{occ}}$ denoting this projector. The basis-independent expression for the inversion topological invariant avoids having to use of the momentum-space representation. In particular, using the form of the inversion operator given by (\ref{eq:inversionoperator2}), one can construct a topological marker\cite{mondragon2019robust} for Eq.~(\ref{eq:inversioninvariant2}) in the position-space basis as follows:
\begin{equation}\begin{split}\label{eq:inversioninvariant3}
\Delta_{\mathcal{X}}(x)=\langle x|\text{Tr}'[\bar{I}_{\mathcal{X}}]|x\rangle,
\end{split}\end{equation}
\noindent where $\{|x\rangle\}$ denotes the position-space basis and $\text{Tr}'$ indicates that the trace is being performed only over the local degrees of freedom within each unit cell. The quantity $\Delta_{\mathcal{X}}(x)$ captures the spatial distribution of the inversion invariant from which we can calculate $\Delta_{\mathcal{X}}=\sum\limits_{x=1}^{N}\Delta_{\mathcal{X}}(x)$.\\
\begin{figure}[ht]
\includegraphics[scale=0.425]{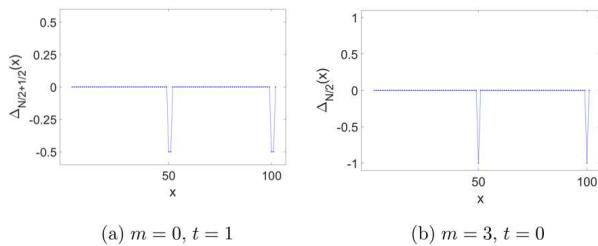}
\caption{Distribution of the inversion topological invariants (a) $\Delta_{\mathcal{X}=\frac{N}{2}+\frac{1}{2}}$ and (b) $\Delta_{\mathcal{X}=\frac{N}{2}}$ in the clean limit. The parameters used are indicated at the bottom of each plot. The subscripts on the $\Delta$ for each plot indicate the inversion center $\mathcal{X}=\frac{N}{2}+\rho$ where $\rho\in\{0,\frac{1}{2}\}.$ These calculations are for a chain of $N=100$ sites with periodic boundary conditions (points appearing after site $N=100$ are labeled $1$, $2$, etc. because of this).}
\label{fig:distributionplots}
\end{figure}
 
In Fig. \ref{fig:distributionplots} we show calculations of the spatially resolved inversion topological invariant given by (\ref{eq:inversioninvariant3}) in the clean limit, which clearly illustrate that the distribution is sharply peaked at the inversion centers; for $\Delta_{\mathcal{X}=\frac{N}{2}}$, the distribution is peaked at  $x=\frac{N}{2}$ and $x=N$, and for $\Delta_{\mathcal{X}=\frac{N}{2}+\frac{1}{2}}$, the distribution is peaked at $x=\frac{N}{2}+\frac{1}{2}$ and $x=N+\frac{1}{2}$. To gain intution about the structure of this distribution we can calculate $\Delta_{\mathcal{X}}(x)$ in two dimerized, flat-band limits of (\ref{eq:realspacehamiltonian}), for which the Wannier functions of the Hamiltonian take a simple form. In the first dimerized limit, in which $m=0$ and $t\neq 0$, the eigenstates  of (\ref{eq:realspacehamiltonian}) (which are also Wannier functions in this limit) are given as $|W_{\mp}(n)\rangle=\frac{1}{\sqrt{2}}(|n+1,A\rangle\mp|n,B\rangle)$, where $n\in\{1,\ldots,N\},$ and $\mp$ indicates occupied/unoccupied eigenstates (states with energies less than the Fermi level $E_{F}=0$ are occupied). These eigenstates are localized and have weight on just two neighboring unit cells $n$ and $n+1$. In this limit, it is possible to provide a simple form for the inversion topological invariant (\ref{eq:inversioninvariant2}) and its distribution in position-space given by (\ref{eq:inversioninvariant3}) (calculation details in SM B).  We find the distribution for each inversion topological invariant in this limit is given by
\begin{equation}\begin{split}\label{eq:distribution1}
\Delta_{\mathcal{X}=\frac{N}{2}+\frac{1}{2}}(n)=-\frac{1}{2}(\delta_{n,1}+\delta_{n,\frac{N}{2}}+\delta_{n,\frac{N}{2}+1}+\delta_{n,N})
\end{split}\end{equation}
\begin{equation}\begin{split}\label{eq:distribution2}
\Delta_{\mathcal{X}=\frac{N}{2}}=-\delta_{n,\frac{N}{2}+\frac{1}{2}}
\end{split}\end{equation}
where $\delta_{m,n}$ is the Kronecker delta, which is only equal to $1$ when $m=n$ and is zero otherwise. Performing the sum over all the lattice sites $n$ in (\ref{eq:distribution1}) and (\ref{eq:distribution2}), results in $\Delta_{\mathcal{X}=\frac{N}{2}}=0$ and $\Delta_{\mathcal{X}=\frac{N}{2}+\frac{1}{2}}=-2$. 

We can establish intuition for why $\Delta_{\mathcal{X}=\frac{N}{2}+\frac{1}{2}}$ is non-zero in this limit by considering the spatial structure of these localized eigenstates. Most of the eigenstates are transformed to a partner eigenstate under inversion symmetry, and the pair will contribute both a $+$ and $-$ eigenvalue such that $\Delta_{\mathcal{X}}$ will receive vanishing contributions near the points in space where those states are localized. In contrast,  there are precisely two  occupied eigenstates that get mapped to themselves under inversion. Indeed, the states centered about positions $x=\frac{N}{2}+\frac{1}{2}$ and $x=N+\frac{1}{2}$: $|W_{-}\left(\frac{N}{2}\right)\rangle$ and $|W_{-}(N)\rangle$, have this property. Each of these are eigenstates of the inversion operator (\ref{eq:inversionoperator2}) with inversion eigenvalue $-1$. Therefore, the distribution of $\Delta_{\mathcal{X}=\frac{N}{2}+\frac{1}{2}}$ is peaked at the sites $x=1,\frac{N}{2},\frac{N}{2}+1,N$ as shown in Fig. \ref{fig:distributionplots} (a). Similar arguments can be applied to the second dimerized limit, $m\neq 0$ and $t=0$, in which the eigenstates are $|W_{\mp}(n)\rangle=\frac{1}{\sqrt{2}}(|n,A\rangle\mp|n,B\rangle),$ which are localized Wannier functions with weight solely on unit cell $n$ for $n\in\{1,\ldots,N\}$. This results in $\Delta_{\mathcal{X}=\frac{N}{2}}=-2$ and $\Delta_{\mathcal{X}=\frac{N}{2}+\frac{1}{2}}=0$, with the distribution of the former peaked at sites $x=\frac{N}{2}$ and $x=N$ with equal weight, as shown in Fig. \ref{fig:distributionplots} (b).\\

 The chiral winding number can also be expressed in a basis-independent representation for eventual evaluation in position space. Let $Q = P_{\text{unocc}} - P_{\text{occ}} = 1 - 2P_{\text{occ}}$ be a spectrally flat Hamiltonian constructed from the projectors corresponding to the unoccupied and occupied subspaces (these are above and below $E_{F}=0$ respectively). The chiral operator $S$ has eigenvalues $\pm 1$ and can be expressed as $S=S_{+}-S_{-}$ where $S_{+}$ and $S_{-}$ correspond to the projectors onto the subspaces labeled by eigenvalues $\pm 1$ respectively. $Q$, being chiral-symmetric, can be decomposed as $Q=S_{+}QS_{-}+S_{-}QS_{+}\equiv Q_{+-}+Q_{-+}$ where $Q_{+-}=S_{+}QS_{-}$ and $Q_{-+}=S_{-}QS_{+}$. Then, the basis-independent form of the chiral winding number is \cite{mondragon2014topological,song2014aiii}
\begin{equation}\begin{split}\label{eq:windingnumber2}
\nu=\frac{1}{N}\text{Tr}[Q_{-+}[X,Q_{+-}]],
\end{split}\end{equation}
where $X$ is the position operator. This formula is only well-defined for open boundary conditions, and in all the calculations for $\nu$, the position operator is expressed in the form $X=\sum\limits_{j=1}^{N}[j(|j,A\rangle\langle j,A|+|j,B\rangle\langle j,B|)]$. We are currently unaware of a basis-independent formula for $\nu$ that is valid for periodic boundary conditions, which is why we use (\ref{eq:windingnumber2}). To illustrate how the winding number $\nu$ can be computed from (\ref{eq:windingnumber2}), we once again consider the dimerized limit where the bulk eigenstates are given as $|W_{\mp}(n)\rangle=\frac{1}{\sqrt{2}}(|n+1,A\rangle\mp|n,B\rangle)$. In this limit one has $Q_{+-}=\sum\limits_{n=1}^{N}|n+1,A\rangle\langle n,B|=Q_{-+}^{\dagger}$ which results in $Q_{-+}[X,Q_{+-}]=\sum\limits_{j=1}^{N}|j,B\rangle\langle j,B|$. Substituting this into (\ref{eq:windingnumber2}) results in $\nu=1$.\\

Finally, we also consider the polarization of the system. The polarization of the system can be expressed in a basis-independent formulation by computing the eigenvalues of the position operator projected onto the occupied states given as $X_{P}=P_{\text{occ}}XP_{\text{occ}}$. When computing the bulk polarization when disorder is present, we use periodic boundary conditions, and hence the position operator takes the exponential form\cite{resta1998quantum}: $X=\sum\limits_{j=1}^{N}[e^{\frac{2\pi i}{N}j}(|j,A\rangle j,A|+|j,B\rangle\langle j,B|)]$. Using the set of eigenvalues $\{\xi_{n}\}$ of $X_{P}$, the polarization is then given as
\begin{equation}\begin{split}\label{eq:polarization}
P_{0}=\sum\limits_{n=1}^{N}\left(\frac{1}{2\pi}\text{Im}\log\xi_{n}\right).
\end{split}\end{equation}Similar to the calculations of the inversion invariant and the winding number, we can calculate the bulk polarization using (\ref{eq:polarization}) in the same dimerized limit as before. This results in
\begin{equation}\begin{split}\label{eq:projectedpositionoperator}
X_{P}=\sum\limits_{n=1}^{N}e^{\frac{2\pi i}{N}\left(j+\frac{1}{2}\right)}\cos\left(\frac{\pi}{N}\right)|W_{-}(n)\rangle\langle W_{-}(n)|.
\end{split}\end{equation}
In the thermodynamic limit $N\to\infty$ the eigenvalues of $X_{P}$ are
\begin{equation}\begin{split}\label{eq:polarizationeigenvalues}
\{\xi_{n}\}_{n=1}^{N}=\left\{e^{\frac{2\pi i}{N}\left(n+\frac{1}{2}\right)}\right\}_{n=1}^{N}.
\end{split}\end{equation}
Substituting this into (\ref{eq:polarization}) results in a bulk polarization of $P_{0}=\frac{1}{2}$.

\section{Disordered Regime}\label{sec:disorderedregime}
Having established the features of (\ref{eq:realspacehamiltonian}) and its topological invariants in the clean limit in both momentum-space and position-space, we now proceed to discuss the effects of disorder on this system. We first explain how inversion-symmetric disorder is applied to this system while taking into consideration the two distinct classes of inversion centers. Then, we present phase diagrams of the inversion topological invariant, chiral winding number, and bulk polarization when disorder is introduced.
\subsection{Adding Disorder}\label{subsec:addingdisorder}
We introduce disorder in (\ref{eq:realspacehamiltonian}), by varying the values of the intra-cell hoppings and inter-cell hoppings throughout the lattice. Thus, the Hamiltonian given by (\ref{eq:realspacehamiltonian}) becomes,
\begin{equation}\begin{split}\label{eq:disorderedHamiltonian1}
H=\sum\limits_{j=1}^{N}\left[m_{j}c_{j}^{\dagger}\sigma_{1}c_{j}+\frac{1}{2}t_{j}c_{j}^{\dagger}(\sigma_{1}-i\sigma_{2})c_{j+1}+\text{h.c.}\right],    
\end{split}\end{equation}
\noindent where
\begin{equation}\begin{split}\label{eq:disorderconfig1}
m_{j}&=m+W_{2}\omega_{j},\\
t_{j}&=t+W_{1}\omega_{j}',
\end{split}\end{equation}
\noindent $W_{2}$ and $W_{1}$ are the intra-cell disorder strength, and inter-cell disorder strength respectively, and $\omega_{j},\omega_{j}'$ are random numbers uniformly distributed in the interval $[-\frac{1}{2},\frac{1}{2}]$.\\

\begin{figure}[t!]
\includegraphics[scale=0.45]{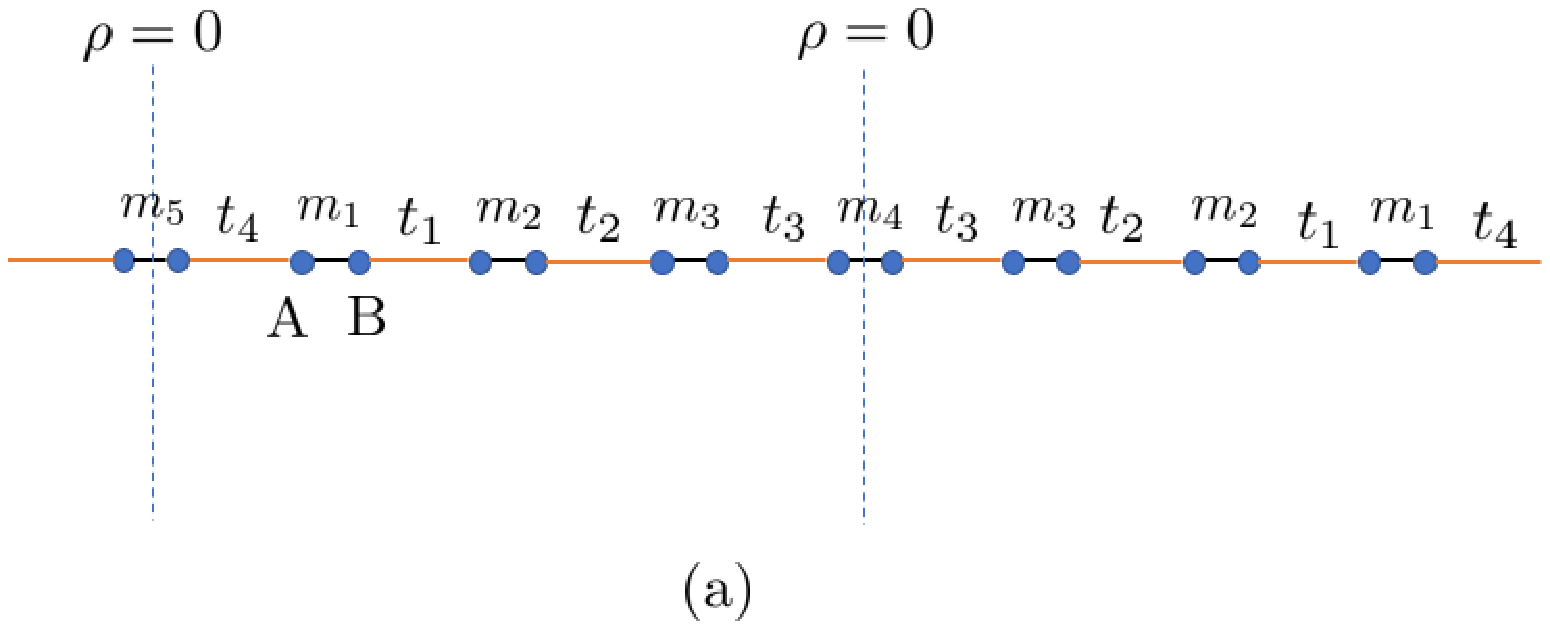}
\includegraphics[scale=0.45]{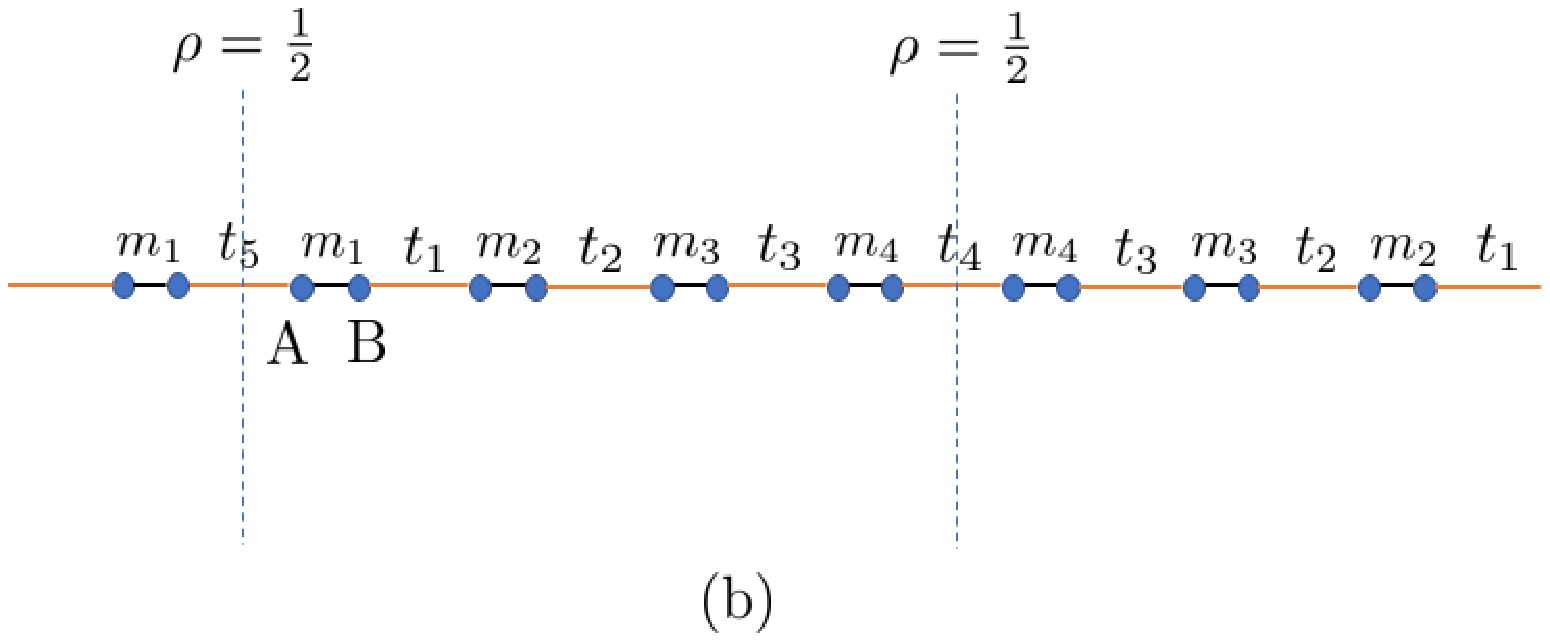}
\caption{Illustration of the distinct choices of inversion center $\mathcal{X}=\frac{N}{2}+\rho$ for the 1D inversion-symmetric chain with periodic boundary conditions for $N=8$ unit cells for a given disorder configuration. For (a) $\rho=0$, $m_{i}=m_{N-i}$ and $t_{i}=t_{N-1-i},$ and for (b) $\rho=\frac{1}{2}$, $m_{i}=m_{N+1-i}$ and $t_{i}=t_{N-i}$. In the figures above, the notation $m_{\frac{N}{2}+1}\equiv m_{N}$ and $t_{\frac{N}{2}+1}\equiv t_{N}$ has been adopted (i.e., $m_{5}\equiv m_{8}$ and $t_{5}\equiv t_{8}$).}
\label{fig:disorderconfigs}
\end{figure}

To preserve the inversion symmetry, the disorder configuration must respect one of the two inequivalent choices of inversion center on the one-dimensional lattice.
Fig. \ref{fig:disorderconfigs} illustrates how the disorder can be modeled so that it satisfies  inversion symmetry. If the inversion center is chosen about $\mathcal{X}=\frac{N}{2}$, this leads to the constraint that $m_{j}=m_{N-j}$ and $t_{j}=t_{N-1-j}$, which can also be expressed as
\begin{equation}\begin{split}\label{eq:disorderconfig2}
\omega_{j}=\omega_{N-j},\\
\omega_{j}'=\omega_{N-1-j}'.
\end{split}\end{equation}
Alternatively, if $\mathcal{X}=\frac{N}{2}+\frac{1}{2}$ is chosen as the inversion center, this will lead to $m_{j}=m_{N+1-j}$ and $t_{j}=t_{N-j}$ or
\begin{equation}\begin{split}\label{eq:disorderconfig3}
\omega_{j}=\omega_{N+1-j},\\
\omega_{j}'=\omega_{N-j}'.
\end{split}\end{equation}
In all the results that follow, we will consider $\mathcal{X}=\frac{N}{2}+\frac{1}{2}$ as the inversion center for our disorder configurations.  Recall that in the clean limit, $\Delta_{\mathcal{X}=\frac{N}{2}+\frac{1}{2}}=2\nu=4P_{0}\hspace{0.05cm}(\text{mod}\hspace{0.1cm}4)$. For the topological phase $|m|<t$, $\left\{\Delta_{\mathcal{X}=\frac{N}{2}+\frac{1}{2}}=-2,\nu=1,P_{0}=\frac{1}{2}\right\},$ whereas for the trivial phase $|m|>t$, $\left\{\Delta_{\mathcal{X}=\frac{N}{2}+\frac{1}{2}}=\nu=P_{0}=0\right\}$. The results for the quantity $\Delta_{\mathcal{X}=\frac{N}{2}}$ for inversion-symmetric disorder configurations around the inversion center $\mathcal{X}=\frac{N}{2}$ are discussed separately in the SM C and F.

\subsection{Phase Diagrams of the Inversion Topological Invariant, Chiral Winding Number, and Bulk Polarization}\label{subsec:phasediagrams}
\indent In this subsection we discuss the phase diagrams of the inversion topological invariant $\Delta_{\mathcal{X}=\frac{N}{2}+\frac{1}{2}}$ the chiral winding number $\nu,$ and the polarization $P_0$ for a disordered system. When performing the numerical calculations, we define the occupied projector $P_{\text{occ}}$ to include all states that lie below $E_{F}=0$. Furthermore, in all the calculations we set $t=1$. All the results were obtained using (\ref{eq:inversioninvariant2}) for the inversion topological invariant $\Delta_{\mathcal{X}=\frac{N}{2}+\frac{1}{2}}$, (\ref{eq:windingnumber2}) for the chiral winding number, and (\ref{eq:polarization}) for the bulk polarization.

In addition to numerically computing the topological invariants in the disordered limit, we also computed the localization length $\Lambda$ of the wavefunction of the end states (e.g. $\psi_{\text{edge}}(x)\sim \exp\left\{-\frac{x}{\Lambda}\right\}$) using a numerical transfer matrix method \cite{mackinnon1983scaling}. Using previous results on a purely chiral symmetric 1D chain\cite{mondragon2014topological}, it is possible to obtain an analytic form for the critical scaling of the localization length in the thermodynamic limit where $N\to\infty$, yielding
\begin{equation}\begin{split}\label{eq:localizationlength}
\Lambda=\left(\left|\text{ln}\left[\frac{|2+W_{1}|^{\frac{1}{W_{1}}+\frac{1}{2}}|2m-W_{2}|^{\frac{m}{W_{2}}-\frac{1}{2}}}{|2-W_{1}|^{\frac{1}{W_{1}}-\frac{1}{2}}|2m+W_{2}|^{\frac{m}{W_{2}}+\frac{1}{2}}}\right]\right|\right)^{-1}.  
\end{split}\end{equation} This equation traces out a critical surface $\mathcal{S}_{c}$ in the space of $(m,W_{1},W_{2})$, along which $\Lambda\to\infty$, delineating the phase boundaries within which the chiral winding number $\nu$ is quantized and has a non-zero value. 

\begin{figure}[t]
\includegraphics[scale=0.425]{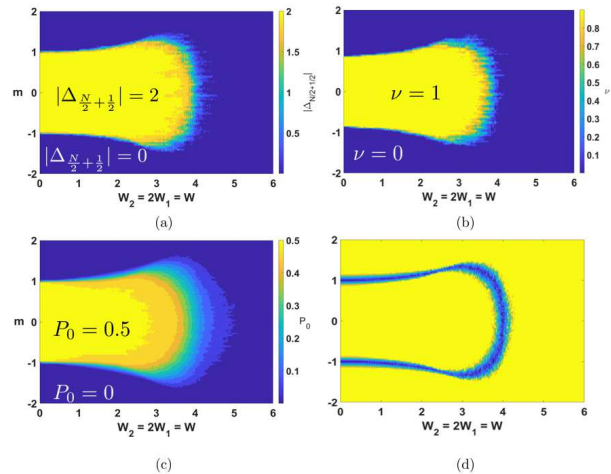}
\caption{Phase diagrams for (a) $\Delta_{\mathcal{X}=\frac{N}{2}+\frac{1}{2}}$, (b) $\nu$, and (c) $P_{0}$, and (d) the phase boundary for $\frac{W_{2}}{W_{1}}=2$ with $t=1$. The phase diagrams for $\Delta_{\mathcal{X}=\frac{N}{2}+\frac{1}{2}}$ and $\nu$ were constructed for $N=500$ sites and disorder averaged over $N_{\text{configs}}=10$ configurations, while the phase diagram for $P_{0}$ was performed over $N=100$ sites and $N_{\text{configs}}=500$ configurations. For (d) the phase boundary is given by the divergence of the localization length, constructed for $N=5000$ sites. In this regime, all three quantities are quantized and non-fluctuating within the phase boundaries.}
\label{fig:numericalphasediagrams1}
\end{figure}

\begin{figure}[ht]
\includegraphics[scale=0.425]{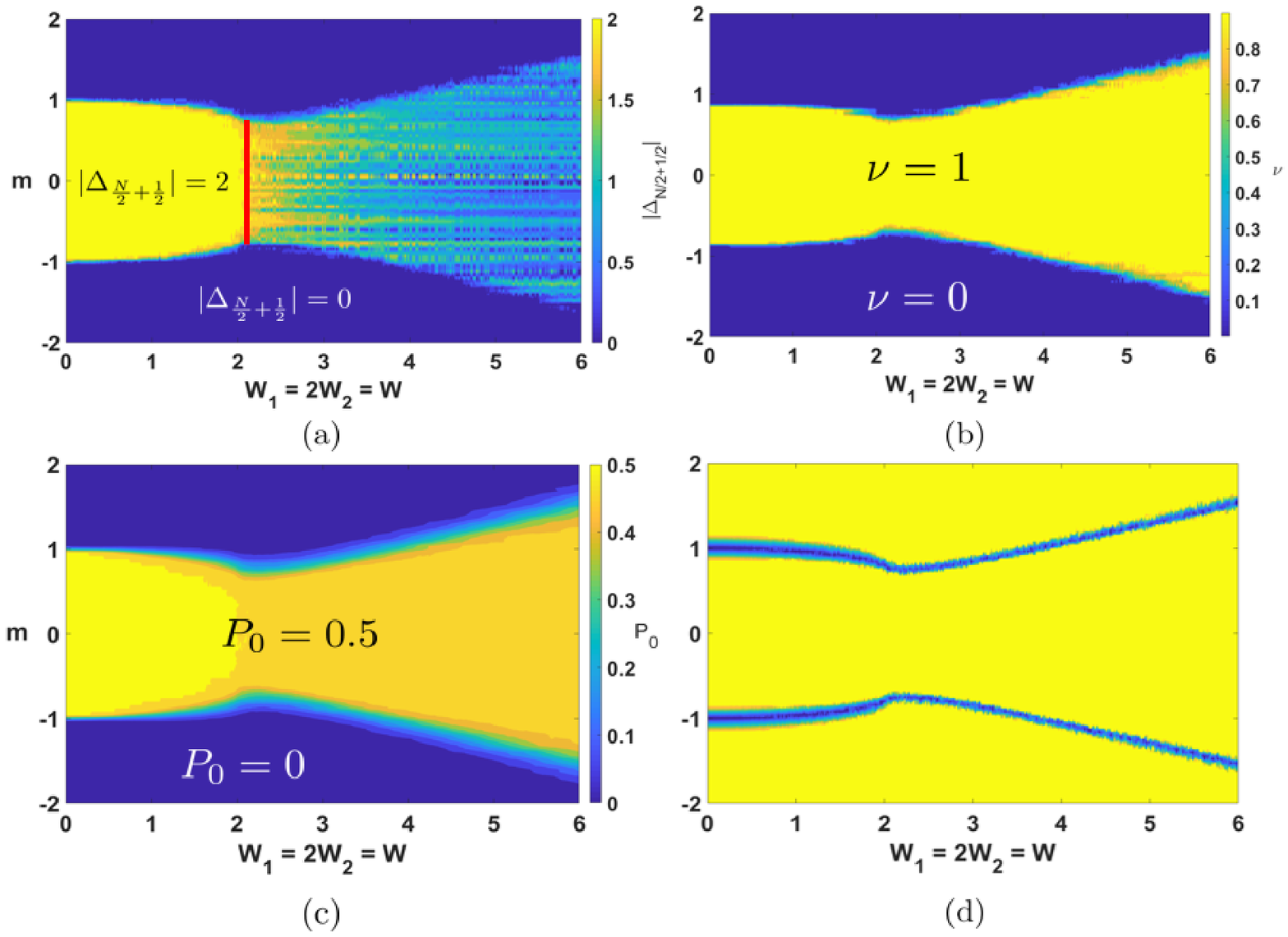}
\caption{Phase diagrams for (a) $\Delta_{\mathcal{X}=\frac{N}{2}+\frac{1}{2}}$, (b) $\nu$, and (c) $P_{0}$, and (d) the phase boundary for $\frac{W_{2}}{W_{1}}=\frac{1}{2}$ with $t=1$. The parameters used for $N$ and $N_{\text{configs}}$ are the same as in Fig. \ref{fig:numericalphasediagrams1}. In this regime, $\Delta_{\mathcal{X}=\frac{N}{2}+\frac{1}{2}}$ begins to deviate from its disorder-averaged value $|\Delta_{\mathcal{X}=\frac{N}{2}+\frac{1}{2}}|=2$ due to the onset of fluctuations for $|W_{1}|>|W_{1}^*|=2$ (indicated by the bold red line), which is also the value the bulk spectral gap closes. However, the $\nu$ and $P_{0}$ remain quantized and nontrivial within their phase boundaries.}
\label{fig:numericalphasediagrams2}
\end{figure}

\begin{figure}[ht]
\includegraphics[scale=0.425]{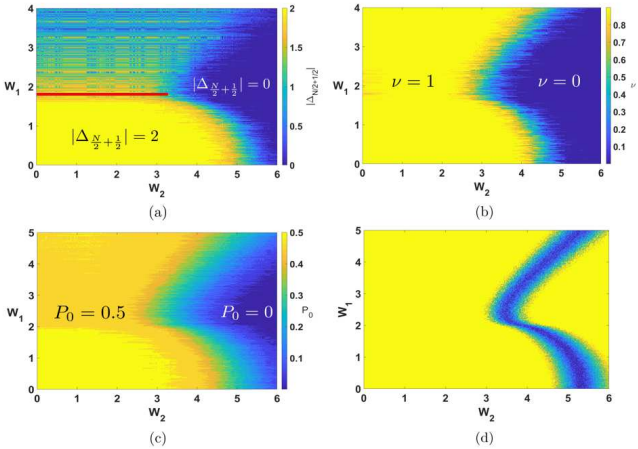}
\caption{Phase diagrams for (a) $\Delta_{\mathcal{X}=\frac{N}{2}+\frac{1}{2}}$, (b) $\nu$, and (c) $P_{0}$, and (d) the phase boundary for $m=0.5$ with $t=1$. The parameters used for $N$ and $N_{\text{configs}}$ are the same as in Fig. \ref{fig:numericalphasediagrams1}. In this regime, $\Delta_{\mathcal{X}=\frac{N}{2}+\frac{1}{2}}$ begins to deviate from its disorder-averaged value $|\Delta_{\mathcal{X}=\frac{N}{2}+\frac{1}{2}}|=2$ due to the onset of fluctuations for $|W_{1}|>|W_{1}^*|=2$ (indicated by the bold red line), which is also the value the bulk spectral gap closes. However, the $\nu$ and $P_{0}$ remain quantized and nontrivial within their phase boundaries.}
\label{fig:numericalphasediagrams3}
\end{figure}
We consider the three phase diagrams for parameter values $\frac{W_{2}}{W_{1}}=2$, $\frac{W_{2}}{W_{1}}=\frac{1}{2}$, and $m=0.5$ shown in Figs. \ref{fig:numericalphasediagrams1}-\ref{fig:numericalphasediagrams3} respectively. These phase diagrams were constructed by computing the disorder-averaged values of the inversion topological invariant $\Delta_{\mathcal{X}=\frac{N}{2}+\frac{1}{2}}$, the chiral winding number $\nu$, and the bulk polarization $P_{0}$. In addition to the phase diagrams, Figs. \ref{fig:numericalphasediagrams1} (d), \ref{fig:numericalphasediagrams2} (d), and \ref{fig:numericalphasediagrams3} (d)  illustrate the phase boundaries of the phase diagram as determined by the critical surface given by (\ref{eq:localizationlength}) where the localization length $\Lambda$ diverges. Fig. \ref{fig:numericalphasediagrams1} illustrates that when $\frac{W_{2}}{W_{1}}=2$, the phase diagrams for $\Delta_{\mathcal{X}=\frac{N}{2}+\frac{1}{2}}$, $\nu$, and $P_{0}$ are identical. Within each region of the phase diagram, the relationship given by (\ref{eq:topologicalinvariantrelationship}) is upheld. The disorder averaged values of $\Delta_{\mathcal{X}=\frac{N}{2}+\frac{1}{2}}$, $\nu$, and $P_{0}$ are quantized and nontrivial up to the phase boundary where $\Lambda$ diverges, past which there is a transition in the values of $\Delta_{\mathcal{X}=\frac{N}{2}+\frac{1}{2}}$, $\nu$, and $P_{0}$ with all three of them becoming trivial and equaling $0$.

In contrast, Fig. \ref{fig:numericalphasediagrams2} illustrates that when $\frac{W_{2}}{W_{1}}=\frac{1}{2}$, the inversion topological invariant is only precisely $\Delta_{\mathcal{X}=\frac{N}{2}+\frac{1}{2}}=-2$ within a smaller 
region of the phase diagram compared to the phase diagrams of the other two invariants. For the inversion invariant phase diagram we find that there are two distinct boundaries surrounding this region: one is the phase boundary separating the topological and trivial phases, and the other is given by a special value of the disorder strength which we denote as the fluctuation onset (FO) value (denoted by the bold red line in Fig. \ref{fig:numericalphasediagrams2} a) and summarized for each phase diagram in Table I). This also holds true in the regime where $m=0.5$ for generic values of $W_{2}$ and $W_{1}$ (the FO value is denoted by the bold red line in Fig. \ref{fig:numericalphasediagrams3} a)). The FO values for each phase diagram are specified in the accompanying Table \ref{table:summary}. Past this value, there is a region of the phase diagram where the disorder-averaged value of the inversion topological invariant is no longer quantized. This occurs because the inversion topological invariant fluctuates between a set of integer values which alters the disorder-averaged value of $\Delta_{\mathcal{X}=\frac{N}{2}+\frac{1}{2}}$ . 

The fluctuations can be explained as follows. Given a disorder configuration at weak disorder, the inversion topological invariant is still sharply peaked at the inversion centers, but also has contributions from sites neighboring the inversion centers. When the disorder strength exceeds the FO value, we find (numerically) that it is possible for $\Delta_{\mathcal{X}=\frac{N}{2}+\frac{1}{2}}$ to change from $-2$ in the clean limit to  $0$ or $+2$ for a given disorder configuration. When disorder averaged this will appear as a non-quantized invariant, but the value is quantized for each individual disorder configuration.  To provide an interpretation for these results we will show explicitly in the next section that the FO value of the disorder strength $|W_{1}^*|=2$ is equal to the value of the disorder strength at which the disorder-averaged gap in the bulk energy spectrum closes. Using this observation, we will analyze the physical reasons why this occurs, and also determine the mean and variance of the inversion topological invariant $\Delta_{\mathcal{X}=\frac{N}{2}+\frac{1}{2}}$ as a function of the disorder strength $|W_{1}|$ for $|W_{1}|>|W_{1}^*|$.
\begin{table}[t!]
\caption{Fluctuation Onset of  $\Delta_{\mathcal{X}=\frac{N}{2}+\frac{1}{2}}$ in Each Phase Diagram}
\centering
\begin{tabular}{c c c}
\hline\hline
Section & FO Value & Fluctuations Present\\  [0.5ex]
\hline
$\frac{W_{2}}{W_{1}} = 2$ & N/A & N/A \\\\
$\frac{W_{2}}{W_{1}} = \frac{1}{2}$ & $|W_{1}^*|=2$ & $|W_{1}|>|W_{1}^*|=2$ \\\\
$m=0.5$ & $|W_{1}^*|=2$ & $|W_{1}|>|W_{1}^*|=2$ \\  [1ex]
\hline
\end{tabular}
\label{table:summary}
\end{table}

\section{Fluctuations of the Inversion Topological Invariant}\label{sec:fluctuations}
As shown in Figs. \ref{fig:numericalphasediagrams2} and \ref{fig:numericalphasediagrams3}, when the disorder strength $|W_{1}|$ exceeds the FO value shown in Table I, the disorder averaged value of the inversion topological invariant $\Delta_{\mathcal{X}=\frac{N}{2}+\frac{1}{2}}$ begins to deviate from its quantized value. Specifically, in Sec. \ref{subsec:phasediagrams} it was stated that the inversion topological invariant fluctuates between the values of $-2,0$ and $2$ based on whether one or both peaks in the distribution of the inversion topological invariant switch sign. We will now quantify these statements by considering the dimerized limits of (\ref{eq:disorderedHamiltonian1}). We will show that the onset of fluctuations is caused by the closing of the disorder-averaged spectral gap also occurs at the FO value of $|W_{1}^*|=2$. We also derive the mean and variance of $\Delta_{\mathcal{X}=\frac{N}{2}+\frac{1}{2}}$ as a function of $|W_{1}|>|W_{1}^*|$.

We consider the limit $m=0$ and $t=1$ with bond disorder strength $W_{1}\neq 0$ and intra-cell disorder strength $W_{2}=0$ (i.e., $\frac{W_{2}}{W_{1}}=0$) in (\ref{eq:disorderedHamiltonian1}). For a fixed $W_{1}$, we consider a set of $N_{\text{configs}}$ disorder configurations and enumerate a collection of $N$ random numbers $\{\omega_{i}^{\prime (n)}\}_{i=1}^{N}$ for each disorder configuration. The subscript $i=1,\ldots,N$ indicates the random number, and the superscript $n=1,\ldots,N_{\text{configs}}$ labels the disorder configuration (note that in (\ref{eq:disorderconfig1})-(\ref{eq:disorderconfig3}) this index was suppressed). This notation fixes the disorder configuration $n$ and indexes a set of $N$ random numbers labeled by $i$. For each disorder configuration $n$, $\{\omega_{i}^{\prime (n)}\}$ is chosen so that it is symmetric about the inversion center $\mathcal{X}=\frac{N}{2}+\frac{1}{2}$, which constrains $\omega_{i}^{\prime (n)}=\omega_{N-i}^{\prime (n)}$ as per (\ref{eq:disorderconfig3}).
 
\begin{figure}[t!]
\includegraphics[scale=0.75]{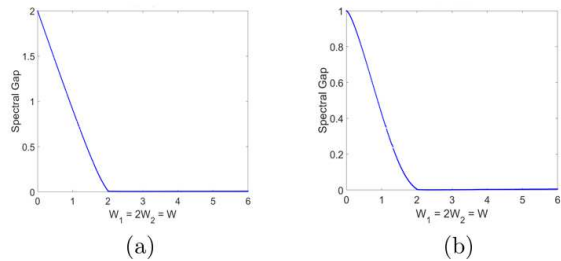}
\caption{Plots of the disorder averaged spectral gap for (a) $m=0$ and (b) $m=0.5$. The plots were constructed for a chain of $N=300$ sites with $t=1$ and $\frac{W_{2}}{W_{1}}=2$, and disorder averaging was performed over $1000$ disorder configurations. For (a) and (b), the disorder averaged spectral gap vanishes at $W_{1}=2$ which is consistent with (\ref{eq:disorderaveragedspectralgap1}).}
\label{fig:disorderaveragedspectralgap}
\end{figure}

In this dimerized limit where all the intra-cell hoppings are set to $0$, the energy eigenvalues can be determined exactly from the Hamiltonian given by (\ref{eq:disorderedHamiltonian1}) and are
\begin{equation}\begin{split}\label{eq:energyeigenvalues1}
\{E_{i,\pm}^{(n)}\}=\{\pm t_{i}^{(n)}\}=\{\pm(1+W_{1}\omega_{i}^{\prime (n)})\},
\end{split}\end{equation}
\noindent  The inversion symmetry implies that each of the $E_{i}^{(n)}$ for $i=1,\ldots,\frac{N}{2}-1$ is two-fold degenerate, while $E_{\frac{N}{2}}^{(n)}$ and $E_{N}^{(n)}$ are generically non-degenerate since the random numbers $\omega_{\frac{N}{2}}'$ and $\omega_{N}'$ are independent. At half-filling, where the Fermi level is set to $E_{\text{F}}=0$, the bulk spectral gap is determined by the difference between the  positive energy eigenvalue and negative energy eigenvalue that are closest to the Fermi level $E_F=0$. For positive $W_{1}$ these energy eigenvalues are $\pm(1+W_{1}\min(\{\omega_{i}^{\prime (n)}\}))$ while for negative $W_{1}$ these are $\pm(1-W_{1}\max(\{\omega_{i}^{\prime (n)}\}))$. Hence, if we let  $\text{SG}^{(n)}$ denote the bulk spectral gap for a disorder configuration $n,$ we find
\begin{equation}\begin{split}\label{eq:spectralgap1}
\text{SG}^{(n)}=2(1+W_{1}\min(\{\omega_{i}^{\prime (n)}\}))\\
\text{SG}^{(n)}=2(1-W_{1}\max(\{\omega_{i}^{\prime (n)}\})),
\end{split}\end{equation}
\noindent for positive and negative $W_{1}$ respectively, with $i=1,\ldots,\frac{N}{2}-1$, $i=\frac{N}{2}$, and $i=N$.  We now perform a disorder average of the bulk spectral gap over $N_{\text{configs}}$ disorder configurations:
\begin{equation}\begin{split}\label{eq:disorderaverage}
\langle\text{SG}\rangle=\frac{1}{N_{\text{configs}}}\sum\limits_{n=1}^{N_{\text{configs}}}\text{SG}^{(n)}.
\end{split}\end{equation}
\noindent In SM C, we show that taking the thermodynamic limit where $N\to\infty$ ensures that the disorder averages $\langle\min(\{\omega_{i}'\}_{i=1}^{N})\rangle$ and $\langle\max(\{\omega_{i}'\}_{i=1}^{N})\rangle$ equal $-\frac{1}{2}$ and $\frac{1}{2}$ respectively. This leads to the following expression for the disorder averaged bulk spectral gap
\begin{equation}\begin{split}\label{eq:disorderaveragedspectralgap1}
\langle\text{SG}\rangle=2-|W_{1}|.
\end{split}\end{equation}
\noindent The disorder averaged spectral gap will vanish when $\langle\text{SG}\rangle=0$ at precisely the FO value of $|W_{1}^*|=2$. This is illustrated in the numerical calculations shown Fig. \ref{fig:disorderaveragedspectralgap}, which also indicate that the relationship holds away from the dimerized limit.

\begin{figure}[t!]
\includegraphics[scale=0.4]{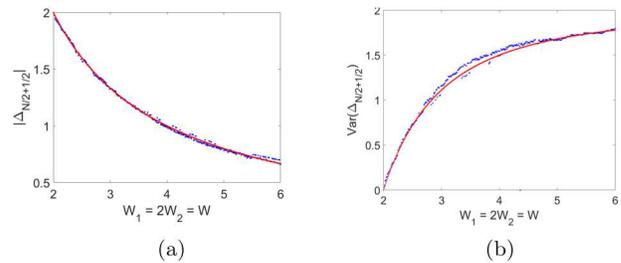}
\caption{Plots of (a) mean (disorder-averaged) inversion topological invariant $\Delta_{\mathcal{X}=\frac{N}{2}+\frac{1}{2}}$ and (b) the variance of $\Delta_{\mathcal{X}=\frac{N}{2}+\frac{1}{2}}$, past the disorder averaged spectral gap closing. Both plots were constructed for a chain of $N=300$ sites with $m=0.5$, $t=1$, and $\frac{W_{2}}{W_{1}}=2$. Each blue dot in (a) is the mean value of the inversion topological invariant at each disorder strength computed over $1000$ disorder configurations, and each blue dot in (b) is the variance of the inversion topological invariant at each disorder strength also computed over $1000$ disorder configurations. The red curve in (a) is the plot of (\ref{eq:meanmain}) and the red curve in (b) is the plot of (\ref{eq:variancemain}).}
\label{fig:variance}
\end{figure}

We have now shown that the fluctuations in the inversion topological invariant $\Delta_{\mathcal{X}=\frac{N}{2}+\frac{1}{2}}$ can be connected to the closing of the disorder averaged spectral gap, but now let us provide some physical intuition. For $|W_{1}|<|W_{1}^*|$, the system is gapped and the occupied states (states with $E<E_{F}=0$) and unoccupied states (states with $E>E_{F}=0$) can be clearly distinguished. For $|W_{1}|>|W_{1}^*|$, there is no longer a well defined gap in the energy spectrum, and hence no clear distinction between the two sets of states. Specifically, states that were previously denoted as occupied states before the gap closed can become unoccupied states and vice versa. As a result, the eigenstates associated with these energies will undergo exchanges between the occupied and unoccupied subspaces. Since the inversion topological invariant is dominated by the occupied eigenstates that are localized at the inversion centers, we examine the energies of the occupied and unoccupied states at $x=\frac{N}{2}$ and $x=N$. The energies of the occupied and unoccupied eigenstates at the inversion centers in the dimerized limit are given by,
\begin{equation}\begin{split}\label{eq:energyeigenvalues2}
E_{\frac{N}{2},\pm}^{(n)}=\pm t_{\frac{N}{2}}^{(n)}=\pm(1+W_{1}\omega_{\frac{N}{2}}^{\prime (n)})\\
E_{N,\pm}^{(n)}=\pm t_{N}^{(n)}=\pm(1+W_{1}\omega_{N}^{\prime (n)}).
\end{split}\end{equation}
\noindent When either $t_{\frac{N}{2}}^{(n)}$ or $t_{N}^{(n)}$ changes sign, there will be one exchange of occupied and unoccupied states. When both $t_{\frac{N}{2}}^{(n)}$ and $t_{N}^{(n)}$ change sign, then two exchanges will occur. This is precisely when $|\omega_{\frac{N}{2}}^{\prime (n)}|>\frac{1}{|W_{1}|}$ and/or when $|\omega_{N}^{\prime (n)}|>\frac{1}{|W_{1}|}$. The maximum value of $|\omega_{\frac{N}{2}}^{\prime (n)}|$ or $|\omega_{N}^{\prime (n)}|$ is $\frac{1}{2}$ (since $\omega_{\frac{N}{2}}^{\prime (n)},\omega_{N}^{\prime (n)}\in\left[-\frac{1}{2},\frac{1}{2}\right]$). This means $\frac{1}{2}>\frac{1}{|W_{1}|}$ which results in $|W_{1}|>|W_{1}^*|=2$. Therefore, for $|W_{1}|>|W_{1}^*|=2$, fluctuations in the inversion topological invariant $\Delta_{\mathcal{X}=\frac{N}{2}+\frac{1}{2}}$ onset simultaneously with the vanishing of the disorder averaged spectral gap. These arguments hold even when $m$ and $W_{2}$ are generically non-zero as we can show numerically as shown in Figs. \ref{fig:disorderaveragedspectralgap} and \ref{fig:variance}, or even analytically if we treat $m$ perturbatively (for details see SM C). We note that exchanges of the occupied and unoccupied states occur throughout the chain when disorder is added, even at sites away from the inversion center. However, the inversion topological invariant obtains its largest nonzero contributions from exchanges that are localized at or near the inversion centers. 

Once the FO disorder strength is reached  the inversion topological invariant $\Delta_{\mathcal{X}=\frac{N}{2}+\frac{1}{2}}$ fluctuates between the values of $-2$, $0$, and $2$ for each disorder configuration. The mean and variance of the distribution of $\Delta_{\mathcal{X}=\frac{N}{2}+\frac{1}{2}}$ can be determined analytically in the thermodynamic limit $N\to\infty$ when the model is tuned to the dimerized limit $m=0$, $t=1$ with disorder strengths $W_{1}\neq 0$ and $W_{2}=0$ ($\frac{W_{2}}{W_{1}}=0$). The resulting expressions for the mean and variance shown below also hold for other values of $m$, $t$, $W_{1}$ and $W_{2}$ away from the dimerized limit, which is illustrated in our numerical results (see Fig. \ref{fig:variance}), and can be demonstrated through a simple perturbation theory analysis (see SM C). The mean is,
\begin{equation}\begin{split}\label{eq:meanmain}
\langle\Delta_{\mathcal{X}=\frac{N}{2}+\frac{1}{2}}\rangle=-\frac{4}{|W_{1}|}\hspace{0.2cm}\text{for}\hspace{0.2cm}|W_{1}|\geq|W_{1}^*|=2,
\end{split}\end{equation}
and the variance is,
\begin{equation}\begin{split}\label{eq:variancemain}
\text{Var}(\Delta_{\mathcal{X}=\frac{N}{2}+\frac{1}{2}})=2-\frac{8}{|W_{1}|^{2}}\hspace{0.2cm}\text{for}\hspace{0.2cm}|W_{1}|\geq|W_{1}^*|=2.
\end{split}\end{equation}
The derivation of the mean and variance of the fluctuations in $\Delta_{\mathcal{X}=\frac{N}{2}+\frac{1}{2}}$ can be found in SM C. At the FO value $|W_{1}^*|=2$, the mean $\langle\Delta_{\mathcal{X}=\frac{N}{2}+\frac{1}{2}}\rangle=-2$ and the variance is $0$, indicating that the inversion topological invariant is still quantized and non-fluctuating at this disorder strength. For $|W_{1}|>|W_{1}^*|$, the disorder averaged value (mean) of the inversion topological invariant deviates from its quantized value of $-2,$ and the variance becomes nonzero.
\section{Establishing Simplified Position-Space Topological Invariants Using a Renormalization Group Procedure at Strong Disorder}\label{sec:simplifiedpositionspaceinvariants}
\indent At this point we have shown from the phase diagrams in Figs. \ref{fig:numericalphasediagrams2} and \ref{fig:numericalphasediagrams3} that the inversion topological invariant can fluctuate at strong disorder unlike the chiral winding number and the bulk polarization, which remain quantized and non-fluctuating within each non-critical region of the phase diagram. We have demonstrated that the onset of fluctuations in the inversion topological invariant is caused by the spectral gap closing where exchanges between the occupied and unoccupied states can occur at the inversion centers. The difference in how the inversion topological invariant, the chiral winding number, and the bulk polarization all behave at strong disorder indicates that the relationship between all three quantities given by (\ref{eq:topologicalinvariantrelationship}) in the clean limit breaks down in the strong disorder limit.

In this section, we will establish that the relationship between the inversion topological invariant and the bulk polarization and separately, that the relationship between the inversion topological invariant and chiral winding number no longer hold at strong disorder. However, we will also show that the relationship between the chiral winding number and bulk polarization persists at strong disorder. We do this by computing the asymptotic ground state of the system through a position-space renormalization group (RG) method that is asymptotically exact in the thermodynamic limit, and in the limit of disorder strength going to infinity \cite{refael2009criticality}. Using the ground state obtained from this RG method, we analytically derive the inversion topological invariant, chiral winding number, and bulk polarization and compare them.

 We first map the Hamiltonian given by (\ref{eq:disorderedHamiltonian1}) to a spin-$\frac{1}{2}$ Hamiltonian defined on a lattice of size $2N$ via the Jordan-Wigner transformation
\begin{equation}\begin{split}\label{eq:JordanWignertransformation}
c_{n,A}=K(2n-1)S_{2n-1}^{-}\hspace{0.5cm}c_{n,B}=K(2n)S_{2n}^{-},
\end{split}\end{equation}
where $S_{i}^{a}$ are spin-$\frac{1}{2}$ variables, and $K(m)=\exp\left\{i\pi\sum\limits_{j=1}^{m-1}S_{j}^{+}S_{j}^{-}\right\}$ is a string operator. These transformations lead to the Hamiltonian
\begin{equation}\begin{split}\label{eq:XXmodel}
&H=\sum\limits_{n=1}^{N}[2t_{n}(S_{2n}^{x}S_{2n+1}^{x}+S_{2n}^{y}S_{2n+1}^{y})\\
&+2m_{n}(S_{2n}^{x}S_{2n-1}^{x}+S_{2n}^{y}S_{2n-1}^{y})].
\end{split}\end{equation}
The Hamiltonian above shows that the exchange couplings $2m_{i}$ occur on the odd bonds, while the exchange couplings $2t_{i}$ occur on the even bonds. We will consider the Hamiltonian in (\ref{eq:XXmodel}) with the disorder configuration about the inversion center $\mathcal{X}=N+\frac{1}{2}$ ($m_{n}=m_{N+1-n}$ and $t_{n}=t_{N-n}$), and treat the system with periodic boundary conditions (note that we denote the inversion center as $\mathcal{X}=N+\frac{1}{2}$ since we are now considering a lattice of size $2N$). The Hamiltonian given by (\ref{eq:XXmodel}) is an inversion-symmetric spin-$\frac{1}{2}$ XX model with random exchange couplings $2t_{n}$ ($2m_{n+1}$) between the even (odd) bonds.

Each step in the position-space RG method consists of replacing a pair of spins that have the strongest exchange interaction by enforcing a spin-singlet state for that pair, and then generating a new and weaker bond between the neighboring spins. However, because of the spatial inversion symmetry, the exchange couplings appearing on one half of the chain will also appear on the other half, while the exchange couplings at the inversion centers ($t_{\frac{N}{2}}$ and $t_{N}$) are arbitrary. As a result, the position-space RG method needs to be handled carefully, both away from the inversion centers and at the inversion centers (details in SM D).

The nature of the RG procedure generates singlets that never cross each other. Every singlet state formed during this RG procedure will be formed by one spin belonging to sublattice A and another spin belonging to sublattice B (which occurs due to the underlying chiral symmetry). There are two types of singlets that form during this procedure: singlets that are inversion-symmetric partners with each other that form away from the inversion centers, and singlets that form across the inversion centers. The end of the RG procedure is reached when there are a total of $N$ singlets formed. We denote the number of singlets formed across the inversion centers as $M$, and the remaining $N-M$ of these singlets are formed away from the inversion centers.  Importantly, due to the inversion symmetry there are $\frac{1}{2}\left(N-M\right)$ inversion-symmetric pairs of these singlets. The $n^{\text{th}}$ singlet is associated to a pair of numbers $d_{n}=\{d_{n1},d_{n2}\}$ which specify the sites of the two spins in the singlet. The asymptotic ground state is:
\begin{equation}\begin{split}\label{eq:groundstate}
|\Omega\rangle=\prod\limits_{i=1}^{\frac{1}{2}\left(N-M\right)}\prod\limits_{j=1}^{M}\left(\frac{1}{\sqrt{2}}(S_{2N+2-2d_{i2}}^{+}-S_{2N+1-2d_{i1}}^{+})\right)\\
\times\left(\frac{1}{\sqrt{2}}(S_{2d_{j}}^{+}-S_{2N+1-2d_{j}}^{+})\right)\\
\times\left(\frac{1}{\sqrt{2}}(S_{2d_{i1}}^{+}-S_{2d_{i2}-1}^{+})\right)|\downarrow\cdots\downarrow\rangle.
\end{split}\end{equation}
\indent Mapping this back to the fermion representation, the ground state can be simplified to
\begin{equation}\begin{split}\label{eq:fermiongroundstate}
|\Omega\rangle=\prod\limits_{i=1}^{N}\left(\frac{1}{\sqrt{2}}(\alpha_{i}c_{d_{i1},B}^{\dagger}-\beta_{i}c_{d_{i2},A}^{\dagger})\right)|0\rangle,
\end{split}\end{equation}
where $\alpha_{i}$ and $\beta_{i}$ are coefficients that have unit modulus. When mapping (\ref{eq:groundstate}) back to the fermion representation, the $\alpha_{i}$ and $\beta_{i}$ are constructed by accumulating the product of coefficients obtained by successively moving Jordan-Wigner string operators towards the vacuum $|0\rangle$ [this is detailed in SM E and appears in multiple steps in the simplification of the ground state, specifically (S74)-(S75), (S80)-(S82), (S93)-(S94), (S103)-(S108) and (S112)-(S113)]. Note that because we have mapped this back to the fermionic representation, this form of the ground state is expressed over a lattice of size $N$. The ground state in this representation is a Slater determinant of single-particle states of the form
\begin{equation}\begin{split}\label{eq:singleparticlestates}
|\psi_{i}\rangle=\frac{1}{\sqrt{2}}(\alpha_{i}c_{d_{i1},B}^{\dagger}-\beta_{i}c_{d_{i2},A}^{\dagger})|0\rangle,
\end{split}\end{equation}
which only have weight on two sites, where the sites can have arbitrary distance from each other. We call this  distance the length of the singlet, and we define it as the difference in unit cell indices, i.e., for (\ref{eq:singleparticlestates}) this is given by $d_{i2}-d_{i1}$. When the ground state is expressed in this form, there are three notable properties that are enforced by the RG procedure:\\

\noindent (1) The singlets formed across the inversion centers are labeled by the index $i$ where $1\leq i\leq M$. The numbers $d_{i1}$ and $d_{i2}$ are expressed in terms of a single number $d_{i}$ such that $d_{i1}\equiv d_{i}$ and $d_{i2}=N+1-d_{i}$.\\

\noindent (2) The singlets formed in pairs away from the inversion centers are labeled by the index $i$ where $M+1\leq i\leq N$. An inversion-symmetric pair of singlets consists of one singlet between sites $d_{i1}$ and $d_{i2}$, and separately, another singlet between sites $N+1-d_{i2}$ and $N+1-d_{i1}$.\\

\noindent (3) In general, no two intervals $[d_{i1},d_{i2}]$ and $[d_{j1},d_{j2}]$ with $i\neq j$ can overlap in such a way that \textit{only one} of the ends of one interval is contained in the other. Otherwise, this violates the non-crossing nature of the singlets.\\

\noindent Using the fermionic basis, the inversion topological invariant simplifies to the following expression 
\begin{equation}\begin{split}\label{eq:inversioninvariant8}
\Delta_{\mathcal{X}=\frac{N}{2}+\frac{1}{2}}=-\frac{1}{2}\sum\limits_{i=1}^{M}[\alpha_{i}\beta_{i}^*+\beta_{i}\alpha_{i}^*].
\end{split}\end{equation}
This expression solely involves the singlets crossing the inversion center. The ground state in the fermionic representation also leads to the following expression for the chiral winding number
\begin{equation}\begin{split}\label{eq:windingnumber3}
\nu=\frac{1}{N}\sum\limits_{i=1}^{N}(d_{i2}-d_{i1}),
\end{split}\end{equation}
which is precisely the same as the simplified form of the chiral winding number derived in \cite{mondragon2014topological} with no inversion symmetry present. Thus, in the strong disorder limit, the chiral winding number $\nu$ takes on a simple form given by the sum of the singlet lengths in the ground state.\\
\begin{figure}[t!]
\includegraphics[scale=0.25]{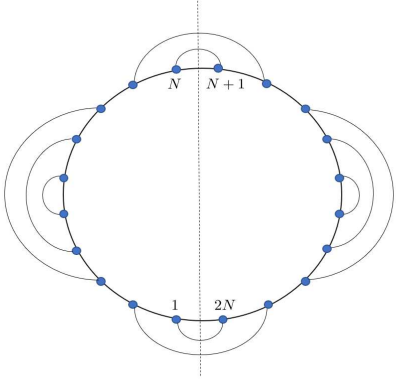}
\caption{Example of an inversion symmetric random singlet ground state after the position-space renormalization group. The lattice has $2N=20$ sites with periodic boundary conditions, and we labeled the sites nearest the inversion centers. The dashed line indicates the inversion center fixed by the disorder configuration. For this example, there are $M=4$ singlets over one of the inversion centers, and $\frac{1}{2}\left(N-M\right)=3$ inversion-symmetric pairs of singlets formed away from the inversion center.}
\label{fig:RGgroundstate}
\end{figure}

Fig. \ref{fig:RGgroundstate} shows an example of a random singlet ground state after implementing the position-space renormalization group. This mapping to the spin model can be used to understand the nature of the topological phase transition. At strong disorder, (\ref{eq:windingnumber3}) shows that the chiral winding number $\nu$ is given by the average singlet length where $d_{i2}-d_{i1}$ is the length of the $i^\text{th}$ singlet for $i=1,\ldots,N$. In the topological phase, the ground state of the system has singlets on the even bonds. As the system approaches criticality, the localization length diverges and singlets are formed over all length scales. Once the system has passed criticality, the singlets are formed over the odd bonds, which corresponds to a trivial phase. Furthermore, comparing (\ref{eq:inversioninvariant8}) and (\ref{eq:windingnumber3}) reveals that the inversion topological invariant $\Delta_{\mathcal{X}=\frac{N}{2}+\frac{1}{2}}$ and the chiral winding number $\nu$ behave differently near criticality (defined by the phase boundary where the localization length diverges) despite the two quantities having similar phase diagrams as evidenced in Figs. \ref{fig:numericalphasediagrams1}-\ref{fig:numericalphasediagrams3}. The inversion topological invariant $\Delta_{\mathcal{X}=\frac{N}{2}+\frac{1}{2}}$  depends only on the set of coefficients $\{\alpha_{i}\}_{i=1}^{M}$ and $\{\beta_{i}\}_{i=1}^{M}$ for the singlets over the inversion centers, whereas  the chiral winding number has an explicit dependence on the \textit{lengths} of all the singlets formed. As mentioned in the previous section, when the spectral gap closes, the energies of the occupied and unoccupied states undergo exchanges, and so do the corresponding states. When the corresponding states undergo exchanges, this alters the set of coefficients $\{\alpha_{i}\}_{i=1}^{M}$ and $\{\beta_{i}\}_{i=1}^{M}$ which in turn, can shift the value of the inversion topological invariant. This is consistent with the phase diagrams shown in Figs. \ref{fig:numericalphasediagrams2}-\ref{fig:numericalphasediagrams3} since at $|W_{1}^*|=2$, the spectral gap closes. Thus from the RG picture, the inversion topological invariant and the chiral winding number behave differently because the inversion topological invariant changes either when the gap closes or when criticality is reached, as opposed to the chiral winding number which only changes when the system becomes critical. This suggests that the relation $\Delta_{\mathcal{X}=\frac{N}{2}+\frac{1}{2}}=2\nu\hspace{0.05cm}(\text{mod}\hspace{0.1cm}4)$ no longer holds at strong disorder.\\
\begin{figure}[ht]
\includegraphics[scale=1.1]{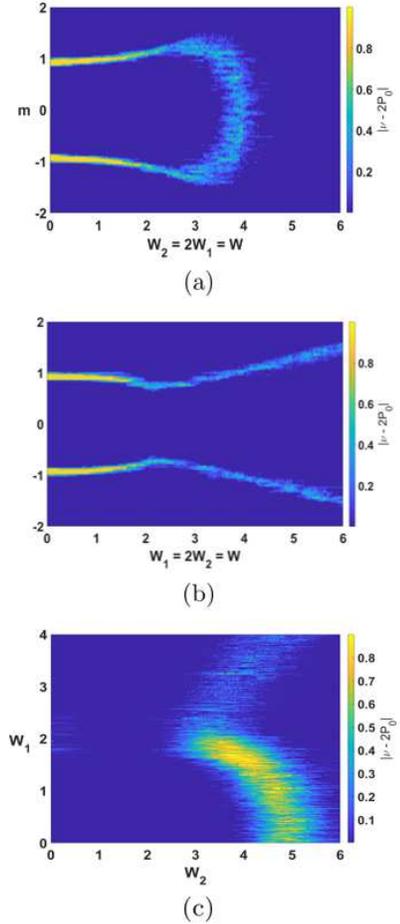}
\caption{Plots of the difference $|\nu-2P_{0}|$ between the winding number $\nu$ and the bulk polarization $P_{0}$. (a) and (b) were constructed for a chain of $N=500$ sites for $\frac{W_{2}}{W_{1}}=2$ and $\frac{W_{2}}{W_{1}}=\frac{1}{2}$ respectively, while (c) was constructed for a chain of $N=400$ sites for $m=0.5$. Each plot was disorder averaged over $10$ configurations.}
\label{fig:numericalphasediagrams4}
\end{figure}

Motivated by the results from the position-space RG procedure, we contrast the behavior of the inversion topological invariant $\Delta_{\mathcal{X}=\frac{N}{2}+\frac{1}{2}},$ and the chiral winding number $\nu,$ with the bulk polarization $P_{0}$. The bulk polarization $P_{0}$ can be computed using the ground state (\ref{eq:fermiongroundstate})
\begin{equation}\begin{split}\label{eq:polarization3}
P_{0}=\frac{1}{N}\sum\limits_{n=1}^{N}\left(\frac{1}{2}(d_{n1}+d_{n2})\right).
\end{split}\end{equation}
This expression for $P_{0}$ illustrates that the contributions to the polarization come from the location of each singlet center. At strong disorder, (\ref{eq:polarization3}) shows that the bulk polarization $P_{0}$ is given by the average of the singlet centers (e.g., for the $i^{\text{th}}$ singlet, its center is given as $\frac{1}{2}(d_{i1}+d_{i2})$). Within the topological phase, the singlet centers are located at the midpoints between lattice sites, which directly corresponds to the singlets formed over even bonds. As criticality is approached, singlets are formed over all length scales and the singlet centers begin to shift. When the system has passed criticality, the singlet centers are located on the lattice sites themselves, which directly corresponds to the singlets formed over odd bonds, signaling the trivial phase. Therefore, because the singlet centers shift at criticality, and not when the spectral gap closes, this signifies a major difference in $P_{0}$ and $\Delta_{\mathcal{X}=\frac{N}{2}+\frac{1}{2}}$. Hence, the relation $\Delta_{\mathcal{X}=\frac{N}{2}+\frac{1}{2}}=4P_{0}\hspace{0.05cm}(\text{mod}\hspace{0.1cm}4)$ does not hold at strong disorder.\\

Finally, to complete the discussion, we now consider the relationship between $\nu$ and $P_{0}$ at strong disorder. Comparing (\ref{eq:windingnumber3}) and (\ref{eq:polarization3}), we note that $N(\nu-2P_{0})=-\sum\limits_{n=1}^{N}2d_{n1}\in 2\mathbb{Z}$, which implies that $N\nu=2NP_{0}\hspace{0.05cm}(\text{mod}\hspace{0.1cm}2)$ and therefore $\nu=2P_{0}\hspace{0.05cm}(\text{mod}\hspace{0.1cm}2)$. Thus, the relation between the chiral winding number and the bulk polarization established in the clean limit holds when strong disorder is present. This is supported by the calculations shown in Fig. \ref{fig:numericalphasediagrams4} which contains plots of the difference $|\nu-2P_{0}|$ for $\frac{W_{2}}{W_{1}}=2$, $\frac{W_{2}}{W_{1}}=\frac{1}{2}$, and $m=0.5$ respectively. Fig. \ref{fig:numericalphasediagrams4} shows that throughout the phase diagram the relationshp holds, but that it seems to break down as the system approaches criticality, but this is likely  an artifact of finite size effects. 

\section{Effects of Filling Adjustment and Removing Disorder at the Inversion Centers}\label{sec:fillingadjustment}
To further illustrate the differences between the inversion topological invariant and the polarization, here we use a very simple violation of translation symmetry to break the clean-limit relationship. To provide a proof of concept we consider the effects of including additional occupied states localized at only the inversion centers. We will show that filling additional states at the inversion centers will cause the value of the inversion topological invariant to change, but will leave the polarization unaffected. In order to do this, we consider a 1D system comprised of $N$ unit cells with periodic boundary conditions in the clean limit, where each unit cell consists of two $s$ orbitals labeled $1$ and $2$. The Hamiltonian has the same form as (\ref{eq:realspacehamiltonian}), replacing the labels $A$ and $B$  with $1$ and $2$ respectively. We consider the inversion center $\mathcal{X}=\frac{N}{2}+\frac{1}{2}$ . When $N$ is even, in the dimerized limit where $m=0$ and $t\neq 0$, we have that $\Delta_{\mathcal{X}=\frac{N}{2}+\frac{1}{2}}=-2$, and the bulk polarization is $P_{0}=\frac{1}{2}$. The eigenstates in this limit are given by
\begin{equation}\begin{split}
\{|W_{\pm}(n)\rangle\}_{n=1}^{N}=\left\{\frac{1}{\sqrt{2}}(|n+1,1\rangle\pm|n,2\rangle)\right\}_{n=1}^{N}.
\end{split}\end{equation}

Now we add an additional $s$ orbital to each of the unit cells located at $\frac{N}{2}$ and $\frac{N}{2}+1$ as well as $1$ and $N$. The additional $s$ orbital is labeled as $3$ in each of these unit cells. In doing so, we introduce the following term to the Hamiltonian,
\begin{equation}\begin{split}
H'=\epsilon\sum\limits_{j=1}^{2}(\Phi_{+,j}^{\dagger}\Phi_{+,j}-\Phi_{-,j}^{\dagger}\Phi_{-,j}).
\end{split}\end{equation}
$\{\Phi_{\pm,j}\}_{j=1}^{2}$ are the states formed by hybridizing the two $s$ orbitals at each of the inversion centers, i.e.,
\begin{equation}\begin{split}
\Phi_{\pm,1}^{\dagger}|0\rangle=\frac{1}{\sqrt{2}}(c_{\frac{N}{2}+1,3}^{\dagger}\pm c_{\frac{N}{2},3}^{\dagger})|0\rangle=|\Phi_{\pm,1}\rangle
\end{split}\end{equation}
and
\begin{equation}\begin{split}
\Phi_{\pm,2}^{\dagger}|0\rangle=\frac{1}{\sqrt{2}}(c_{N,3}^{\dagger}\pm c_{1,3}^{\dagger})|0\rangle=|\Phi_{\pm,2}\rangle,
\end{split}\end{equation}
where the subscript $\pm$ indicates unoccupied and occupied states respectively. The additional states have energies $\pm\epsilon$. The addition of this term to the Hamiltonian breaks the translation symmetry, but preserves both the chiral and inversion symmetries. 
Indeed the states $\{|\Phi_{\pm,j}\rangle\}_{j=1}^{2}$ are eigenstates of the inversion operator $I_{\mathcal{X}=\frac{N}{2}+\frac{1}{2}}$:
\begin{equation}\begin{split}
I_{\mathcal{X}=\frac{N}{2}+\frac{1}{2}}|\Phi_{\pm,j}\rangle=\pm|\Phi_{\pm,j}\rangle,
\end{split}\end{equation}
where the inversion operator $I_{\mathcal{X}=\frac{N}{2}+\frac{1}{2}}$ has been modified to account for the additional states $|\Phi_{\pm}\rangle$ so that $[I_{\mathcal{X}=\frac{N}{2}+\frac{1}{2}},H]=0$:
\begin{equation}\begin{split}
&I_{\mathcal{X}=\frac{N}{2}+\frac{1}{2}}=\sum\limits_{n=1}^{N}(c_{N+1-n,1}^{\dagger}c_{n,2}+c_{N+1-n,2}^{\dagger}c_{n,1})\\
&+c_{\frac{N}{2},3}^{\dagger}c_{\frac{N}{2}+1,3}+c_{\frac{N}{2}+1,3}^{\dagger}c_{\frac{N}{2},3}+c_{N,3}^{\dagger}c_{1,3}+c_{1,3}^{\dagger}c_{N,3}.
\end{split}\end{equation}
Similarly, the chiral operator $S$ can be modified to account for these additional states so that $\{S,H\}=0$:
\begin{equation}\begin{split}
&S=\sum\limits_{j=1}^{N}(c_{j,1}^{\dagger}c_{j,1}-c_{j,2}^{\dagger}c_{j,2})\\
&+c_{1,3}^{\dagger}c_{1,3}-c_{\frac{N}{2},3}^{\dagger}c_{\frac{N}{2},3}+c_{\frac{N}{2}+1,3}^{\dagger}c_{\frac{N}{2}+1,3}-c_{N,3}^{\dagger}c_{N,3}.
\end{split}\end{equation}
The occupied projector now includes an additional term for the additional occupied states $|\Phi_{-,j}\rangle$ and is given by
\begin{equation}\begin{split}
P_{\text{occ}}=\sum\limits_{n=1}^{N}W_{n,-}^{\dagger}W_{n,-}+\sum\limits_{j=1}^{2}\Phi_{-,j}^{\dagger}\Phi_{-,j},
\end{split}\end{equation}
where $W_{\pm,n}^{\dagger}|0\rangle=\frac{1}{\sqrt{2}}(c_{n+1,1}^{\dagger}\pm c_{n,2}^{\dagger})|0\rangle=|W_{\pm}(n)\rangle$. Evaluating $P_{\text{occ}}I_{\mathcal{X}=\frac{N}{2}+\frac{1}{2}}P_{\text{occ}}$ and taking the trace yields
\begin{equation}\begin{split}
\Delta_{\mathcal{X}=\frac{N}{2}+\frac{1}{2}}=\text{Tr}[P_{\text{occ}}I_{\mathcal{X}=\frac{N}{2}+\frac{1}{2}}P_{\text{occ}}]=-4.
\end{split}\end{equation}
Hence, occupying the additional $s$ orbitals at both inversion centers changes the value of the inversion topological invariant $\Delta_{\mathcal{X}=\frac{N}{2}+\frac{1}{2}}$. Before filling the additional $s$ orbitals, the spectrum of $P_{\text{occ}}I_{\mathcal{X}=\frac{N}{2}+\frac{1}{2}}P_{\text{occ}}$ consisted of two more eigenstates that had inversion eigenvalues of $-1$ than eigenstates with inversion eigenvalues of $+1$. When taking the trace, eigenstates away from the inversion centers form inversion-symmetric pairs with inversion eigenvalues opposite each other, so their contributions cancel each other. The only non-zero contributions come from the eigenstates at the inversion centers, each of which have inversion eigenvalue $-1$, yielding $\Delta_{\mathcal{X}=\frac{N}{2}+\frac{1}{2}}=-2$ in the clean limit. Once the additional $s$ orbitals were occupied, two more negative inversion eigenvalues were included in the spectrum of $P_{\text{occ}}I_{\mathcal{X}=\frac{N}{2}+\frac{1}{2}}P_{\text{occ}}$, shifting the value of $\Delta_{\mathcal{X}=\frac{N}{2}+\frac{1}{2}}$ by $-2$. Hence, the value of the inversion topological invariant can be shifted by occupying additional states at the inversion centers.\\
\indent Unlike the inversion topological invariant, the value of the bulk polarization $P_{0}$ remains unaffected by such an addition. To illustrate this, we note that the position operator for periodic boundaries with the addition of the $s$ orbitals is
\begin{equation}\begin{split}
&X=\sum\limits_{n=1}^{N}e^{\frac{2\pi i}{N}n}(c_{n,1}^{\dagger}c_{n,1}+c_{n,2}^{\dagger}c_{n,2})\\
&-\left(c_{\frac{N}{2},3}^{\dagger}c_{\frac{N}{2},3}+e^{\frac{2\pi i}{N}}c_{\frac{N}{2}+1,3}c_{\frac{N}{2}+1,3}\right)\\
&+\left(c_{N,3}^{\dagger}c_{N,3}+e^{\frac{i\pi}{N}}c_{1,3}^{\dagger}c_{1,3}\right).
\end{split}\end{equation}
Evaluating the projected position operator $X_{P}=P_{\text{occ}}XP_{\text{occ}}$ in the thermodynamic limit $N\to\infty$ gives
\begin{equation}\begin{split}
&P_{\text{occ}}XP_{\text{occ}}=\sum\limits_{n=1}^{N}e^{\frac{2\pi i}{N}\left(n+\frac{1}{2}\right)}W_{n,-}^{\dagger}W_{n,-}\\
&+e^{\frac{2\pi i}{N}\left(\frac{N}{2}+\frac{1}{2}\right)}\Phi_{-,1}^{\dagger}\Phi_{-,1}+e^{\frac{2\pi i}{N}\left(N+\frac{1}{2}\right)}\Phi_{-,2}^{\dagger}\Phi_{-,2}.
\end{split}\end{equation}
The spectrum of $P_{\text{occ}}XP_{\text{occ}}$ is given by $\{\xi_{n}\}_{n=1}^{N+2}=\left\{e^{\frac{2\pi i}{N}\left(n+\frac{1}{2}\right)}\right\}_{n=1}^{N}\cup\left\{e^{\frac{2\pi i}{N}\left(\frac{N}{2}+\frac{1}{2}\right)},e^{\frac{2\pi i}{N}\left(N+\frac{1}{2}\right)}\right\}$, where, compared to the result at the end of Sec. \ref{subsec:positionspacerepresentation}, there are additional eigenvalues with arguments (after multiplying by $\frac{N}{2\pi}$) of $\frac{N}{2}+\frac{1}{2}$ and $N+\frac{1}{2}$ in the spectrum of $P_{\text{occ}}XP_{\text{occ}}$. These correspond to the additional eigenstates filled by the two coupled $s$ orbitals at each inversion center. We compute the polarization $P_{0}$ using (\ref{eq:polarization}) which gives $P_{0}=\frac{1}{2}$. Thus, the bulk polarization $P_{0}$ is unchanged under the filling of additional $s$ orbitals at the inversion centers.\\
\begin{figure}[t!]
\includegraphics[scale=0.425]{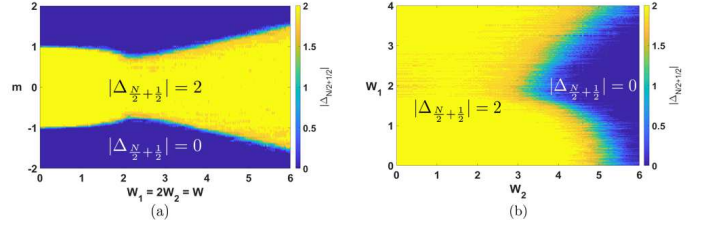}
\caption{Phase diagrams of the inversion topological invariant $\Delta_{\mathcal{X}=\frac{N}{2}+\frac{1}{2}}$ when no disorder is placed at the inversion centers for \textit{all} disorder configurations (i.e., $\omega_{\frac{N}{2}}^{\prime (n)}=\omega_{N}^{\prime (n)}=0$, but $\omega_{i}^{\prime (n)}\neq 0\hspace{0.2cm}\forall i\neq\frac{N}{2},N$ and $\forall n\in\{1,\ldots,N_{\text{configs}}\}$). The phase diagram in (a) is for the regime $\frac{W_{2}}{W_{1}}=\frac{1}{2}$ and in (b) is for when $m=0.5.$ One can directly compare with Figs. \ref{fig:numericalphasediagrams2} (a) and \ref{fig:numericalphasediagrams3} (a) where the inversion topological invariant experienced fluctuations past $|W_{1}|>|W_{1}^*|=2$. With no disorder placed at the inversion centers, the fluctuations are removed completely. These plots were constructed for a chain of $N=500$ sites disorder averaged over $10$ configurations.}
\label{fig:numericalphasediagrams5}
\end{figure}

It is worthwhile to note that just as how filling additional states at the inversion centers causes the value of the inversion topological invariant $\Delta_{\mathcal{X}=\frac{N}{2}+\frac{1}{2}}$ to change, removing disorder from the inversion centers can also eliminate the fluctuations present in the phase diagrams, and stabilize the inversion invariant as shown in Fig. \ref{fig:numericalphasediagrams5}. Specifically, if for every disorder configuration $n=1,\ldots,N_{\text{configs}}$, we fix $\omega_{\frac{N}{2}}^{\prime (n)}=\omega_{N}^{\prime (n)}=0$, then from (\ref{eq:energyeigenvalues2}) we have that the energy eigenvalues at the inversion centers are $E_{\frac{N}{2},\pm}^{(n)}=\pm t_{\frac{N}{2}}^{(n)}=\pm 1$ and $E_{N,\pm}^{(n)}=\pm t_{N}^{(n)}=\pm 1$. Hence, no exchanges can ever occur at the inversion centers and $\Delta_{\mathcal{X}=\frac{N}{2}+\frac{1}{2}}=-2$. Note that disorder is still applied to all of the sites away from the inversion centers (i.e. $\omega_{i}^{\prime (n)}\neq 0\hspace{0.2cm}\forall i\neq\frac{N}{2},N$). This results in the inversion topological invariant becoming non-fluctuating, as shown in Fig. \ref{fig:numericalphasediagrams5}, further illustrating how the inversion topological invariant can be altered by adjusting the states at the inversion centers.\\
\section{Conclusion}\label{sec:conclusion}
In this work, we have given a complete picture of the disordered 1D inversion-symmetric chain with chiral symmetry. We showed that with disorder that preserves inversion symmetry, the inversion topological invariant exhibits a similar phase diagram to other topological invariants such as the chiral winding number and bulk polarization. However, these phase diagrams differ when the bulk energy gap closes, causing the inversion topological invariant to deviate from its quantized value since occupied and unoccupied states at the inversion centers exchange at half-filling. This results in fluctuations that occur between a fixed set of values depending on whether one or both states at the inversion centers exchange for a given set of disorder configurations, thereby changing the inversion eigenvalues of these states. The mean and variance of these fluctuations past the closing of the spectral gap were determined in limits where either only the intercell or intracell hoppings are nonzero. The results from a position-space RG calculation illustrate how singlet states that are formed across the inversion centers determine the value of the inversion topological invariant, as opposed to the chiral winding number which has a contribution from each of the states throughout the 1D chain. This property also extends to situations where additional states are applied to the inversion centers, in which only the inversion topological invariant changes but quantities such as the bulk polarization remain invariant.\\
\indent Previous experimental realizations of disordered chiral-symmetric BDI and AIII chains have been performed in ultra-cold atomic systems \cite{meier2018observation}. Given that the model studied in this work is still fundamentally based in the chiral-symmetric BDI class, our findings could potentially be realized experimentally in ultra-cold atomic systems. We note that the Hamiltonian in (\ref{eq:disorderedHamiltonian1}) has been previously studied in \cite{meier2018observation}, but only with uncorrelated, chiral-symmetric disorder. Using a spectroscopic Hamiltonian engineered by driving lasers into a weakly-trapped Bose-Einstein condensate of $^{87}\text{Rb}$ atoms, the spatial periodicity of the interference pattern between the lasers results in a set of discrete momentum states that can be considered as effective sites of a synthetic lattice that replicates the Hamiltonian in (\ref{eq:disorderedHamiltonian1}). The tunneling energies $m_{j}$ and $t_{j}$ in (\ref{eq:disorderconfig1}) are produced in this setup by simultaneously driving many two-photon Bragg transitions between the applied laser fields, and can be precisely controlled by manipulating the amplitudes and phases of the laser fields. This allows for controllable disorder to be introduced into the model. Therefore, with this experimental setup, it would be possible to produce the inversion-symmetric disorder configurations given by (\ref{eq:disorderconfig2}) and (\ref{eq:disorderconfig3}). Because the disorder is also chiral-symmetric, the mean chiral displacement given by the expectation value of the chiral displacement operator $\mathcal{C}=2\langle SX\rangle$\cite{cardano2017detection} can be utilized to probe the chiral winding number $\nu$, where $S$ is the chiral operator given by (\ref{eq:chiraloperator2}) and $X$ is the position operator for open boundary conditions. When disorder-averaged, the mean chiral displacement converges to the chiral winding number $\nu$. Therefore, when inversion-symmetric disorder is present, it should be viable to determine the chiral winding number $\nu$ using the mean chiral displacement in the same manner as in \cite{meier2018observation}.\\
\indent Our results provide a pathway to exploring the bulk nature of TCPs when crystalline-symmetry preserving disorder is present, and indicate that examining the behavior of the states at the fixed points of the symmetry group on the position-space lattice is vital to understanding the stability of TCPs. By utilizing projected symmetry operators to formulate topological crystalline invariants in position space, it is possible to extend this work to higher dimensions where one can study disordered systems preserving rotation symmetries or mirror symmetries.\\
\indent Furthermore, there is still an open question of how including interactions while maintaining inversion-symmetric disorder can affect the results obtained in this work (e.g., how it can affect the ground state of the topological crystalline phase). Such a scenario has been studied previously in many-body localized (MBL) systems\cite{iadecola2018quantum} but has not been considered in TCPs. Therefore, it would be interesting to study the effects of interactions in addition to point-group symmetric disorder on TCPs. \\
\section{Acknowledgments}\label{sec:acknowledgements}
S.V. thanks Penghao Zhu and Oleg Dubinkin for insightful discussions. S.V. is supported by the NSF Graduate Research Fellowship Program under Grant No. DGE - 1746047. B.B. acknowledges the support of the Alfred P. Sloan foundation, and the National Science Foundation under grant DMR-1945058. T.L.H. thanks the US Office of Naval Research (ONR) Multidisciplinary University Research Initiative (MURI) grant N00014-20-1-2325 on Robust Photonic Materials with High-Order Topological Protection and the US National Science Foundation (NSF) Emerging Frontiers in Research and Innovation (EFRI) grant EFMA-1641084 for support. This work made use of the Illinois Campus Cluster Program (ICCP) in conjunction with the National Center for Supercomputing Applications (NCSA) and which is supported by funds from the University of Illinois at Urbana-Champaign.

\bibliographystyle{apsrev4-1}
\bibliography{Disorder_Inversion_Symmetric_Chain_Bibliography}

\clearpage

\widetext

\section*{Supplemental Material (SM)}
\beginsupplement
\section*{SM A: Proof of the relation between the bulk polarization and the inversion topological invariant in the clean limit}\label{suppsec:proofofequivalence}

\noindent The bulk polarization in 1D inversion symmetric insulators is related to the inversion eigenvalues at the invariant momenta $k_{x}^{\text{inv}}=0,\pi$. Denote the total number of occupied bands to be $N$, where $N$ is even. This relation is given by \cite{hughes2011inversion,turner2012quantized}

\begin{equation}\begin{split}\label{eq:polarizationandinversioninvariant1}
(-1)^{2P_{0}}=\prod\limits_{n=1}^{N}\xi_{n,\mathcal{X}}(0)\xi_{n,\mathcal{X}}(\pi).
\end{split}\end{equation}

\noindent where $\xi_{n,\mathcal{X}=x+\rho}(k_{x}^{\text{inv}})=\pm 1$ denotes the inversion eigenvalue for the occupied band $n$ for inversion center $\mathcal{X}=x+\rho$ (where $x$ denotes a choice of lattice site in $\{1,\ldots,N\}$ and $\rho\in\left\{0,\frac{1}{2}\right\}$) at each inversion invariant momentum $k_{x}=0$ and $k_{x}=\pi$. This can be expressed as

\begin{equation}\begin{split}\label{eq:polarizationandinversioninvariant2}
(-1)^{2P_{0}}=\left(\prod\limits_{n=1}^{N}\xi_{n,\mathcal{X}}(0)\right)\left(\prod\limits_{n=1}^{N}\xi_{n,\mathcal{X}}(\pi)\right)=(-1)^{n_{\mathcal{X}}^{(-)}(0)+n_{\mathcal{X}}^{(-)}(\pi)}.
\end{split}\end{equation}

\noindent where $n_{\mathcal{X}}^{(\pm)}(k_{x}^{\text{inv}})$ denotes the number of occupied bands that have inversion eigenvalue $\pm 1$. Note that $n_{\mathcal{X}}^{(+)}(0)+n_{\mathcal{X}}^{(-)}(0)=n_{\mathcal{X}}^{(+)}(\pi)+n_{\mathcal{X}}^{(-)}(\pi)=N$. Therefore the above can be expressed as

\begin{equation}\begin{split}\label{eq:polarizationandinversioninvariant3}
(-1)^{2P_{0}}=(-1)^{\frac{1}{2}(2(n_{\mathcal{X}}^{(-)}(0)+n_{\mathcal{X}}^{(-)}(\pi)))}=(-1)^{N-\frac{1}{2}(n_{\mathcal{X}}^{(+)}(0)-n_{\mathcal{X}}^{(-)}(0))-\frac{1}{2}(n_{\mathcal{X}}^{(+)}(\pi)-n_{\mathcal{X}}^{(-)}(\pi))}.
\end{split}\end{equation}

\noindent Since $N$ is even one has

\begin{equation}\begin{split}\label{eq:polarizationandinversioninvariant4}
&2P_{0}=\frac{1}{2}((n_{\mathcal{X}}^{(+)}(0)-n_{\mathcal{X}}^{(-)}(0))+(n_{\mathcal{X}}^{(+)}(\pi)-n_{\mathcal{X}}^{(-)}(\pi)))\\
&=\frac{1}{2}\sum\limits_{k_{x}=0,\pi}[n_{\mathcal{X}}^{(+)}(k_{x})-n_{\mathcal{X}}^{(-)}(k_{x})]=\frac{1}{2}\Delta_{\mathcal{X}}\hspace{0.05cm}(\text{mod}\hspace{0.1cm}2),
\end{split}\end{equation}

\noindent At this point, we consider the inversion centers $\mathcal{X}=x$ and $\mathcal{X}=x+\frac{1}{2}$ separately. For $\mathcal{X}=x$, one has $\Delta_{\mathcal{X}=x}=0$ for $|m|<t$ (topological phase) and $\Delta_{\mathcal{X}=x}=-2$ for $|m|>t$ (trivial phase), whereas $P_{0}=\frac{1}{2}$ for $|m|<t$ and $P_{0}=0$ for $|m|>t$. This establishes the relation $\Delta_{\mathcal{X}=x}=4P_{0}\hspace{0.05cm}(\text{mod}\hspace{0.1cm}2)$. However, for $\mathcal{X}=x+\frac{1}{2}$, one has $\Delta_{\mathcal{X}=x+\frac{1}{2}}=-2$ for $|m|<t$ and $\Delta_{\mathcal{X}=x+\frac{1}{2}}=0$ for $|m|>t$, establishing the relation,

\begin{equation}\begin{split}\label{eq:polarizationandinversioninvariant5}
\Delta_{\mathcal{X}=x+\frac{1}{2}}=4P_{0}\hspace{0.05cm}(\text{mod}\hspace{0.1cm}4).
\end{split}\end{equation}

\noindent Since $\nu=2P_{0}\hspace{0.05cm}(\text{mod}\hspace{0.10cm}2)$ (where $\nu=1$ for $|m|<t$ and $\nu=0$ for $|m|>t$), it follows that $2\nu=4P_{0}\hspace{0.05cm}(\text{mod}\hspace{0.1cm}4)$ and therefore, $\Delta_{\mathcal{X}=x+\frac{1}{2}}=2\nu\hspace{0.05cm}(\text{mod}\hspace{0.1cm}4)$. This leads to the relation given by (\ref{eq:topologicalinvariantrelationship}) in the main text.

\section*{SM B: Analytic expression for the distribution of the inversion topological invariant in the dimerized limits}\label{suppsec:analyticdistribution}

\noindent For the case when $m=0$, which is one of the dimerized limits of the Hamiltonian (9) in the main text, the eigenstates for the occupied and unoccupied states (denoted by $-$ and $+$ respectively) are,

\begin{equation}\begin{split}
|W_{\mp}(n)\rangle=\frac{1}{\sqrt{2}}(|n+1,A\rangle\mp|n,B\rangle)
\end{split}\end{equation}

\noindent Thus, in the $m=0$ limit, the eigenstates are localized at neighboring sites $n$ and $n+1$. One can express the $\{|n,\sigma\rangle\}$ in terms of the Wannier basis $\{|W_{n}\rangle\}$,

\begin{equation}\begin{split}
|n,B\rangle=\frac{1}{\sqrt{2}}(|W_{+}(n)\rangle-|W_{-}(n)\rangle)
\end{split}\end{equation}

\begin{equation}\begin{split}
|n+1,A\rangle=\frac{1}{\sqrt{2}}(|W_{+}(n)\rangle+|W_{-}(n)\rangle)
\end{split}\end{equation}

\noindent We consider $\mathcal{X}=\frac{N}{2}+\rho$ where $\rho=\{0,\frac{1}{2}\}$ as the choice of inversion center. The inversion operators can be written in the $\{|n,\sigma\rangle\}$ basis where $n=1,\ldots,N$ and $\sigma=A,B$,

\begin{equation}\begin{split}
&I_{\mathcal{X}=\frac{N}{2}+\frac{1}{2}}=\sum\limits_{n=1}^{N}(|N-n+1,A\rangle\langle n,B|+|N-n+1,B\rangle\langle n,A|)\\
&I_{\mathcal{X}=\frac{N}{2}}=\sum\limits_{n=1}^{N}(|N-n,A\rangle\langle n,B|+|N-n,B\rangle\langle n,A|)
\end{split}\end{equation}

\noindent The projector over the occupied states is given in the Wannier basis by $P_{\text{occ}}=\sum\limits_{n=1}^{N}|W_{-}(n)\rangle\langle W_{-}(n)|$, and the projected inversion operators for each of the inversion centers are given by,

\begin{equation}\begin{split}
&P_{\text{occ}}I_{\mathcal{X}=\frac{N}{2}+\frac{1}{2}}P_{\text{occ}}=-\frac{1}{2}\sum\limits_{n=1}^{N}[|W_{-}(N-n)\rangle\langle W_{-}(n)|+|W_{-}(N-n+1)\rangle\langle W_{-}(n-1)|]\\
&P_{\text{occ}}I_{\mathcal{X}=\frac{N}{2}}P_{\text{occ}}=-\frac{1}{2}\sum\limits_{n=1}^{N}[|W_{-}(N-n-1)\rangle\langle W_{-}(n)|+|W_{-}(N-n)\rangle\langle W_{-}(n-1)|]
\end{split}\end{equation}

\noindent To determine the distribution of each invariant, we use $\Delta_{\mathcal{X}}(n)=\langle n|\text{Tr}'[P_{\text{occ}}I_{\mathcal{X}}P_{\text{occ}}]|n\rangle=\sum\limits_{\sigma=\{A,B\}}\langle n,\sigma|P_{\text{occ}}I_{\mathcal{X}}P_{\text{occ}}|n,\sigma\rangle$. We denote $\text{Tr}'$ to indicate that the trace is being performed over the local degrees of freedom ($\sigma=\{A,B\}$) within each unit cell. The invariant as a function of the lattice sites $n$ for the $m=0$ phase is therefore (for $N$ even),

\begin{equation}\begin{split}
\Delta_{\mathcal{X}=\frac{N}{2}+\frac{1}{2}}(n)=-\frac{1}{2}(\delta_{n,1}+\delta_{n,\frac{N}{2}}+\delta_{n,\frac{N}{2}+1}+\delta_{n,N})
\end{split}\end{equation}

\begin{equation}\begin{split}
\Delta_{\mathcal{X}=\frac{N}{2}}(n)=-\frac{1}{2}(\delta_{n,\frac{N}{2}-\frac{1}{2}}+\delta_{n,\frac{N}{2}+\frac{1}{2}})
\end{split}\end{equation}

\noindent which means the value of the invariant for each choice of the inversion center is,

\begin{equation}\begin{split}
\Delta_{\mathcal{X}=\frac{N}{2}+\frac{1}{2}}=\sum\limits_{n=1}^{N}\Delta_{\mathcal{X}=x+\frac{1}{2}}(n)=-\frac{1}{2}\sum\limits_{n=1}^{N}(\delta_{n,1}+\delta_{n,\frac{N}{2}}+\delta_{n,\frac{N}{2}+1}+\delta_{n,N})=-2
\end{split}\end{equation}

\begin{equation}\begin{split}
\Delta_{\mathcal{X}=\frac{N}{2}}=\sum\limits_{n=1}^{N}\Delta_{\mathcal{X}=x}(n)=0
\end{split}\end{equation}

\noindent One can also consider the second dimerized limit in which the inter-cell hopping is set to $t=0$. The corresponding eigenstates are,

\begin{equation}\begin{split}
|W_{\mp}(n)\rangle=\frac{1}{\sqrt{2}}(\mp|n,A\rangle+|n,B\rangle)
\end{split}\end{equation}

\noindent This means,

\begin{equation}\begin{split}
&|n,A\rangle=\frac{1}{\sqrt{2}}(|W_{+}(n)\rangle-|W_{-}(n)\rangle)\\
&|n,B\rangle=\frac{1}{\sqrt{2}}(|W_{+}(n)\rangle+|W_{-}(n)\rangle)
\end{split}\end{equation}

\noindent Therefore,

\begin{equation}\begin{split}
&P_{\text{occ}}I_{\mathcal{X}=\frac{N}{2}+\frac{1}{2}}P_{\text{occ}}=-\sum\limits_{n=1}^{N}|W_{-}(N-n+1)\rangle\langle W_{-}(n)|\\
&P_{\text{occ}}I_{\mathcal{X}=\frac{N}{2}}P_{\text{occ}}=-\sum\limits_{n=1}^{N}|W_{-}(N-n)\rangle\langle W_{-}(n)|
\end{split}\end{equation}

\noindent Once again using $\Delta_{\mathcal{X}}(n)=\langle n|\text{Tr}'[P_{\text{occ}}I_{\mathcal{X}}P_{\text{occ}}]|n\rangle=\sum\limits_{\sigma=\{A,B\}}\langle n,\sigma|P_{\text{occ}}I_{\mathcal{X}}P_{\text{occ}}|n,\sigma\rangle$, the invariant as a function of the lattice sites $n$ in this dimerized limit is therefore (for $N$ even),

\begin{equation}\begin{split}
\Delta_{\mathcal{X}=\frac{N}{2}+\frac{1}{2}}(n)=-\delta_{n,\frac{N}{2}+\frac{1}{2}}
\end{split}\end{equation}

\begin{equation}\begin{split}
\Delta_{\mathcal{X}=\frac{N}{2}}(n)=-\delta_{n,\frac{N}{2}}-\delta_{n,N}
\end{split}\end{equation}

\noindent which means the value of the invariant for each choice of the inversion center is,

\begin{equation}\begin{split}
\Delta_{\mathcal{X}=\frac{N}{2}+\frac{1}{2}}=\sum\limits_{n=1}^{N}\Delta_{\mathcal{X}=\frac{N}{2}+\frac{1}{2}}(n)=0
\end{split}\end{equation}

\begin{equation}\begin{split}
\Delta_{\mathcal{X}=\frac{N}{2}}=\sum\limits_{n=1}^{N}\Delta_{\mathcal{X}=\frac{N}{2}}(n)=-\sum\limits_{n=1}^{N}(\delta_{n,\frac{N}{2}}+\delta_{n,N})=-2
\end{split}\end{equation}

\section*{SM C: Derivation of the Statistics of the inversion topological invariant}\label{suppsec:statisticsderivation}

\noindent In this section, we provide the details on disorder averages taken in the thermodynamic limit as well as explicit calculations of the probabilities $P(\Delta_{\mathcal{X}=\frac{N}{2}+\frac{1}{2}}=-2)$, $P(\Delta_{\mathcal{X}=\frac{N}{2}+\frac{1}{2}}=0)$, and $P(\Delta_{\mathcal{X}=\frac{N}{2}+\frac{1}{2}}=2)$ that are necessary to compute the mean and variance of the fluctuations of the inversion topological invariant.\\

\noindent We will first show that the disorder average of the distribution of the minimums of the set of $N$ random numbers equals $-\frac{1}{2}$ in the thermodynamic limit where $N\to\infty$, and likewise for the maximums which equals $\frac{1}{2}$ (i.e. $\langle\min(\{\omega_{i}\}_{i=1}^{N})\rangle=-\frac{1}{2}$ and $\langle\max(\{\omega_{i}\}_{i=1}^{N})\rangle=\frac{1}{2}$, the definition of disorder average is provided in (\ref{eq:disorderaverage}) in the main text). In the main text, we consider the set of random numbers on each disorder configuration uniformly sampled from the interval $\left[-\frac{1}{2},\frac{1}{2}\right]$ however this result can easily be generalized to the uniform distribution from the interval $[a,b]$ so we instead provide proof for the claim that $\langle\min(\{\omega_{i}\}_{i=1}^{N})\rangle=a$ and $\langle\max(\{\omega_{i}\}_{i=1}^{N})\rangle=b$ in the thermodynamic limit $N\to\infty$.\\

\noindent In probability theory, for a given configuration $n$ we consider $\{\omega_{i}^{(n)}\}=\{\omega_{1}^{(n)},\ldots,\omega_{N}^{(n)}\}$ to be a set of independent identically distributed (i.i.d.) random variables. We take $n=1,\ldots,N_{\text{configs}}$ where $N_{\text{configs}}$ denotes the number of disorder configurations. Given some value $y$, the probability $P(\min(\{\omega_{i}^{(n)}\}_{i=1}^{N})\leq y)$ implies that at least one of $\omega_{i}^{(n)}$ is less than or equal to $y$. This means that the probability $P(\min(\{\omega_{i}^{(n)}\}_{i=1}^{N})\leq y)$ must be equal to one minus the probability $P(\omega_{1}^{(n)}>y,\ldots,\omega_{N}^{(n)}>y)$ or
\begin{equation}\begin{split}\label{eq:probability1}
P(\min(\{\omega_{i}^{(n)}\}_{i=1}^{N})\leq y)=1-P(\omega_{1}^{(n)}>y,\ldots,\omega_{N}^{(n)}>y)
\end{split}\end{equation}
\noindent The cumulative distribution function (CDF) of the random variable $\omega_{i}^{(n)}$ is defined to be the probability $P(\omega_{i}^{(n)}\leq y)$. For the uniform distribution of random variables sampled from the interval $[a,b]$, the CDF is given as
\begin{equation}
P(\omega_{i}^{(n)}\leq y)=\begin{cases} 
      0 & y< a \\\\
      \frac{y-a}{b-a} & a\leq y\leq b \\\\
      1 & y> b
   \end{cases}
\end{equation}
\noindent Note that since the set of random variables $\{\omega_{i}^{(n)}\}_{i=1}^{N}$ are independent, it follows straightforwardly that
\begin{equation}\label{eq:probability2}
P(\omega_{1}^{(n)}>y,\ldots,\omega_{N}^{(n)}>y)=\prod\limits_{i=1}^{N}P(\omega_{i}^{(n)}>y)=\prod\limits_{i=1}^{N}(1-P(\omega_{i}^{(n)}\leq y))=\begin{cases} 
      0 & y< a \\\\
      \left(\frac{b-y}{b-a}\right)^{N} & a\leq y\leq b \\\\
      1 & y> b
   \end{cases}
\end{equation}
\noindent Substituting (\ref{eq:probability2}) into (\ref{eq:probability1}) reveals
\begin{equation}\begin{split}
 P(\min(\{\omega_{i}^{(n)}\}_{i=1}^{N})\leq y)=\begin{cases} 
      0 & y< a \\\\
      1-\left(\frac{b-y}{b-a}\right)^{N} & a\leq y\leq b \\\\
      1 & y> b
   \end{cases}  
\end{split}\end{equation}
This is precisely the CDF of the distribution of $\min(\{\omega_{i}^{(n)}\}_{i=1}^{N})$ for the disorder configurations $n=1,\ldots,N_{\text{configs}}$. The probability density function (PDF) for this distribution can be derived from the CDF as follows,
\begin{equation}
p(y)\equiv\frac{dP(\min(\{\omega_{i}^{(n)}\}_{i=1}^{N})\leq y)}{dy}=\begin{cases} 
      0 & y< a \\\\
      \frac{N}{(b-a)^{N}}(b-y)^{N-1} & a\leq y\leq b \\\\
      1 & y> b
   \end{cases}  
\end{equation}
At this point, using the PDF one can obtain the expectation value of the distribution of minimums over the $n=1,\ldots,N_{\text{configs}}$ disorder configurations. Note that this expectation value over disorder configurations is the definition of disorder average given by (\ref{eq:disorderaverage}) in the main text. This means,
\begin{equation}\begin{split}
\langle\min(\{\omega_{i}\}_{i=1}^{N})\rangle=\int\limits_{-\infty}^{\infty}yp(y)\hspace{0.05cm}dy=\frac{N}{(b-a)^{N}}\int\limits_{a}^{b}y(b-y)^{N-1}\hspace{0.05cm}dy=\frac{N}{(b-a)^{N}}\int\limits_{0}^{b-a}(b-y)y^{N-1}\hspace{0.05cm}dy=\frac{1}{1+\frac{1}{N}}\left(\frac{b}{N}+a\right)
\end{split}\end{equation}
Taking the thermodynamic limit $N\to\infty$ one clearly sees that $\langle\min(\{\omega_{i}\}_{i=1}^{N})\rangle=a$. In this case, $a=-\frac{1}{2}$ and $b=\frac{1}{2}$ hence $\langle\min(\{\omega_{i}\}_{i=1}^{N})\rangle=-\frac{1}{2}$.\\

\noindent Repeating this argument for the distribution of maximums, we note that given some value $y$, the probability $P(\max(\{\omega_{i}^{(n)}\})\leq y)$ implies that every $\omega_{i}^{(n)}$ for $i=1,\ldots,N$ is less than or equal to $y$. This means that the probability $P(\max(\{\omega_{i}^{(n)}\})\leq y)$ is simply,
\begin{equation}
P(\max(\{\omega_{i}^{(n)}\}_{i=1}^{N})\leq y)=P(\omega_{1}^{(n)}\leq y,\ldots,\omega_{N}^{(n)}\leq y)=\prod\limits_{i=1}^{n}P(\omega_{i}^{(n)}\leq y)=\begin{cases} 
      0 & y< a \\\\
      \left(\frac{y-a}{b-a}\right)^{N} & a\leq y\leq b \\\\
      1 & y> b
   \end{cases}
\end{equation}
which is the CDF of the distribution of $\max(\{\omega_{i}^{(n)}\}_{i=1}^{N})$ for the disorder configurations $n=1,\ldots,N_{\text{configs}}$. The PDF is,
\begin{equation}
q(y)=\frac{dP(\max(\{\omega_{i}^{(n)}\}_{i=1}^{N})\leq y)}{dy}=\begin{cases} 
      0 & y< a \\\\
      \frac{N}{(b-a)^{N}}(y-a)^{N-1} & a\leq y\leq b \\\\
      1 & y> b
   \end{cases}
\end{equation}
\noindent Hence,
\begin{equation}\begin{split}
\langle\max(\{\omega_{i}\}_{i=1}^{N})\rangle=\int\limits_{-\infty}^{\infty}yq(y)\hspace{0.05cm}dy=\frac{N}{(b-a)^{N}}\int\limits_{a}^{b}y(y-a)^{N-1}\hspace{0.05cm}dy=\frac{N}{(b-a)^{N}}\int\limits_{0}^{b-a}(y+a)y^{N-1}\hspace{0.05cm}dy=\frac{1}{1+\frac{1}{N}}\left(\frac{a}{N}+b\right)
\end{split}\end{equation}
\noindent It follows that in the limit $N\to\infty$ that $\langle\max(\{\omega_{i}\}_{i=1}^{N})\rangle=b$. Since $a=-\frac{1}{2}$ and $b=\frac{1}{2}$, this means $\langle\max(\{\omega_{i}\}_{i=1}^{N})\rangle=\frac{1}{2}$.\\

\noindent We next provide the derivation of the probabilities $P(\Delta_{\mathcal{X}=\frac{N}{2}+\frac{1}{2}}=-2)$, $P(\Delta_{\mathcal{X}=\frac{N}{2}+\frac{1}{2}}=0)$, and $P(\Delta_{\mathcal{X}=\frac{N}{2}+\frac{1}{2}}=2)$ involved in computing the mean and variance.  Before proceeding, there are a couple of things to note in the following derivations. For simplicity, we will work out the cases for positive $W_{1}$ so that $\text{sgn}(W_{1})=1$. The expressions are identical for negative $W_{1}$, so we will express the following in terms of $|W_{1}|$. These expressions only characterize the behavior of the inversion topological invariant once the disorder averaged spectral gap has closed, hence they are only valid for $|W_{1}|>|W_{1}^*|=2$. Finally, we suppress the disorder configuration index $n$ for simplicity since we are interested in computing the mean and variance.\\

\noindent Since the disorder amplitudes at the inversion centers map to themselves under inversion, $\omega_{\frac{N}{2}}'$ and $\omega_{N}'$ can generically take on different values (i.e., $\omega_{i}'=\omega_{N-i}'$ so $\omega_{\frac{N}{2}}'$ and $\omega_{N}'$ can take on different values in general, this is illustrated in Fig. \ref{fig:disorderconfigs} of the main text). For $\Delta_{\mathcal{X}=\frac{N}{2}+\frac{1}{2}}=0$ only one exchange must occur. This means we must have either $t_{\frac{N}{2}}<0$ \textit{or} $t_{N}<0$. From (\ref{eq:energyeigenvalues2}) in the main text, this means $\omega_{\frac{N}{2}}'\in\left[-\frac{1}{2},-\frac{1}{W_{1}}\right)$ and $\omega_{N}'\in\left(-\frac{1}{W_{1}},\frac{1}{2}\right]$, \textit{or} vice versa. Hence
\begin{equation}\begin{split}\label{eq:oneexchange}
P(\text{one exchange})=P(\Delta_{\mathcal{X}=\frac{N}{2}+\frac{1}{2}}=0)\\
=P\left(\omega_{\frac{N}{2}}'\in\left[-\frac{1}{2},-\frac{1}{W_{1}}\right)\right)P\left(\omega_{N}'\in\left(-\frac{1}{W_{1}},\frac{1}{2}\right]\right)\\
+P\left(\omega_{\frac{N}{2}}'\in\left(-\frac{1}{W_{1}},\frac{1}{2}\right]\right)P\left(\omega_{N}'\in\left[-\frac{1}{2},-\frac{1}{W_{1}}\right)\right)\\
=2\left(\frac{1}{2}-\frac{1}{W_{1}}\right)\left(\frac{1}{2}+\frac{1}{W_{1}}\right)=\frac{1}{2}-\frac{2}{|W_{1}|^{2}}.
\end{split}\end{equation}
\noindent For $\Delta_{\mathcal{X}=\frac{N}{2}+\frac{1}{2}}=2$, two exchanges must occur. This means \textit{both} $t_{\frac{N}{2}}<0$ \textit{and} $t_{N}<0$. Therefore, $\omega_{\frac{N}{2}}',\omega_{N}'\in\left[-\frac{1}{2},-\frac{1}{W_{1}}\right)$ resulting in the following expression,
\begin{equation}\begin{split}\label{eq:twoexchanges}
P(\text{two exchanges})=P(\Delta_{\mathcal{X}=\frac{N}{2}+\frac{1}{2}}=2)\\
=P\left(\omega_{\frac{N}{2}}'\in\left[-\frac{1}{2},-\frac{1}{W_{1}}\right)\right)P\left(\omega_{N}'\in\left[-\frac{1}{2},-\frac{1}{W_{1}}\right)\right)\\
=\left(\frac{1}{2}-\frac{1}{W_{1}}\right)\left(\frac{1}{2}-\frac{1}{W_{1}}\right)=\left(\frac{1}{2}-\frac{1}{|W_{1}|}\right)^{2}.
\end{split}\end{equation}
\noindent Finally, for $\Delta_{\mathcal{X}=\frac{N}{2}+\frac{1}{2}}=-2$ there must be no exchanges, so $t_{\frac{N}{2}}>0$ \textit{and} $t_{N}>0$. This means $\omega_{\frac{N}{2}}',\omega_{N}'\in\left(-\frac{1}{W_{1}},\frac{1}{2}\right]$ hence
\begin{equation}\begin{split}\label{eq:zeroexchanges}
P(\text{zero exchanges})=P(\Delta_{\mathcal{X}=\frac{N}{2}+\frac{1}{2}}=-2)\\
=P\left(\omega_{\frac{N}{2}}'\in\left(-\frac{1}{W_{1}},\frac{1}{2}\right]\right)P\left(\omega_{N}'\in\left(-\frac{1}{W_{1}},\frac{1}{2}\right]\right)\\
=\left(\frac{1}{2}+\frac{1}{W_{1}}\right)\left(\frac{1}{2}+\frac{1}{W_{1}}\right)=\left(\frac{1}{2}+\frac{1}{|W_{1}|}\right)^{2}
\end{split}\end{equation}
and it follows that $P(\Delta_{\mathcal{X}=\frac{N}{2}+\frac{1}{2}}=-2)+P(\Delta_{\mathcal{X}=\frac{N}{2}+\frac{1}{2}}=0)+P(\Delta_{\mathcal{X}=\frac{N}{2}+\frac{1}{2}}=2)=1$. Given expressions (\ref{eq:oneexchange})-(\ref{eq:zeroexchanges}), one can compute the mean and variance of the inversion topological invariant $\Delta_{\mathcal{X}=\frac{N}{2}+\frac{1}{2}}$ at a given disorder strength $W_{1}$ for a set of disorder configurations. This yields the mean to be,
\begin{equation}\begin{split}\label{eq:mean}
\langle\Delta_{\mathcal{X}=\frac{N}{2}+\frac{1}{2}}\rangle=2\left(\frac{1}{2}-\frac{1}{|W_{1}|}\right)^{2}-2\left(\frac{1}{2}+\frac{1}{|W_{1}|}\right)^{2}\\=-\frac{4}{|W_{1}|},
\end{split}\end{equation}
and the variance to be,
\begin{equation}\begin{split}\label{eq:variance}
\text{Var}(\Delta_{\mathcal{X}=\frac{N}{2}+\frac{1}{2}})=\langle\Delta_{\mathcal{X}=\frac{N}{2}+\frac{1}{2}}^{2}\rangle-\langle\Delta_{\mathcal{X}=\frac{N}{2}+\frac{1}{2}}\rangle^{2}\\
=4\left(\left(\frac{1}{2}-\frac{1}{|W_{1}|}\right)^{2}-\left(\frac{1}{2}+\frac{1}{|W_{1}|}\right)^{2}\right)-\frac{16}{|W_{1}|^{2}}\\
=2-\frac{8}{|W_{1}|^{2}}.
\end{split}\end{equation}
At $|W_{1}^*|=2$, the mean $\langle\Delta_{\mathcal{X}=\frac{N}{2}+\frac{1}{2}}\rangle=-2$ and the variance is $0$, indicating that the inversion topological invariant is still quantized at this disorder strength. For $|W_{1}^*|>2$, the value of the invariant deviates from this value. The probabilities given by (\ref{eq:oneexchange})-(\ref{eq:zeroexchanges}) are expressed in terms of $\frac{1}{2}\pm\frac{1}{W_{1}}$ depending on whether $W_{1}$ is negative or positive. These expressions appear as exponents in the localization length given by (24) of the main text.\\

\noindent The analysis performed in (\ref{eq:energyeigenvalues1})-(\ref{eq:energyeigenvalues2}) in the main text as well as in (\ref{eq:oneexchange})-(\ref{eq:variance}) are for the limit in which $m=0$, $t=1$, and $\frac{W_{2}}{W_{1}}=0$. However, the results derived in (\ref{eq:oneexchange})-(\ref{eq:variance}) also hold for $m\neq 0$ and $\frac{W_{2}}{W_{1}}\neq 0$. This can be best illustrated using degenerate perturbation theory. When $m=0$ and $\frac{W_{2}}{W_{1}}=0$, we define,

\begin{equation}\begin{split}
H_{0}=\sum\limits_{n=1}^{N}\left[\frac{1}{2}t_{n}c_{n}^{\dagger}(\sigma_{1}-i\sigma_{2})c_{n+1}+\text{h.c}\right].
\end{split}\end{equation}

\noindent where $t_{n}=t+W_{1}\omega_{n}'$ where $\omega_{n}'=\omega_{N-n}'$ due to the inversion symmetry about $\mathcal{X}=\frac{N}{2}+\frac{1}{2}$. Because of the inversion symmetry, each eigenstate with energy $E_{n,\pm}=\pm t_{n}$ for $n=1,\ldots,\frac{N}{2}-1$ is two-fold degenerate, whereas the eigenstates with energies $E_{\frac{N}{2},\pm}=\pm t_{\frac{N}{2}}$ and $E_{N,\pm}=\pm t_{N}$ are non-degenerate. We now introduce intra-cell hopping $m$ and intra-cell disorder strength $W_{2}$ with

\begin{equation}\begin{split}
H_{1}=\sum\limits_{n=1}^{N}m_{n}c_{n}^{\dagger}\sigma_{1}c_{n}.
\end{split}\end{equation}

\noindent where $m_{n}=m+W_{2}\omega_{n}$ where $\omega_{n}=\omega_{N+1-n}$ due to the inversion symmetry about $\mathcal{X}=\frac{N}{2}+\frac{1}{2}$. Treating $H_{1}$ as a perturbation, we determine the correction to the energies at the inversion centers through second order (we restore the disorder configuration index $n\in\{1,\ldots,N_{\text{configs}}\}$),

\begin{equation}\begin{split}\label{eq:2ndordercorrectionalternate}
E_{\frac{N}{2},\pm}^{(n)}\approx\pm\left(t_{\frac{N}{2}}^{(n)}+\frac{\left(m_{\frac{N}{2}}^{(n)}\right)^{2}}{2\left(t_{\frac{N}{2}}^{(n)}+t_{\frac{N}{2}-1}^{(n)}\right)}+\frac{\left(m_{\frac{N}{2}}^{(n)}\right)^{2}}{2\left(t_{\frac{N}{2}}^{(n)}-t_{\frac{N}{2}-1}^{(n)}\right)}\right)\\\\
E_{N,\pm}^{(n)}\approx\pm\left(t_{N}^{(n)}+\frac{\left(m_{1}^{(n)}\right)^{2}}{2\left(t_{N}^{(n)}+t_{1}^{(n)}\right)}+\frac{\left(m_{1}^{(n)}\right)^{2}}{2\left(t_{N}^{(n)}-t_{1}^{(n)}\right)}\right).
\end{split}\end{equation}

\noindent The corrections appearing at second order have the form $\frac{m^{2}}{\Delta}$ where $\Delta$ is the energy difference. Each energy at the inversion center has two corrections of this type: one with energy difference between two occupied states or two unoccupied states $\left(\text{e.g., }\Delta\sim\pm\left(t_{\frac{N}{2}}^{(n)}-t_{\frac{N}{2}-1}^{(n)}\right)\text{ and }\Delta\sim\pm\left(t_{N}^{(n)}-t_{1}^{(n)}\right)\right)$ and another with energy difference between one occupied state and one unoccupied state $\left(\text{e.g., }\Delta\sim\pm\left(t_{\frac{N}{2}}^{(n)}+t_{\frac{N}{2}-1}^{(n)}\right)\text{ and }\Delta\sim\pm\left(t_{N}^{(n)}+t_{1}^{(n)}\right)\right)$. Upon combining the terms, the corrections take the particularly simple form shown in (\ref{eq:2ndordercorrection}),

\begin{equation}\begin{split}\label{eq:2ndordercorrection}
E_{\frac{N}{2},\pm}^{(n)}\approx\pm t_{\frac{N}{2}}^{(n)}\left(1+\frac{\left(m_{\frac{N}{2}}^{(n)}\right)^{2}}{\left(t_{\frac{N}{2}}^{(n)}\right)^{2}-\left(t_{\frac{N}{2}-1}^{(n)}\right)^{2}}\right)\\\\
E_{N,\pm}^{(n)}\approx\pm t_{N}^{(n)}\left(1+\frac{\left(m_{1}^{(n)}\right)^{2}}{\left(t_{N}^{(n)}\right)^{2}-\left(t_{1}^{(n)}\right)^{2}}\right).
\end{split}\end{equation}

\noindent Fluctuations in $\Delta_{\mathcal{X}=\frac{N}{2}+\frac{1}{2}}$ occur only when the occupied and unoccupied states exchange, which is when the energy eigenvalues $E_{\frac{N}{2},\pm}^{(n)}$ and/or $E_{N,\pm}^{(n)}$ change sign. From (\ref{eq:2ndordercorrection}), one has that even if $m\neq 0$ and $W_{2}\neq 0$, for a given disorder configuration $n\in\{1,\ldots,N_{\text{configs}}\}$, exchanges can still only occur when either $t_{\frac{N}{2}}^{(n)}$ or $t_{N}^{(n)}$ change sign, which is only possible when $|W_{1}|>|W_{1}^*|=2$.\\

\noindent The arguments provided above also hold for the inversion topological invariant $\Delta_{\mathcal{X}=\frac{N}{2}}$ when the inversion center chosen for the disorder configurations is $\mathcal{X}=\frac{N}{2}$ in the dimerized limit $m\neq 0$ and $t=0$ with inter-cell disorder hopping strength $W_{1}=0$ but intra-cell disorder hopping strength $W_{2}\neq 0$ (i.e., $\frac{W_{2}}{W_{1}}=\pm\infty$). Computing the energy eigenvalues from the Hamiltonian in (20) in this regime for a fixed disorder configuration $n\in\{1,\ldots,N_{\text{configs}}\}$ yields,
\begin{equation}\begin{split}\label{eq:energyeigenvalues3}
\{E_{i,\pm}^{(n)}\}=\{\pm m_{i}^{(n)}\}=\{\pm(m+W_{2}\omega_{i}^{(n)})\}
\end{split}\end{equation}
for $i=1,\ldots,N$. Since $\mathcal{X}=\frac{N}{2}$ is the inversion center, the disorder configuration $\{\omega_{i}^{(n)}\}$ is constrained to satisfy the relation $\omega_{i}^{(n)}=\omega_{N-i}^{(n)}$, which results in each $E_{i}^{(n)}$ for $i=1,\ldots,\frac{N}{2}-1$ to be two-fold degenerate with generically non-degenerate energies $E_{\frac{N}{2}}^{(n)}$ and $E_{N}^{(n)}$. Following the same arguments as laid out previously, the disorder averaged bulk spectral gap is 
\begin{equation}\begin{split}\label{eq:disorderaveragedspectralgap2}
\langle\text{SG}\rangle=2m-|W_{2}|.
\end{split}\end{equation}
In this regime, the disorder averaged bulk spectral gap vanishes at precisely $|W_{2}^*|=2m$. Expressions (\ref{eq:oneexchange})-(\ref{eq:variance}) are exactly the same as before in this regime but with $W_{1}$ replaced with $\frac{W_{2}}{m}$. Furthermore, (\ref{eq:2ndordercorrection}) also holds with $m\leftrightarrow t$. This results in the probabilities $P(\Delta_{\mathcal{X}=\frac{N}{2}}=-2)$, $P(\Delta_{\mathcal{X}=\frac{N}{2}}=0)$, and $P(\Delta_{\mathcal{X}=\frac{N}{2}}=2)$ being expressed in terms of $\frac{1}{2}\pm\frac{m}{W_{2}}$ depending on whether $W_{2}$ is negative or positive, and are also the exponents in (\ref{eq:localizationlength}). 

\section*{SM D: Derivation of the Ground State for the Inversion-Symmetric Random Spin-$\frac{1}{2}$ XX Model}\label{suppsec:derivationofRGgroundstate}

\noindent In this section we describe how to apply the position-space renormalization group (RG) method to the inversion-symmetric random spin-$\frac{1}{2}$ chain. Each RG step consists of decimating a pair of spins that have the strongest exchange interaction by enforcing a spin-singlet state for that pair and generating a new and weaker bond between the neighboring spins. However, due to the inversion symmetry, the exchange couplings away from the inversion centers appearing on one half of the chain will also appear on the other side, while the exchange couplings at the inversion centers map to themselves under inversion. Therefore, we consider the exchange couplings away from the inversion centers separately from the exchange couplings at the inversion centers when applying the RG procedure.\\

\noindent The inversion-symmetric random spin-$\frac{1}{2}$ XX model (with periodic boundary conditions) we consider is given as follows,

\begin{equation}\begin{split}
H=\sum\limits_{n=1}^{N}[2t_{n}(S_{2n}^{x}S_{2n+1}^{x}+S_{2n}^{y}S_{2n+1}^{y})+2m_{n}(S_{2n}^{x}S_{2n-1}^{x}+S_{2n}^{y}S_{2n-1}^{y})]
\end{split}\end{equation}

\noindent with $m_{n}=m_{N+1-n}$ and $t_{n}=t_{N-n}$ (i.e., we have chosen the $\mathcal{X}=\frac{N}{2}+\frac{1}{2}$ inversion center in the original Hamiltonian). Implementing this spatial inversion symmetry, the Hamiltonian above can be written in an alternative manner,

\begin{equation}\begin{split}\label{eq:inversionsymmetricXXmodel}
&H=\sum\limits_{n=1}^{\frac{N}{2}-1}[2t_{n}(S_{2n}^{x}S_{2n+1}^{x}+S_{2n}^{y}S_{2n+1}^{y}+S_{2N-2n}^{x}S_{2N+1-2n}^{x}+S_{2N-2n}^{y}S_{2N+1-2n}^{y})]\\
&+[2t_{\frac{N}{2}}(S_{N}^{x}S_{N+1}^{x}+S_{N}^{y}S_{N+1}^{y})+2t_{\frac{N}{2}+1}(S_{2N}^{x}S_{1}^{x}+S_{2N}^{y}S_{1}^{y})]\\
&+\sum\limits_{n=1}^{\frac{N}{2}}[2m_{n}(S_{2n}^{x}S_{2n-1}^{x}+S_{2n}^{y}S_{2n-1}^{y}+S_{2N+2-2n}^{x}S_{2N+1-2n}^{x}+S_{2N+2-2n}^{y}S_{2N+1-2n}^{y})]
\end{split}\end{equation}

\noindent We will illustrate all the possible cases that can occur in the first step of the RG procedure itself. The nature of the position-space RG procedure ensures that the structure of (\ref{eq:inversionsymmetricXXmodel}) is preserved throughout the process.\\

\noindent Starting away from the inversion centers, there are two possibilities for the first RG step. There is either an intra-cell hopping $m_{i}$ or an inter-cell hopping $t_{i}$ that is the strongest exchange coupling, where $1\leq i\leq\frac{N}{2}-1$. In the case where $m_{i}$ is the strongest exchange coupling, the Hamiltonian is $\mathcal{H}_{0}(m_{i})$ for $m_{i}$. In order to perform the decimation procedure, the neighboring sites with couplings $t_{i-1}$ and $t_{i}$ are treated perturbatively, with the perturbation given by $V(t_{i-1},t_{i})$.

\begin{equation}\begin{split}\label{eq:hamiltonian1}
\mathcal{H}_{0}(m_{i})=2m_{i}(S_{2i}^{x}S_{2i-1}^{x}+S_{2i}^{y}S_{2i-1}^{y}+S_{2N+2-2i}^{x}S_{2N+1-2i}^{x}+S_{2N+2-2i}^{y}S_{2N+1-2i}^{y})
\end{split}\end{equation}

\begin{equation}\begin{split}\label{eq:perturbation1}
V(t_{i-1},t_{i})=2t_{i-1}(S_{2i-2}^{x}S_{2i-1}^{x}+S_{2i-2}^{y}S_{2i-1}^{y}+S_{2N+2-2i}^{x}S_{2N+3-2i}^{x}+S_{2N+2-2i}^{y}S_{2N+3-2i}^{y})\\\\
+2t_{i}(S_{2i}^{x}S_{2i+1}^{x}+S_{2i}^{y}S_{2i+1}^{y}+S_{2N-2i}^{x}S_{2N+1-2i}^{x}+S_{2N-2i}^{y}S_{2N+1-2i}^{y})
\end{split}\end{equation}

\noindent Similarly, the Hamiltonian $\mathcal{H}_{0}(t_{i})$ and its perturbation $V(m_{i-1},m_{i})$ for its nearest-neighbor couplings are given as follows,

\begin{equation}\begin{split}\label{eq:hamiltonian2}
\mathcal{H}_{0}(t_{i})=2t_{i}(S_{2i}^{x}S_{2i+1}^{x}+S_{2N-2i}^{x}S_{2N+1-2i}^{x}+S_{2i}^{y}S_{2i+1}^{y}+S_{2N-2i}^{y}S_{2N+1-2i}^{y})
\end{split}\end{equation}

\begin{equation}\begin{split}\label{eq:perturbation2}
V(m_{i},m_{i+1})=2m_{i}(S_{2i}^{x}S_{2i-1}^{x}+S_{2i}^{y}S_{2i-1}^{y}+S_{2N+2-2i}^{x}S_{2N+1-2i}^{x}+S_{2N+2-2i}^{y}S_{2N+1-2i}^{y})\\\\
+2m_{i+1}(S_{2i+2}^{x}S_{2i+1}^{x}+S_{2i+2}^{y}S_{2i+1}^{y}+S_{2N-2i}^{x}S_{2N-1-2i}^{x}+S_{2N-2i}^{y}S_{2N-1-2i}^{y})
\end{split}\end{equation}

\noindent To perform the decimation procedure, we first determine the ground state of $\mathcal{H}_{0}$. For (\ref{eq:hamiltonian1}), this will take the form,

\begin{equation}\begin{split}
|\psi^{(0)}\rangle=|\psi_{x<2i-1}\rangle\left(\frac{1}{\sqrt{2}}(|\uparrow\rangle_{2i-1}|\downarrow\rangle_{2i}-|\downarrow\rangle_{2i-1}|\uparrow\rangle_{2i})\right)|\psi_{2i<x<2N+1-2i}\rangle\\
\times\left(\frac{1}{\sqrt{2}}(|\uparrow\rangle_{2N+1-2i}|\downarrow\rangle_{2N+2-2i}-|\downarrow\rangle_{2N+1-2i}|\uparrow\rangle_{2N+2-2i})\right)|\psi_{x>2N+2-2i}\rangle
\end{split}\end{equation}

\noindent and for (\ref{eq:hamiltonian2}),

\begin{equation}\begin{split}
|\psi^{(0)}\rangle=|\psi_{x<2i}\rangle\left(\frac{1}{\sqrt{2}}(|\uparrow\rangle_{2i}|\downarrow\rangle_{2i+1}-|\downarrow\rangle_{2i}|\uparrow\rangle_{2i+1})\right)|\psi_{2i+1<x<2N-2i}\rangle\\
\times\left(\frac{1}{\sqrt{2}}(|\uparrow\rangle_{2N-2i}|\downarrow\rangle_{2N+1-2i}-|\downarrow\rangle_{2N-2i}|\uparrow\rangle_{2N+1-2i})\right)|\psi_{x>2N+1-2i}\rangle
\end{split}\end{equation}

\noindent One can derive the effective Hamiltonian by performing degenerate perturbation theory on either (\ref{eq:hamiltonian1}) and (\ref{eq:perturbation1}) or (\ref{eq:hamiltonian2}) and (\ref{eq:perturbation2}) in the following manner,

\begin{equation}\begin{split}
\mathcal{H}_{\text{eff}}(m_{\frac{N}{2}})=\mathcal{H}_{0}(m_{\frac{N}{2}})+P_{0}V(t_{\frac{N}{2}-1},t_{\frac{N}{2}})\left(\sum\limits_{n=0}^{\infty}\left(\sum\limits_{m>0}\frac{P_{m}V(t_{\frac{N}{2}-1},t_{\frac{N}{2}})}{E_{m}-E_{0}}\right)^{n}\right)P_{0}   
\end{split}\end{equation}

\noindent where $P_{0}$ is the projector onto the ground state subspace and is given by $P_{0}=|\psi^{(0)}\rangle\langle\psi^{(0)}|$, and $P_{m}$ for $m>0$ are the projectors onto the subspaces for each excited state. This will give a new correction to the effective Hamiltonian $\mathcal{H}_{\text{eff}}$ to second order in degenerate perturbation theory. For (\ref{eq:hamiltonian1}) and (\ref{eq:perturbation1}) this is,

\begin{equation}\begin{split}\label{eq:effectivehamiltonian1}
\mathcal{H}_{\text{eff}}(m_{i})\approx\frac{4t_{i-1}t_{i}}{m_{i}}(S_{2i-2}^{x}S_{2i+1}^{x}+S_{2i-2}^{y}S_{2i+1}^{y}+S_{2N-2i}^{x}S_{2N+3-2i}^{x}+S_{2N-2i}^{y}S_{2N+3-2i}^{y})
\end{split}\end{equation}

\noindent and for (\ref{eq:hamiltonian2}) and (\ref{eq:perturbation2}) this is,

\begin{equation}\begin{split}\label{eq:effectivehamiltonian2}
\mathcal{H}_{\text{eff}}(t_{i})\approx\frac{4m_{i}m_{i+1}}{t_{i}}(S_{2i-1}^{x}S_{2i+2}^{x}+S_{2i-1}^{y}S_{2i+2}^{y}+S_{2N-1-2i}^{x}S_{2N+2-2i}^{x}+S_{2N-1-2i}^{y}S_{2N+2-2i}^{y})
\end{split}\end{equation}

\noindent In both cases, a new weaker coupling is generated between the next neighboring spins, with the previous two couplings having been decimated. Away from the inversion centers, (\ref{eq:effectivehamiltonian1}) and (\ref{eq:effectivehamiltonian2}) represents the expression of the effective Hamiltonian as the decimation process ensues.\\

\noindent If the first decimation step of the RG procedure occurs at one of the two inversion centers, then two possible cases arise. The decimation procedure is identical at both inversion centers so we will focus on the inversion center between sites $N$ and $N+1$ (the other inversion center is between $1$ and $2N$).\\

\noindent The first case is when $t_{\frac{N}{2}}$ is the strongest exchange coupling. The Hamiltonian for this exchange coupling and its corresponding perturbation are given as follows,

\begin{equation}\begin{split}\label{eq:hamiltonian3}
\mathcal{H}_{0}(t_{\frac{N}{2}})=2t_{\frac{N}{2}}(S_{N}^{x}S_{N+1}^{x}+S_{N}^{y}S_{N+1}^{y})
\end{split}\end{equation}

\begin{equation}\begin{split}\label{eq:perturbation3}
V(m_{\frac{N}{2}})=2m_{\frac{N}{2}}(S_{N}^{x}S_{N-1}^{x}+S_{N}^{y}S_{N-1}^{y}+S_{N+2}^{x}S_{N+1}^{x}+S_{N+2}^{y}S_{N+1}^{y})
\end{split}\end{equation}

\noindent A singlet state is enforced between sites $N$ and $N+1$ for $\mathcal{H}_{0}(t_{\frac{N}{2}})$,

\begin{equation}\begin{split}
|\psi^{(0)}\rangle=|\psi_{x<N}\rangle\frac{1}{\sqrt{2}}(|\uparrow\rangle_{N}|\downarrow\rangle_{N+1}-|\downarrow\rangle_{N}|\uparrow\rangle_{N+1})|\psi_{x>N+1}\rangle
\end{split}\end{equation}

\noindent and performing degenerate perturbation theory yields the following effective Hamiltonian to second order,

\begin{equation}\begin{split}\label{eq:effectivehamiltonian3}
\mathcal{H}_{\text{eff}}(t_{\frac{N}{2}})\approx\frac{4m_{\frac{N}{2}}^{2}}{t_{\frac{N}{2}}}(S_{N-1}^{x}S_{N+2}^{x}+S_{N-1}^{y}S_{N+2}^{y})
\end{split}\end{equation}

\noindent which means an effective coupling is formed between sites $N-1$ and $N+2$.\\

\noindent The other case is when $m_{\frac{N}{2}}$ is the strongest coupling. In this case, the Hamiltonian and its perturbation are,

\begin{equation}\begin{split}\label{eq:hamiltonian4}
\mathcal{H}_{0}(m_{\frac{N}{2}})=2m_{\frac{N}{2}}(S_{N}^{x}S_{N-1}^{x}+S_{N}^{y}S_{N-1}^{y}+S_{N+2}^{x}S_{N+1}^{x}+S_{N+2}^{y}S_{N+1}^{y})
\end{split}\end{equation}

\begin{equation}\begin{split}\label{eq:perturbation4}
V(t_{\frac{N}{2}-1},t_{\frac{N}{2}})=2t_{\frac{N}{2}-1}(S_{N-2}^{x}S_{N-1}^{x}+S_{N-2}^{y}S_{N-1}^{y}+S_{N+2}^{x}S_{N+3}^{x}+S_{N+2}^{y}S_{N+3}^{y})+2t_{\frac{N}{2}}(S_{N}^{x}S_{N+1}^{x}+S_{N}^{y}S_{N+1}^{y})
\end{split}\end{equation}

\noindent It is important to note here that the perturbation involved in this case has three terms as opposed to two terms in the previous case, or four terms when the perturbation is away from the inversion center. The ground state for $\mathcal{H}_{0}(m_{\frac{N}{2}})$ is given as,

\begin{equation}\begin{split}
|\psi^{(0)}\rangle=|\psi_{x<N-1}\rangle\left(\frac{1}{\sqrt{2}}(|\uparrow\rangle_{N-1}|\downarrow\rangle_{N}-|\downarrow\rangle_{N-1}|\uparrow\rangle_{N})\right)\left(\frac{1}{\sqrt{2}}|\uparrow\rangle_{N+1}|\downarrow\rangle_{N+2}-|\downarrow\rangle_{N+1}|\uparrow\rangle_{N+2})\right)|\psi_{x>N+2}\rangle
\end{split}\end{equation}

\noindent The leading correction appears at $n=3$ and the effective Hamiltonian has the form,

\begin{equation}\begin{split}\label{eq:effectivehamiltonian4}
\mathcal{H}_{\text{eff}}(m_{\frac{N}{2}})\approx\frac{4t_{\frac{N}{2}-1}^{2}t_{\frac{N}{2}}}{m_{\frac{N}{2}}}(S_{N-2}^{x}S_{N+3}^{x}+S_{N-2}^{y}S_{N+3}^{y})
\end{split}\end{equation}

\noindent Unlike the other cases, the effective Hamiltonian appears at third order due to the three terms appearing in the perturbation given by (\ref{eq:perturbation4}) as opposed to two or four terms in the other cases.\\

\noindent At each step of the RG procedure, this decimation procedure via perturbation theory continues, enforcing singlets that respect the inversion symmetry on pairs of sites that contain the strongest exchange coupling for that RG step. Furthermore, at each RG step, the strongest exchange coupling is either away from the inversion or at one of the inversion centers. At the end of this procedure, a total of $N$ singlets will be present in the ground state. We will denote $M$ to be the number of singlets that form across the inversion centers, and hence, $\frac{1}{2}(N-M)$ denote the number of singlets that have formed in pairs away from the inversion center. From this procedure, the ground state takes the final form,

\begin{equation}\begin{split}\label{eq:suppRGgroundstate}
|\Omega\rangle=\prod\limits_{i=1}^{\frac{1}{2}(N-M)}\prod\limits_{j=1}^{M}\left(\frac{1}{\sqrt{2}}(S_{2N+2-2d_{i2}}^{+}-S_{2N+1-2d_{i1}}^{+})\right)\left(\frac{1}{\sqrt{2}}(S_{2d_{j}}^{+}-S_{2N+1-2d_{j}}^{+})\right)\left(\frac{1}{\sqrt{2}}(S_{2d_{i1}}^{+}-S_{2d_{i2}-1}^{+})\right)|\downarrow\cdots\downarrow\rangle
\end{split}\end{equation}

\section*{SM E: Expressing the RG Ground State in the Fermionic Basis}\label{suppsec:simplificationofgroundstate}

\noindent In this section, using the inverse Jordan-Wigner transformation we express the ground state given by (\ref{eq:suppRGgroundstate}) in the fermionic basis. This proof is similar to the one presented in \cite{mondragon2014topological}, where the ground state is purely chiral-symmetric. However, the inclusion of inversion symmetry changes many of the details laid out in the proof detailed in \cite{mondragon2014topological}, and hence we review it in its entirety in this section. Finally, we derive the expressions for the inversion topological invariant and the bulk polarization using the expression for the RG ground state in the fermionic basis. We note that the expression for the chiral winding number and its derivation is exactly the same as in \cite{mondragon2014topological}.\\

\noindent Using the inverse Jordan-Wigner transformation, the ground state in (\ref{eq:suppRGgroundstate}) can be expressed as follows,

\begin{equation}\begin{split}\label{eq:unsimplifiedfermionicgroundstate}
&|\Omega\rangle=\prod\limits_{i=1}^{\frac{1}{2}(N-M)}\prod\limits_{j=1}^{M}\left(\frac{1}{\sqrt{2}}(c_{N+1-d_{i2},B}^{\dagger}\tilde{K}(N+1-d_{i2},1)-c_{N+1-d_{i1},A}^{\dagger}\tilde{K}(N+1-d_{i1},0))\right)\\\\
&\times\left(\frac{1}{\sqrt{2}}(c_{d_{j},B}^{\dagger}\tilde{K}(d_{j},1)-c_{N+1-d_{j},A}^{\dagger}\tilde{K}(N+1-d_{j},0))\right)\times\left(\frac{1}{\sqrt{2}}(c_{d_{i1},B}^{\dagger}\tilde{K}(d_{i1},1)-c_{d_{i2},A}^{\dagger}\tilde{K}(d_{i2},0))\right)|0\rangle
\end{split}\end{equation}

\noindent where $\tilde{K}(m,\lambda)=\exp\left\{i\pi\lambda c_{m,B}^{\dagger}c_{m,B}\right\}\exp\left\{i\pi\sum\limits_{j=1}^{m-1}(c_{j,A}^{\dagger}c_{j,A}+c_{j,B}^{\dagger}c_{j,B})\right\}$ is the string operator defined in the fermionic basis, where $\lambda\in\{0,1\}$. Without loss of generality, we will assume $1\leq d_{j}\leq\frac{N}{2}$. To express (\ref{eq:unsimplifiedfermionicgroundstate}) in a simpler form, define the following quantities,

\begin{equation}\begin{split}
Z_{1}(i)=\frac{1}{\sqrt{2}}(c_{d_{i1},B}^{\dagger}\tilde{K}(d_{i1},1)-c_{d_{i2},A}^{\dagger}\tilde{K}(d_{i2},0))
\end{split}\end{equation}

\begin{equation}\begin{split}
Z_{2}(i)=\frac{1}{\sqrt{2}}(c_{d_{i},B}^{\dagger}\tilde{K}(d_{i},1)-c_{N+1-d_{i},A}^{\dagger}\tilde{K}(N+1-d_{i},0))
\end{split}\end{equation}

\begin{equation}\begin{split}
Z_{3}(i)=\frac{1}{\sqrt{2}}(c_{N+1-d_{i2},B}^{\dagger}\tilde{K}(N+1-d_{i2},1)-c_{N+1-d_{i1},A}^{\dagger}\tilde{K}(N+1-d_{i1},0))
\end{split}\end{equation}

\noindent It will also prove useful to define the following quantities,

\begin{equation}\begin{split}\label{eq:tildeZ1}
\tilde{Z}_{1}(i)=\frac{1}{\sqrt{2}}(\alpha_{i}^{(1)}c_{d_{i1},B}^{\dagger}-\beta_{i}^{(1)}c_{d_{i2},A}^{\dagger})
\end{split}\end{equation}

\begin{equation}\begin{split}\label{eq:tildeZ2}
\tilde{Z}_{2}(i)=\frac{1}{\sqrt{2}}(\alpha_{i}^{(2)}c_{d_{i},B}^{\dagger}-\beta_{i}^{(2)}c_{N+1-d_{i},A}^{\dagger})
\end{split}\end{equation}

\begin{equation}\begin{split}\label{eq:tildeZ3}
\tilde{Z}_{3}(i)=\frac{1}{\sqrt{2}}(\alpha_{i}^{(3)}c_{N+1-d_{i2},B}^{\dagger}-\beta_{i}^{(3)}c_{N+1-d_{i1},A}^{\dagger})
\end{split}\end{equation}

\noindent where $\alpha_{i}$ and $\beta_{i}$ have unit modulus and their particular values depend on the configuration of the singlets. Then the ground state in (\ref{eq:unsimplifiedfermionicgroundstate}) can be expressed as follows,

\begin{equation}\begin{split}\label{eq:proof1}
|\Omega\rangle=\prod\limits_{i=1}^{\frac{1}{2}(N-M)}\prod\limits_{j=1}^{M}Z_{3}(i)Z_{2}(j)Z_{1}(i)|0\rangle
\end{split}\end{equation} 

\noindent In order to get the ground state into a simplified form, the goal is to move all the string operators to the right and have them act on the vacuum state $|0\rangle$. The goal therefore, is to express the ground state in the following form,

\begin{equation}\begin{split}\label{eq:proof2}
|\Omega\rangle=\prod\limits_{i=1}^{\frac{1}{2}(N-M)}\prod\limits_{j=1}^{M}\tilde{Z}_{3}(i)\tilde{Z}_{2}(j)\tilde{Z}_{1}(i)|0\rangle
\end{split}\end{equation} 

\noindent First, consider $Z_{1}(i)$ acting on the vacuum $|0\rangle$. Before proceeding, we note that the terms in this product can be organized according to singlet length, since each term commutes with one another before any simplification. We will use the convention that singlets smaller in length always appear to the right of singlets larger in length.

\begin{equation}\begin{split}\label{eq:proof3}
\prod\limits_{i=1}^{\frac{1}{2}(N-M)}Z_{1}(i)|0\rangle=\left(\prod\limits_{i=1}^{\frac{1}{2}(N-M)-1}Z_{1}(i)\right)\left(\frac{1}{\sqrt{2}}(c_{d_{\frac{1}{2}(N-M),1},B}^{\dagger}\tilde{K}(d_{\frac{1}{2}(N-M),1},1)-c_{d_{\frac{1}{2}(N-M),2},A}^{\dagger}\tilde{K}(d_{\frac{1}{2}(N-M),2},0))\right)|0\rangle
\end{split}\end{equation}

\noindent Note that $\tilde{K}(m,\lambda)|0\rangle=|0\rangle$. Hence,

\begin{equation}\begin{split}\label{eq:proof4}
\prod\limits_{i=1}^{\frac{1}{2}(N-M)}Z_{1}(i)|0\rangle=\left(\prod\limits_{i=1}^{\frac{1}{2}(N-M)-1}Z_{1}(i)\right)\left(\frac{1}{\sqrt{2}}(c_{d_{\frac{1}{2}(N-M),1},B}^{\dagger}-c_{d_{\frac{1}{2}(N-M),2},A}^{\dagger})\right)|0\rangle
\end{split}\end{equation}

\noindent which is just,

\begin{equation}\begin{split}\label{eq:proof5}
\left(\prod\limits_{i=1}^{\frac{1}{2}(N-M)-1}Z_{1}(i)\right)\tilde{Z}_{1}\left(\frac{1}{2}(N-M)\right)|0\rangle
\end{split}\end{equation}

\noindent Now consider the following,

\begin{equation}\begin{split}\label{eq:proof6}
\left(\prod\limits_{i=1}^{\frac{1}{2}(N-M)-2}Z_{1}(i)\right)\left(\frac{1}{\sqrt{2}}(c_{d_{\frac{1}{2}(N-M)-1,1},B}^{\dagger}\tilde{K}(d_{\frac{1}{2}(N-M)-1,1},1)-c_{d_{\frac{1}{2}(N-M)-1,2},A}^{\dagger}\tilde{K}(d_{\frac{1}{2}(N-M)-1,2},0))\right)\tilde{Z}_{1}\left(\frac{1}{2}(N-M)\right)|0\rangle
\end{split}\end{equation}

\noindent Recall that the singlets have been organized according to length. Hence, there are three possible cases that arise when the string operators $\tilde{K}(d_{\frac{1}{2}(N-M)-1,1},1)$ and $\tilde{K}(d_{\frac{1}{2}(N-M)-1,2},0)$ act on $\tilde{Z}_{1}\left(\frac{1}{2}(N-M)\right)$:\\

\noindent 1. \textit{The interval $d_{\frac{1}{2}(N-M)}$ lies to the right of the interval $d_{\frac{1}{2}(N-M)-1}$}: In this case, neither $\tilde{K}(d_{\frac{1}{2}(N-M)-1,2},0)$ and $\tilde{K}(d_{\frac{1}{2}(N-M)-1,1},1)$ will have number operators in their exponentials with corresponding creation operators in $\tilde{Z}_{1}\left(\frac{1}{2}(N-M)\right)$, which means that the $K$ operators commute with $\tilde{Z}_{1}\left(\frac{1}{2}(N-M)\right)$. Hence we can write (\ref{eq:proof7}).\\

\noindent 2. \textit{The interval $d_{\frac{1}{2}(N-M)}$ lies inside of the interval $d_{\frac{1}{2}(N-M)-1}$}: Then one, and only one, of $\tilde{K}(d_{\frac{1}{2}(N-M)-1,2},0)$ and $\tilde{K}(d_{\frac{1}{2}(N-M)-1,1},1)$ will have the number operators associated with both of the creation operators in $\tilde{Z}_{1}\left(\frac{1}{2}(N-M)\right)$. The state $\tilde{Z}_{1}\left(\frac{1}{2}(N-M)\right)|0\rangle$ is thus an eigenstate of both $\tilde{K}$ operators, with eigenvalues $\pm 1$, depending on which $\tilde{K}$ has operators shared with $\tilde{Z}_{1}\left(\frac{1}{2}(N-M)\right)$. Hence, we have (\ref{eq:proof7}).\\

\noindent 3. \textit{The interval $d_{\frac{1}{2}(N-M)}$ lies to the left $d_{\frac{1}{2}(N-M)-1}$}: Then both $\tilde{K}(d_{\frac{1}{2}(N-M)-1,2},0)$ and $\tilde{K}(d_{\frac{1}{2}(N-M)-1,1},1)$ will have the number operators associated with \textit{both} of the creation operators in $\tilde{Z}_{1}\left(\frac{1}{2}(N-M)\right)$. The state $\tilde{Z}_{1}\left(\frac{1}{2}(N-M)\right)|0\rangle$ is thus an eigenstate of both $\tilde{K}$ operators, with eigenvalues $-1$. Hence, we have (\ref{eq:proof7}).\\

\noindent All three of the cases described above lead to the following simplification of (\ref{eq:proof6}),

\begin{equation}\begin{split}\label{eq:proof7}
\left(\prod\limits_{i=1}^{\frac{1}{2}(N-M)-2}Z_{1}(i)\right)\left(\frac{1}{\sqrt{2}}(\alpha_{\frac{1}{2}(N-M)-1}^{(1)}c_{d_{\frac{1}{2}(N-M)-1,1},B}^{\dagger}-\beta_{\frac{1}{2}(N-M)-1}^{(1)}c_{d_{\frac{1}{2}(N-M)-1,2},A}^{\dagger})\right)\tilde{Z}_{1}\left(\frac{1}{2}(N-M)\right)|0\rangle
\end{split}\end{equation}

\noindent which can be expressed as,

\begin{equation}\begin{split}\label{eq:proof8}
\left(\prod\limits_{i=1}^{\frac{1}{2}(N-M)-2}Z_{1}(i)\right)\tilde{Z}_{1}\left(\frac{1}{2}(N-M)-1\right)\tilde{Z}_{1}\left(\frac{1}{2}(N-M)\right)|0\rangle
\end{split}\end{equation}

\noindent This is the first inductive step. For the $m^{\text{th}}$ step of the inductive argument, we now assume that the following expression holds, 

\begin{equation}\begin{split}\label{eq:proof9}
\left(\prod\limits_{i=1}^{\frac{1}{2}(N-M)-m}Z_{1}(i)\right)\left(\prod\limits_{j=\frac{1}{2}(N-M)-m+1}^{\frac{1}{2}(N-M)}\tilde{Z}_{1}(j)\right)|0\rangle
\end{split}\end{equation}

\noindent and we must now show that this implies the $(m+1)^{\text{th}}$ case, namely that

\begin{equation}\begin{split}\label{eq:proof10}
\left(\prod\limits_{i=1}^{\frac{1}{2}(N-M)-m-1}Z_{1}(i)\right)\left(\prod\limits_{j=\frac{1}{2}(N-M)-m}^{\frac{1}{2}(N-M)}\tilde{Z}_{1}(j)\right)|0\rangle
\end{split}\end{equation}

\noindent To illustrate this, we again write,

\begin{equation}\begin{split}\label{eq:proof11}
\left(\prod\limits_{i=1}^{\frac{1}{2}(N-M)-m-1}Z_{1}(i)\right)\left(Z_{1}\left(\frac{1}{2}(N-M)-m\right)\right)\left(\prod\limits_{j=\frac{1}{2}(N-M)-m+1}^{\frac{1}{2}(N-M)}\tilde{Z}_{1}(j)\right)|0\rangle
\end{split}\end{equation}

\noindent Group the factors $\prod\limits_{j=\frac{1}{2}(N-M)-m+1}^{\frac{1}{2}(N-M)}\tilde{Z}_{1}(j)=B_{1}B_{2}B_{3}$ in the following way:

\noindent 1. \textit{$B_{1}$ is made out of singlets with intervals that lie to the right of the interval $d_{\frac{1}{2}(N-M)-m}$}: In this case, neither $\tilde{K}(d_{\frac{1}{2}(N-M)-m,2},0)$ nor $\tilde{K}(d_{\frac{1}{2}(N-M)-m,1},1)$ will have number operators in their exponentials with corresponding creation operators in $B_{1}$, which means that the $\tilde{K}$ operators commute with $B_{1}$.\\

\noindent 2. \textit{$B_{2}$ is made out of singlets with intervals that lie inside of the interval $d_{\frac{1}{2}(N-M)-m}$}: Then one, and only one, of $\tilde{K}(d_{\frac{1}{2}(N-M)-m,2},0)$ and $\tilde{K}(d_{\frac{1}{2}(N-M)-m,1},1)$ will have \textit{all} of the number operators associated with the creation operators in $B_{2}$. Let us denote the number of factors in $B_{2}$ by $\rho_{B_{2}}$.\\

\noindent 3. \textit{$B_{3}$ is made out of singlets with intervals such that $d_{\frac{1}{2}(N-M)-m}$ lies to the left of them}: Then both $\tilde{K}(d_{\frac{1}{2}(N-M)-m,2},0)$ and $\tilde{K}(d_{\frac{1}{2}(N-M)-m,1},1)$ will have \textit{all} of the number operators associated with the creation operators in $B_{3}$. Let us denote the number of factors in $B_{3}$ by $\rho_{B_{3}}$.\\

\noindent Hence, depending on which case occurs with the $B_{2}$ singlets, we will either get,

\begin{equation}\begin{split}\label{eq:proof12}
\left(\prod\limits_{i=1}^{\frac{1}{2}(N-M)-m-1}Z_{1}(i)\right)\left(\frac{(-1)^{\rho_{B_{3}}}}{\sqrt{2}}(c_{d_{\frac{1}{2}(N-M)-m,1},B}^{\dagger}(-1)^{\rho_{B_{2}}}-c_{d_{\frac{1}{2}(N-M)-m,2},A}^{\dagger})\right)B_{1}B_{2}B_{3}|0\rangle
\end{split}\end{equation}

\noindent or,

\begin{equation}\begin{split}\label{eq:proof13}
\left(\prod\limits_{i=1}^{\frac{1}{2}(N-M)-m-1}Z_{1}(i)\right)\left(\frac{(-1)^{\rho_{B_{3}}}}{\sqrt{2}}(c_{d_{\frac{1}{2}(N-M)-m,1},B}^{\dagger}-c_{d_{\frac{1}{2}(N-M)-m,2},A}^{\dagger}(-1)^{\rho_{B_{2}}})\right)B_{1}B_{2}B_{3}|0\rangle
\end{split}\end{equation}

\noindent In both cases we end up getting the following,

\begin{equation}\begin{split}\label{eq:proof14}
\left(\prod\limits_{i=1}^{\frac{1}{2}(N-M)-m-1}Z_{1}(i)\right)\left(\frac{1}{\sqrt{2}}(\alpha_{\frac{1}{2}(N-M)-m}^{(1)}c_{d_{\frac{1}{2}(N-M)-m,1},B}^{\dagger}-\beta_{\frac{1}{2}(N-M)-m}^{(1)}c_{d_{\frac{1}{2}(N-M)-m,2},A}^{\dagger})\right)\left(\prod\limits_{j=\frac{1}{2}(N-M)-m+1}^{\frac{1}{2}(N-M)}\tilde{Z}_{1}(j)\right)|0\rangle
\end{split}\end{equation}

\noindent which ends up yielding (\ref{eq:proof10}),

\begin{equation}\begin{split}\label{eq:proof15}
\left(\prod\limits_{i=1}^{\frac{1}{2}(N-M)-m-1}Z_{1}(i)\right)\left(\prod\limits_{j=\frac{1}{2}(N-M)-m}^{\frac{1}{2}(N-M)}\tilde{Z}_{1}(j)\right)|0\rangle
\end{split}\end{equation}

\noindent As a consequence, by successive application of the procedure, we can finally arrive at the first part of the desired result, 

\begin{equation}\begin{split}\label{eq:proof16}
\prod\limits_{i=1}^{\frac{1}{2}(N-M)}\tilde{Z}_{1}(i)|0\rangle
\end{split}\end{equation}

\noindent We have now shown that (\ref{eq:proof1}) can now be expressed as follows,

\begin{equation}\begin{split}\label{eq:proof17}
|\Omega\rangle=\prod\limits_{i=1}^{\frac{1}{2}(N-M)}\prod\limits_{j=1}^{M}Z_{3}(i)Z_{2}(j)\tilde{Z}_{1}(i)|0\rangle
\end{split}\end{equation}

\noindent We now consider the following product of terms,

\begin{equation}\begin{split}\label{eq:proof18}
\prod\limits_{i=1}^{\frac{1}{2}(N-M)}\prod\limits_{j=1}^{M}Z_{3}(i)Z_{2}(j)\tilde{Z}_{1}(i)|0\rangle
\end{split}\end{equation}

\noindent Note that there are two bond-centered inversion centers: one between sites $N$ and $N+1$ and the other one between sites $1$ and $N$. Denote the number of singlets formed across the first inversion center $M_{1}$ and the number of singlets formed across the second inversion center $M-M_{1}$. This product is partitioned then into two parts,

\begin{equation}\begin{split}\label{eq:proof19}
\prod\limits_{i=1}^{\frac{1}{2}(N-M)}\prod\limits_{k=M_{1}+1}^{M}\prod\limits_{j=1}^{M_{1}}Z_{3}(i)Z_{2}(k)Z_{2}(j)\tilde{Z}_{1}(i)|0\rangle
\end{split}\end{equation}

\noindent Prior to any simplification, the terms in the product of $Z_{2}(j)$, $Z_{2}(k)$, and $Z_{3}(i)$ can be organized according to singlet length. Hence, we write the following,

\begin{equation}\begin{split}\label{eq:proof20}
\prod\limits_{k=M_{1}+1}^{M}\prod\limits_{i=1}^{\frac{1}{2}(N-M)}\prod\limits_{j=1}^{M_{1}}Z_{2}(k)Z_{3}(i)Z_{2}(j)\tilde{Z}_{1}(i)|0\rangle
\end{split}\end{equation}

\noindent First, we focus on the following portion of the product in (\ref{eq:proof20}),

\begin{equation}\begin{split}\label{eq:proof21}
\prod\limits_{i=1}^{\frac{1}{2}(N-M)}\prod\limits_{j=1}^{M_{1}}Z_{2}(j)\tilde{Z}_{1}(i)|0\rangle
\end{split}\end{equation}

\noindent which can be expressed as follows,

\begin{equation}\begin{split}\label{eq:proof22}
\left(\prod\limits_{j=1}^{M_{1}-1}Z_{2}(j)\right)\left(\frac{1}{\sqrt{2}}(c_{d_{M_{1}},B}^{\dagger}\tilde{K}(d_{M_{1}},1)-c_{N+1-d_{M_{1}},A}^{\dagger}\tilde{K}(N+1-d_{M_{1}},0))\right)\left(\prod\limits_{i=1}^{\frac{1}{2}(N-M)}\tilde{Z}_{1}(i)\right)|0\rangle
\end{split}\end{equation}

\noindent Since the $M_{1}^{\text{th}}$ singlet is located at the first inversion center between sites $N$ and $N+1$, both $\tilde{K}(N+1-d_{M_{1}},0)$ and $\tilde{K}(d_{M_{1}},1)$ contain \textit{all} of the number operators associated with the creation operators in $\prod\limits_{i=1}^{\frac{1}{2}(N-M)}\tilde{Z}_{1}(i)$. Therefore (\ref{eq:proof22}) becomes,

\begin{equation}\begin{split}\label{eq:proof23}
\left(\prod\limits_{j=1}^{M_{1}-1}Z_{2}(j)\right)\left(\frac{(-1)^{\frac{1}{2}(N-M)}}{\sqrt{2}}(c_{d_{M_{1}},B}^{\dagger}-c_{N+1-d_{M_{1}},A}^{\dagger})\right)\left(\prod\limits_{i=1}^{\frac{1}{2}(N-M)}\tilde{Z}_{1}(i)\right)|0\rangle
\end{split}\end{equation}

\noindent Thus we have,

\begin{equation}\begin{split}\label{eq:proof24}
\left(\prod\limits_{j=1}^{M_{1}-1}Z_{2}(j)\right)\tilde{Z}_{2}(M_{1})\left(\prod\limits_{i=1}^{\frac{1}{2}(N-M)}\tilde{Z}_{1}(i)\right)|0\rangle
\end{split}\end{equation}

\noindent Now consider the next term in the product,

\begin{equation}\begin{split}\label{eq:proof25}
\left(\prod\limits_{j=1}^{M_{1}-2}Z_{2}(j)\right)\left(\frac{1}{\sqrt{2}}(c_{d_{M_{1}-1},B}^{\dagger}\tilde{K}(d_{M_{1}-1},1)-c_{N+1-d_{M_{1}-1},A}^{\dagger}\tilde{K}(N+1-d_{M_{1}-1},0))\right)\tilde{Z}_{2}(M_{1})\left(\prod\limits_{i=1}^{\frac{1}{2}(N-M)}\tilde{Z}_{1}(i)\right)|0\rangle
\end{split}\end{equation}

\noindent Recall that the singlets in this product have been organized according to length, so the interval $[d_{M_{1}},N+1-d_{M_{1}}]$ lies inside the interval $[d_{M_{1}-1},N+1-d_{M_{1}-1}]$. This means one of the operators, and only one, of $\tilde{K}(d_{M_{1}-1},1)$ or $\tilde{K}(N+1-d_{M_{1}-1},0)$ contains all of the number operators associated with the creation operators in $\prod\limits_{i=1}^{\frac{1}{2}(N-M)}\tilde{Z}_{1}(i)$ \textit{and} the creation operators in $\tilde{Z}_{2}(M_{1})$, while the other only contains the number operators associated with the creation operators in $\prod\limits_{i=1}^{\frac{1}{2}(N-M)}\tilde{Z}_{1}(i)$. Therefore, $\tilde{Z}_{2}(M_{1})\left(\prod\limits_{i=1}^{\frac{1}{2}(N-M)}\tilde{Z}_{1}(i)\right)|0\rangle$ is an eigenstate of both $\tilde{K}$ operators. Therefore, we can express this as,

\begin{equation}\begin{split}\label{eq:proof26}
\left(\prod\limits_{j=1}^{M_{1}-2}Z_{2}(j)\right)\left(\frac{1}{\sqrt{2}}(\alpha_{M_{1}-1}^{(2)}c_{d_{M_{1}-1},B}^{\dagger}-\beta_{M_{1}-1}^{(2)}c_{N+1-d_{M_{1}-1},A}^{\dagger})\right)\tilde{Z}_{2}(M_{1})\left(\prod\limits_{i=1}^{\frac{1}{2}(N-M)}\tilde{Z}_{1}(i)\right)|0\rangle
\end{split}\end{equation}

\noindent or,

\begin{equation}\begin{split}\label{eq:proof27}
\left(\prod\limits_{j=1}^{M_{1}-2}Z_{2}(j)\right)\tilde{Z}_{2}(M_{1}-1)\tilde{Z}_{2}(M_{1})\left(\prod\limits_{i=1}^{\frac{1}{2}(N-M)}\tilde{Z}_{1}(i)\right)|0\rangle
\end{split}\end{equation}

\noindent This is the first inductive step. For the $m^{\text{th}}$ step of the inductive argument, we now assume the following expression holds,

\begin{equation}\begin{split}\label{eq:proof28}
\left(\prod\limits_{j=1}^{M_{1}-m}Z_{2}(j)\right)\left(\prod\limits_{k=M_{1}-m+1}^{M_{1}}\tilde{Z}_{2}(k)\right)\left(\prod\limits_{i=1}^{\frac{1}{2}(N-M)}\tilde{Z}_{1}(i)\right)|0\rangle
\end{split}\end{equation}

\noindent and we must now show that this implies the $(m+1)^{\text{th}}$ case, namely that

\begin{equation}\begin{split}\label{eq:proof29}
\left(\prod\limits_{j=1}^{M_{1}-m-1}Z_{2}(j)\right)\left(\prod\limits_{k=M_{1}-m}^{M_{1}}\tilde{Z}_{2}(k)\right)\left(\prod\limits_{i=1}^{\frac{1}{2}(N-M)}\tilde{Z}_{1}(i)\right)|0\rangle
\end{split}\end{equation}

\noindent To illustrate this, we again write,

\begin{equation}\begin{split}\label{eq:proof30}
\left(\prod\limits_{j=1}^{M_{1}-m-1}Z_{2}(j)\right)\left(Z_{2}(M_{1}-m)\right)\left(\prod\limits_{k=M_{1}-m+1}^{M_{1}}\tilde{Z}_{2}(k)\right)\left(\prod\limits_{i=1}^{\frac{1}{2}(N-M)}\tilde{Z}_{1}(i)\right)|0\rangle
\end{split}\end{equation}

\noindent Recall that the singlets have been organized according to length and the singlets in the product of $Z_{2}$'s are formed over the first inversion center. Hence, each interval $[d_{j},N+1-d_{j}]$ lies within the successive interval $[d_{j+1},N+1-d_{j+1}]$ for $M_{1}-m\leq j\leq M_{1}$. Therefore, as before, one and only one of $\tilde{K}(N+1-d_{M_{1}-m},0)$ and $\tilde{K}(d_{M_{1}-m},1)$ contains all the number operators associated with the creation operators in $\left(\prod\limits_{k=M_{1}-m+1}^{M_{1}}\tilde{Z}_{2}(k)\right)\left(\prod\limits_{i=1}^{\frac{1}{2}(N-M)}\tilde{Z}_{1}(i)\right)|0\rangle$ while the other one only contains all the number operators associated with the creation operators in $\left(\prod\limits_{i=1}^{\frac{1}{2}(N-M)}\tilde{Z}_{1}(i)\right)|0\rangle$. Therefore we get the desired result in (\ref{eq:proof29}). As a consequence, by successive application of the procedure, we can finally arrive at the next part of the desired result,

\begin{equation}\begin{split}\label{eq:proof31}
\prod\limits_{i=1}^{\frac{1}{2}(N-M)}\prod\limits_{j=1}^{M_{1}}\tilde{Z}_{2}(j)\tilde{Z}_{1}(i)|0\rangle
\end{split}\end{equation}

\noindent We now have the following,

\begin{equation}\begin{split}\label{eq:proof32}
|\Omega\rangle=\prod\limits_{k=M_{1}+1}^{M}\prod\limits_{i=1}^{\frac{1}{2}(N-M)}\prod\limits_{j=1}^{M_{1}}Z_{2}(k)Z_{3}(i)\tilde{Z}_{2}(j)\tilde{Z}_{1}(i)|0\rangle
\end{split}\end{equation}

\noindent Proceeding further into this product, we note that the inductive proof laid out for the state $\prod\limits_{i=1}^{\frac{1}{2}(N-M)}Z_{1}(i)|0\rangle$ in (\ref{eq:proof3})-(\ref{eq:proof16}) is \textit{exactly the same} for the state $\prod\limits_{i=1}^{\frac{1}{2}(N-M)}Z_{3}(i)\tilde{Z}_{2}(j)\tilde{Z}_{1}(i)|0\rangle$. Therefore, we arrive at the third part of the desired result, 

\begin{equation}\begin{split}\label{eq:proof33}
|\Omega\rangle=\prod\limits_{k=M_{1}+1}^{M}\prod\limits_{i=1}^{\frac{1}{2}(N-M)}\prod\limits_{j=1}^{M_{1}}Z_{2}(k)\tilde{Z}_{3}(i)\tilde{Z}_{2}(j)\tilde{Z}_{1}(i)|0\rangle
\end{split}\end{equation}

\noindent This can be rewritten in slightly different notation,

\begin{equation}\begin{split}\label{eq:proof34}
|\Omega\rangle=\left(\prod\limits_{k=M_{1}+1}^{M}Z_{2}(k)\right)\left(\prod\limits_{i=1}^{\frac{1}{2}(N-M)}\prod\limits_{j=1}^{M_{1}}\tilde{Z}_{3}(i)\tilde{Z}_{2}(j)\tilde{Z}_{1}(i)\right)|0\rangle
\end{split}\end{equation}

\noindent We now consider the full product. The product $\prod\limits_{k=M_{1}+1}^{M}Z_{2}(k)$ consists of singlets formed across the second inversion center between sites $N$ and $1$. Prior to any simplification, the terms in the product of $Z_{2}(k)$'s are organized according to singlet length. Therefore, we express the above as follows,

\begin{equation}\begin{split}\label{eq:proof35}
|\Omega\rangle=\left(\prod\limits_{k=M_{1}+1}^{M-1}Z_{2}(k)\right)\left(\frac{1}{\sqrt{2}}(c_{d_{M},B}^{\dagger}\tilde{K}(d_{M},1)-c_{N+1-d_{M},A}^{\dagger}\tilde{K}(N+1-d_{M},0))\right)\left(\prod\limits_{i=1}^{\frac{1}{2}(N-M)}\prod\limits_{j=1}^{M_{1}}\tilde{Z}_{3}(i)\tilde{Z}_{2}(j)\tilde{Z}_{1}(i)\right)|0\rangle
\end{split}\end{equation}

\noindent Since the $M^{\text{th}}$ singlet is located at the second inversion center between sites $1$ and $N$, one (and only one) of $\tilde{K}(N+1-d_{M},0)$ and $\tilde{K}(d_{M},1)$ contains \textit{all} of the number operators associated with the creation operators in $\prod\limits_{i=1}^{\frac{1}{2}(N-M)}\prod\limits_{j=1}^{M_{1}}\tilde{Z}_{3}(i)\tilde{Z}_{2}(j)\tilde{Z}_{1}(i)$. Therefore, (\ref{eq:proof35}) becomes

\begin{equation}\begin{split}\label{eq:proof36}
|\Omega\rangle=\left(\prod\limits_{k=M_{1}+1}^{M-1}Z_{2}(k)\right)\left(\frac{1}{\sqrt{2}}(c_{d_{M},B}^{\dagger}-(-1)^{\frac{1}{2}(N-M)+M_{1}}c_{N+1-d_{M},A}^{\dagger})\right)\left(\prod\limits_{i=1}^{\frac{1}{2}(N-M)}\prod\limits_{j=1}^{M_{1}}\tilde{Z}_{3}(i)\tilde{Z}_{2}(j)\tilde{Z}_{1}(i)\right)|0\rangle
\end{split}\end{equation}

\noindent which can be expressed as,

\begin{equation}\begin{split}\label{eq:proof37}
|\Omega\rangle=\left(\prod\limits_{k=M_{1}+1}^{M-1}Z_{2}(k)\right)\left(\frac{1}{\sqrt{2}}(\alpha_{M}^{(2)}c_{d_{M},B}^{\dagger}-\beta_{M}^{(2)}c_{N+1-d_{M},A}^{\dagger})\right)\left(\prod\limits_{i=1}^{\frac{1}{2}(N-M)}\prod\limits_{j=1}^{M_{1}}\tilde{Z}_{3}(i)\tilde{Z}_{2}(j)\tilde{Z}_{1}(i)\right)|0\rangle
\end{split}\end{equation}

\noindent or,

\begin{equation}\begin{split}\label{eq:proof38}
|\Omega\rangle=\left(\prod\limits_{k=M_{1}+1}^{M-1}Z_{2}(k)\right)\tilde{Z}_{2}(M)\left(\prod\limits_{i=1}^{\frac{1}{2}(N-M)}\prod\limits_{j=1}^{M_{1}}\tilde{Z}_{3}(i)\tilde{Z}_{2}(j)\tilde{Z}_{1}(i)\right)|0\rangle
\end{split}\end{equation}

\noindent Now consider the next term in this product,

\begin{equation}\begin{split}\label{eq:proof39}
|\Omega\rangle=\left(\prod\limits_{k=M_{1}+1}^{M-2}Z_{2}(k)\right)\left(Z_{2}(M-1)\right)\left(\tilde{Z}_{2}(M)\right)\left(\prod\limits_{i=1}^{\frac{1}{2}(N-M)}\prod\limits_{j=1}^{M_{1}}\tilde{Z}_{3}(i)\tilde{Z}_{2}(j)\tilde{Z}_{1}(i)\right)|0\rangle
\end{split}\end{equation}

\noindent Recall that the singlets in this product have been organized according to length, so the interval $[d_{M},N+1-d_{M}]$ lies inside the interval $[d_{M-1},N+1-d_{M-1}]$. This means one of the operators, and only one, of $\tilde{K}(d_{M-1},1)$ and $\tilde{K}(N+1-d_{M-1},0)$ contains \textit{all} of the number operators associated with the creation operators in $\left(\prod\limits_{i=1}^{\frac{1}{2}(N-M)}\prod\limits_{j=1}^{M_{1}}\tilde{Z}_{3}(i)\tilde{Z}_{2}(j)\tilde{Z}_{1}(i)\right)|0\rangle$ \textit{but only one of the number operators associated with the creation operator in $\tilde{Z}_{2}(M)$}. The other only contains one of the number operators associated with the creation operator in $\tilde{Z}_{2}(M)$. Thus, we can express this result as,

\begin{equation}\begin{split}\label{eq:proof40}
|\Omega\rangle=\left(\prod\limits_{k=M_{1}+1}^{M-2}Z_{2}(k)\right)\left(\frac{1}{\sqrt{2}}(\alpha_{M-1}^{(2)}c_{d_{M-1},B}^{\dagger}-\beta_{M-1}^{(2)}c_{N+1-d_{M-1},A}^{\dagger})\right)\tilde{Z}_{2}(M)\left(\prod\limits_{i=1}^{\frac{1}{2}(N-M)}\prod\limits_{j=1}^{M_{1}}\tilde{Z}_{3}(i)\tilde{Z}_{2}(j)\tilde{Z}_{1}(i)\right)|0\rangle
\end{split}\end{equation}

\noindent or,

\begin{equation}\begin{split}\label{eq:proof41}
\left(\prod\limits_{k=M_{1}+1}^{M-2}Z_{2}(k)\right)\tilde{Z}_{2}(M-1)\tilde{Z}_{2}(M)\left(\prod\limits_{i=1}^{\frac{1}{2}(N-M)}\prod\limits_{j=1}^{M_{1}}\tilde{Z}_{3}(i)\tilde{Z}_{2}(j)\tilde{Z}_{1}(i)\right)|0\rangle
\end{split}\end{equation}

\noindent This is the first inductive step. For the $m^{\text{th}}$ step of the inductive argument, we now assume the following expression holds,

\begin{equation}\begin{split}\label{eq:proof42}
\left(\prod\limits_{k=M_{1}+1}^{M-m}Z_{2}(k)\right)\left(\prod\limits_{l=M-m+1}^{M}\tilde{Z}_{2}(l)\right)\left(\prod\limits_{i=1}^{\frac{1}{2}(N-M)}\prod\limits_{j=1}^{M_{1}}\tilde{Z}_{3}(i)\tilde{Z}_{2}(j)\tilde{Z}_{1}(i)\right)|0\rangle
\end{split}\end{equation}

\noindent and we must now show that this implies the $(m+1)^{\text{th}}$ case, namely that

\begin{equation}\begin{split}\label{eq:proof43}
\left(\prod\limits_{k=M_{1}+1}^{M-m}Z_{2}(k)\right)\left(\prod\limits_{l=M-m+1}^{M}\tilde{Z}_{2}(l)\right)\left(\prod\limits_{i=1}^{\frac{1}{2}(N-M)}\prod\limits_{j=1}^{M_{1}}\tilde{Z}_{3}(i)\tilde{Z}_{2}(j)\tilde{Z}_{1}(i)\right)|0\rangle
\end{split}\end{equation}

\noindent To illustrate this, we again write,

\begin{equation}\begin{split}\label{eq:proof44}
\left(\prod\limits_{k=M_{1}+1}^{M-m-1}Z_{2}(k)\right)\left(Z_{2}(M-m)\right)\left(\prod\limits_{l=M-m+1}^{M}\tilde{Z}_{2}(l)\right)\left(\prod\limits_{i=1}^{\frac{1}{2}(N-M)}\prod\limits_{j=1}^{M_{1}}\tilde{Z}_{3}(i)\tilde{Z}_{2}(j)\tilde{Z}_{1}(i)\right)|0\rangle
\end{split}\end{equation}

\noindent Recall that the singlets have been organized according to length and the singlets in the product of $Z_{2}$'s are formed over the second inversion center. Hence, each interval $[d_{j},N+1-d_{j}]$ lies within the successive interval $[d_{j+1},N+1-d_{j+1}]$ for $M-m\leq j\leq M$. Therefore, both operators $\tilde{K}(N+1-d_{M-m},0)$ and $\tilde{K}(d_{M-m},1)$ contains \textit{only one} of the number operators associated with \textit{one of} the creation operators in \textit{each factor} of $\prod\limits_{l=M-m+1}^{M}\tilde{Z}_{2}(l)$. Furthermore, one, and only one, of the operators $\tilde{K}(N+1-d_{M-m},0)$ and $\tilde{K}(d_{M-m},1)$ contain all of the number operators associated with all of the creation operators in $\prod\limits_{i=1}^{\frac{1}{2}(N-M)}\prod\limits_{j=1}^{M_{1}}\tilde{Z}_{3}(i)\tilde{Z}_{2}(j)\tilde{Z}_{1}(i)$. Therefore, we get the following desired result,

\begin{equation}\begin{split}\label{eq:proof45}
\left(\prod\limits_{k=M_{1}+1}^{M-m-1}Z_{2}(k)\right)\left(\frac{1}{\sqrt{2}}(\alpha_{M-m}^{(2)}c_{d_{M-m},B}^{\dagger}-\beta_{M-m}^{(2)}c_{N+1-d_{M-m},A}^{\dagger})\right)\left(\prod\limits_{l=M-m+1}^{M}\tilde{Z}_{2}(l)\right)\left(\prod\limits_{i=1}^{\frac{1}{2}(N-M)}\prod\limits_{j=1}^{M_{1}}\tilde{Z}_{3}(i)\tilde{Z}_{2}(j)\tilde{Z}_{1}(i)\right)|0\rangle
\end{split}\end{equation}

\noindent which gives (\ref{eq:proof43}). As a consequence, by successive application of the procedure, we can finally arrive at the final result,

\begin{equation}\begin{split}\label{eq:proof46}
|\Omega\rangle=\prod\limits_{k=M_{1}+1}^{M}\prod\limits_{i=1}^{\frac{1}{2}(N-M)}\prod\limits_{j=1}^{M_{1}}\tilde{Z}_{2}(k)\tilde{Z}_{3}(i)\tilde{Z}_{2}(j)\tilde{Z}_{1}(i)|0\rangle
\end{split}\end{equation}

\noindent Combining the product of singlets over the two inversion centers back into one product, we obtain the final result of (\ref{eq:proof2}),

\begin{equation}\begin{split}\label{eq:proof47}
|\Omega\rangle=\prod\limits_{i=1}^{\frac{1}{2}(N-M)}\prod\limits_{j=1}^{M}\tilde{Z}_{3}(i)\tilde{Z}_{2}(j)\tilde{Z}_{1}(i)|0\rangle
\end{split}\end{equation}

\noindent Substituting the expressions for $\tilde{Z}_{1}(i)$, $\tilde{Z}_{2}(j)$ and $\tilde{Z}_{3}(i)$ given by (\ref{eq:tildeZ1})-(\ref{eq:tildeZ3}) into the above yields,

\begin{equation}\begin{split}\label{eq:proof48}
|\Omega\rangle=\prod\limits_{i=1}^{\frac{1}{2}(N-M)}\prod\limits_{j=1}^{M}\left(\frac{1}{\sqrt{2}}(\alpha_{i}^{(3)}c_{N+1-d_{i2},B}^{\dagger}-\beta_{i}^{(3)}c_{N+1-d_{i1},A}^{\dagger})\right)\left(\frac{1}{\sqrt{2}}(\alpha_{j}^{(2)}c_{d_{j},B}^{\dagger}-\beta_{j}^{(2)}c_{N+1-d_{j},A}^{\dagger})\right)\\
\times\left(\frac{1}{\sqrt{2}}(\alpha_{i}^{(1)}c_{d_{i1},B}^{\dagger}-\beta_{i}^{(1)}c_{d_{i2},A}^{\dagger})\right)|0\rangle
\end{split}\end{equation}

\noindent As (\ref{eq:proof48}) shows, the terms appearing in the product in the ground state are all singlets of the same type. So the different terms in the product appearing in the expression for the ground state can simply be consolidated into just one product involving some redefinition of notation,

\begin{equation}\begin{split}\label{eq:suppsimplifiedgroundstate}
|\Omega\rangle=\prod\limits_{i=1}^{N}\left(\frac{1}{\sqrt{2}}(\alpha_{i}c_{d_{i1},B}^{\dagger}-\beta_{i}c_{d_{i2},A}^{\dagger})\right)|0\rangle
\end{split}\end{equation}

\noindent with the following restrictions (note that the $i^{\text{th}}$ singlet is defined by its interval $[d_{i1},d_{i2}]$):\\

\noindent (1) The singlets formed across the inversion centers are labeled by the index $i$ where $1\leq i\leq M$. The numbers $d_{i1}$ and $d_{i2}$ are expressed in terms of a single number $d_{i}$ such that $d_{i1}\equiv d_{i}$ and $d_{i2}=N+1-d_{i}$.\\

\noindent (2) The singlets formed in pairs away from the inversion centers are labeled by the index $i$ where $M+1\leq i\leq N$. An inversion-symmetric pair of singlets consists of one singlet between sites $d_{i1}$ and $d_{i2}$ and separately, another singlet between sites $N+1-d_{i2}$ and $N+1-d_{i1}$.\\

\noindent (3) In general, no two intervals $[d_{i1},d_{i2}]$ and $[d_{j1},d_{j2}]$ with $i\neq j$ can overlap in such a way that \textit{only one} of the ends of one interval is contained in the other. (Otherwise, this violates the non-crossing nature of the singlets.)\\

\noindent In total, there are $N$ singlets formed from the decimation process, but these restrictions reduce this down to $\frac{1}{2}(N-M)+M=\frac{1}{2}(N+M)$ unique states.\\

\noindent In the form given by (\ref{eq:suppsimplifiedgroundstate}), it is clear that this form of the ground state is a Slater determinant constructed from single-particle states,

\begin{equation}\begin{split}\label{eq:suppsingleparticlestates}
|\psi_{i}\rangle=\frac{1}{\sqrt{2}}(\alpha_{i}c_{d_{i1},B}^{\dagger}-\beta_{i}c_{d_{i2},A}^{\dagger})|0\rangle
\end{split}\end{equation}

\noindent The inversion topological invariant given the choice of inversion center being bond centered (i.e., $\mathcal{X}=x+\rho$ where $\rho=\frac{1}{2}$) can be computed as follows,

\begin{equation}\begin{split}
\Delta_{\mathcal{X}=\frac{N}{2}+\frac{1}{2}}=\text{Tr}[P_{\text{occ}}I_{\mathcal{X}=\frac{N}{2}+\frac{1}{2}}P_{\text{occ}}]
\end{split}\end{equation}

\noindent where $P_{\text{occ}}$ is the projector constructed from the occupied single-particle states given by (\ref{eq:suppsingleparticlestates}). The inversion operator $I_{\mathcal{X}=\frac{N}{2}+\frac{1}{2}}$ is expressed as follows,

\begin{equation}\begin{split}
I_{\mathcal{X}=\frac{N}{2}+\frac{1}{2}}=\sum\limits_{n=1}^{N}[|N+1-n,A\rangle\langle n,B|+|N+1-n,B\rangle\langle n,A|]
\end{split}\end{equation}

\noindent The diagonal part of $P_{\text{occ}}I_{\mathcal{X}=\frac{N}{2}+\frac{1}{2}}P_{\text{occ}}$ is given as follows,

\begin{equation}\begin{split}
\text{Diag}[P_{\text{occ}}I_{\mathcal{X}=\frac{N}{2}+\frac{1}{2}}P_{\text{occ}}]=-\frac{1}{4}\sum\limits_{i,j=1}^{N}\sum\limits_{n=1}^{N}[\alpha_{i}\beta_{i}^*\delta_{d_{i2},N+1-n}\delta_{d_{j1},n}|d_{i1},B\rangle\langle d_{j1},B|+\alpha_{j}\beta_{j}^*\delta_{d_{i2},N+1-n}\delta_{d_{j1},n}|d_{i2},A\rangle\langle d_{j2},A|\\
+\beta_{i}\alpha_{i}^*\delta_{d_{i1},N+1-n}\delta_{d_{j2},n}|d_{i2},A\rangle\langle d_{j2},A|+\beta_{j}\alpha_{j}^*\delta_{d_{i1},N+1-n}\delta_{d_{j2},n}|d_{i1},B\rangle\langle d_{j1},B|]
\end{split}\end{equation}

\noindent We consider only the diagonal part since that is what contributes to the inversion topological invariant when computing the trace. Performing the trace gives,

\begin{equation}\begin{split}\label{eq:RGinversioninvariant1}
\Delta_{\mathcal{X}=\frac{N}{2}+\frac{1}{2}}=\text{Tr}[P_{\text{occ}}I_{\mathcal{X}=x+\frac{1}{2}}P_{\text{occ}}]=-\frac{1}{2}\sum\limits_{i=1}^{N}\sum\limits_{n=1}^{N}[\alpha_{i}\beta_{i}^*\delta_{n,N+1-d_{i2}}\delta_{n,d_{i1}}+\beta_{i}\alpha_{i}^*\delta_{n,N+1-d_{i1}}\delta_{n,d_{i2}}]
\end{split}\end{equation}

\noindent Given the restrictions laid out in the previous section, we split the sum over $i$ into two sums in (\ref{eq:RGinversioninvariant1}),

\begin{equation}\begin{split}\label{eq:RGinversioninvariant2}
\Delta_{\mathcal{X}=\frac{N}{2}+\frac{1}{2}}=-\frac{1}{2}\sum\limits_{i=1}^{M}\sum\limits_{n=1}^{N}[\alpha_{i}\beta_{i}^*\delta_{n,d_{i}}\delta_{n,d_{i}}+\beta_{i}\alpha_{i}^*\delta_{n,N+1-d_{i}}\delta_{n,N+1-d_{i}}]\\
-\frac{1}{2}\sum\limits_{i=M+1}^{N}\sum\limits_{n=1}^{N}[\alpha_{i}\beta_{i}^*\delta_{n,N+1-d_{i2}}\delta_{n,d_{i1}}+\beta_{i}\alpha_{i}^*\delta_{n,N+1-d_{i1}}\delta_{n,d_{i2}}]
\end{split}\end{equation}

\noindent Note that from the restrictions, for $M+1\leq i\leq N$, $d_{i2}\neq N+1-d_{i1}$. This means the sum from $M+1\leq i\leq N$ in the second term in (\ref{eq:RGinversioninvariant2}) will be equal to zero, so the only contribution to the sum is for $1\leq i\leq M$, which are the indices for the singlets formed across the inversion center. Therefore,

\begin{equation}\begin{split}\label{eq:RGinversioninvariant3}
\Delta_{\mathcal{X}=\frac{N}{2}+\frac{1}{2}}=-\frac{1}{2}\sum\limits_{i=1}^{M}[\alpha_{i}\beta_{i}^*+\beta_{i}\alpha_{i}^*]
\end{split}\end{equation}

\noindent When no disorder is present, $m_{i}=m$ and $t_{i}=t$ for all $1\leq i\leq N$. Consider the limit where $m=0$ and $t$ is arbitrary. There is one singlet across each inversion center, so $M=2$ and $d_{1}=\frac{N}{2}$ and $d_{2}=1$ (the ends are given by $N+1-d_{1}=\frac{N}{2}+1$ and $N+1-d_{2}=N$ respectively). Furthermore, $\alpha_{i}=\beta_{i}=1$ for $i=1,2$. Thus, (\ref{eq:RGinversioninvariant3}) reduces to $\Delta_{\mathcal{X}=\frac{N}{2}+\frac{1}{2}}=-2$ as expected in this limit.\\

\noindent The polarization $P_{0}$ can be computed explicitly since the position operator is simply $X=\sum\limits_{j=1}^{N}[e^{\frac{2\pi i}{N}j}(|j,A\rangle\langle j,A|+|j,B\rangle\langle j,B|)]$ since the system has periodic boundary conditions. The projected position operator $X_{P}=P_{\text{occ}}XP_{\text{occ}}$ is (in the thermodynamic limit $N\to\infty$),

\begin{equation}\begin{split}
X_{P}=\sum\limits_{i=1}^{N}\left[e^{\frac{i\pi}{N}(d_{i1}+d_{i2})}\left(\frac{1}{\sqrt{2}}(\alpha_{i}|d_{i1},B\rangle-\beta_{i}|d_{i2},A\rangle)\right)\left(\frac{1}{\sqrt{2}}(\alpha_{i}^*\langle d_{i1},B|-\beta_{i}^*\langle d_{i2},A|)\right)\right]=\sum\limits_{i=1}^{N}e^{\frac{i\pi}{N}(d_{i1}+d_{i2})}|\psi_{i}\rangle\langle\psi_{i}|
\end{split}\end{equation}

\noindent Hence, the eigenvalues of $X_{P}$ are simply the collection of singlet centers $\{\xi_{n}\}_{n=1}^{N}=\left\{\frac{1}{2}(d_{n1}+d_{n2})\right\}_{n=1}^{N}$ computed from $\frac{N}{2\pi}\text{Im}\log(X_{P})$. The polarization $P_{0}$ is taken to be the average over the singlet centers. Thus,

\begin{equation}\begin{split}
P_{0}=\frac{1}{N}\sum\limits_{n=1}^{N}\left(\frac{1}{2}(d_{n1}+d_{n2})\right)
\end{split}\end{equation}

\section*{SM F: Additional Plots for the Inversion Topological Invariant $\Delta_{\mathcal{X}=\frac{N}{2}}$}\label{suppsec:additionalplots}

\begin{figure*}[ht]
\includegraphics[scale=1.35]{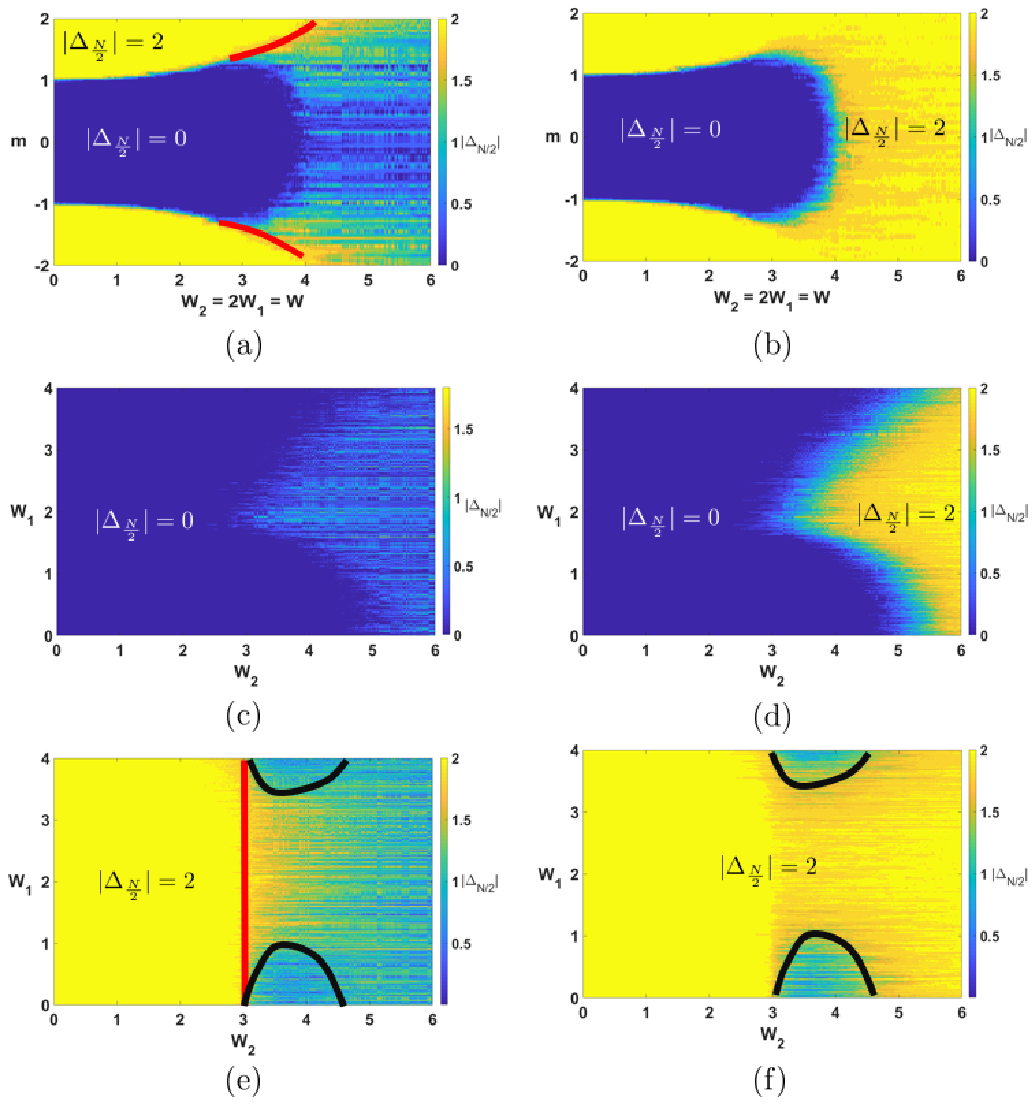}
\caption{Phase diagrams of the inversion topological invariant $\Delta_{\mathcal{X}=\frac{N}{2}}$. On the left, from top to bottom are the phase diagrams for when $\frac{W_{2}}{W_{1}}=2$ ((a)-(b)), $m=0.5$ ((c)-(d)), and $m=1.5$ ((e)-(f)). The inversion topological invariant begins to experience fluctuations past the closing of the disorder averaged spectral gap $|W_{2}|>|W_{2}^*|=2m$ (indicated by the bold red lines). When disorder is removed from the inversion centers for all disorder configurations (i.e., $\omega_{\frac{N}{2}}=\omega_{N}=0$), the inversion topological invariant is non-fluctuating past $|W_{2}^*|=2m$, which is shown on the right for each respective phase diagram. In the plots (e)-(f), the bold black curves indicate the phase boundaries corresponding to the divergence of the localization length. The first row of plots was constructed for $N=500$ sites while the last two rows of plots were for $N=400$ sites, and all of them were disorder averaged over $10$ configurations.}
\label{suppfig:numericalphasediagrams6}
\end{figure*}

\begin{figure*}[ht]
\includegraphics[scale=1.1]{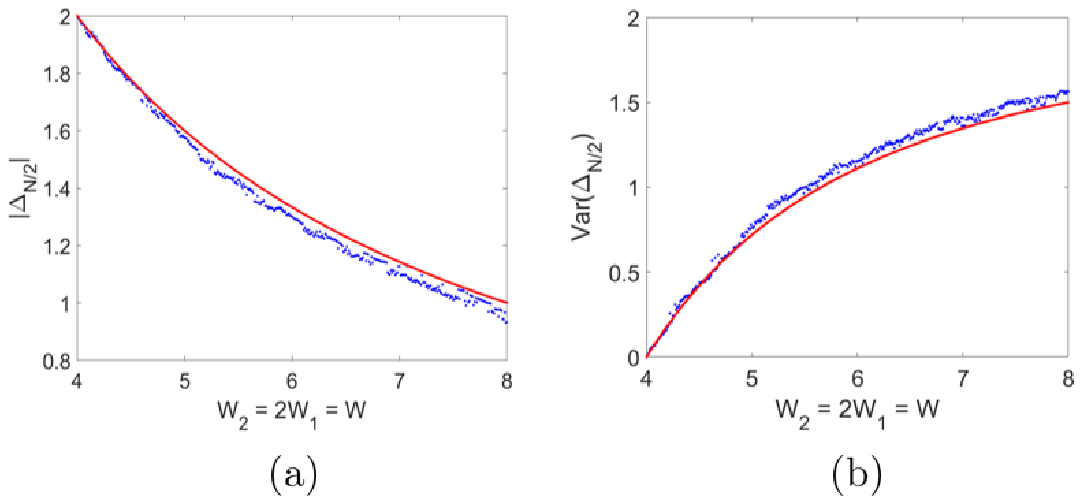}
\caption{Plots of the inversion topological invariant (a) $\Delta_{\mathcal{X}=\frac{N}{2}}$, and (b) the variance of $\Delta_{\mathcal{X}=\frac{N}{2}}$ past the disorder averaged spectral gap closing. All the plots were constructed for a chain of $N=300$ sites with $t=1$. Each blue dot on the top plot is the mean value of the inversion topological invariant at each disorder strength computed over $1000$ disorder configurations, and each blue dot on the bottom plot is the variance of the inversion topological invariant at each disorder strength also computed over $1000$ disorder configurations. The red curve in (a) is the plot of (\ref{eq:mean}) and the red curve in (b) is the plot of (\ref{eq:variance}) (with $\frac{W_{2}}{m}$ instead of $W_{1}$).}
\label{suppfig:variance2}
\end{figure*}

\begin{figure*}[ht]
\includegraphics[scale=1.1]{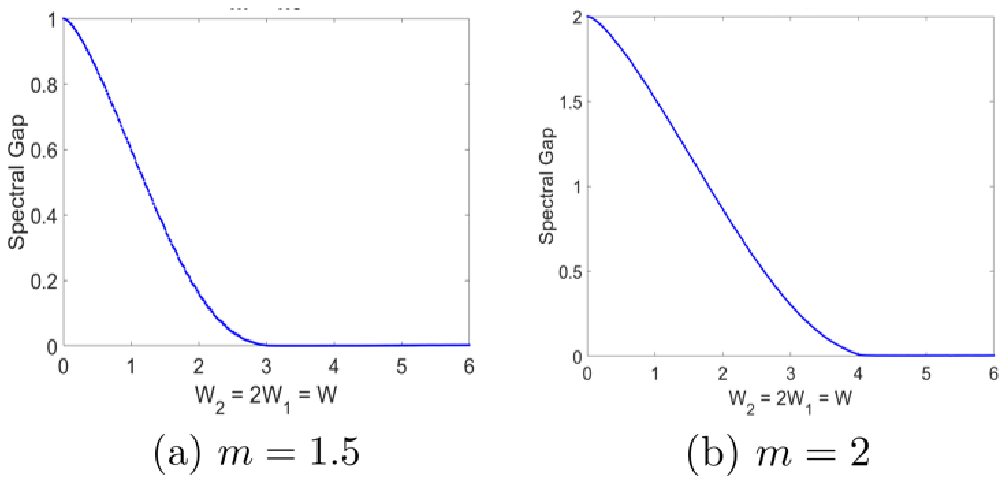}
\caption{Plots of the disorder averaged spectral gap for (a) $m=1.5$ and (b) $m=2$. The plots were constructed for a chain of $N=300$ sites with $t=1$, and disorder averaging was performed over $1000$ disorder configurations. For $m=1.5$ and $m=2$, the disorder averaged spectral gap vanishes at $W_{2}=2(1.5)=3$ and $W_{2}=2(2)=4$ respectively, which is consistent with (\ref{eq:disorderaveragedspectralgap2}).}
\label{suppfig:disorderaveragedspectralgap2}
\end{figure*}

\noindent In this section we provide the remaining phase diagrams for the inversion topological invariant $\Delta_{\mathcal{X}=\frac{N}{2}}$ in Fig. \ref{suppfig:numericalphasediagrams6}. In addition to this, we also show plots of the mean and variance of the fluctuations for $\Delta_{\mathcal{X}=\frac{N}{2}}$ for $|W_{2}|>|W_{2}^*|=2m$ shown in Fig. \ref{suppfig:variance2}. For the three cases considered in the main text, the results given by (\ref{eq:meanmain}) and (\ref{eq:variancemain}) are identical for the disorder configurations inversion symmetric about $\mathcal{X}=\frac{N}{2}$ in the dimerized limit $m\neq 0$, $t=0$ with $\frac{W_{2}}{W_{1}}=\pm\infty$ by replacing $W_{1}$ with $\frac{W_{2}}{m}$ as shown in SM B. In addition to this, we provide the plots of the disorder-averaged spectral gaps for $m=1.5$ and $m=2$ in Fig. \ref{suppfig:disorderaveragedspectralgap2} to corroborate the result that the spectral gap vanishes at $|W_{2}^*|=2m$ as determined by (\ref{eq:disorderaveragedspectralgap2}) in SM C. \\


\end{document}